\documentclass[11pt]{article}

\pdfoutput=1
\usepackage{graphics}
\usepackage{subfigure}
\usepackage{epsfig}
\usepackage{latexsym}
\usepackage{graphicx}
\usepackage{color}
\usepackage{amsmath,amssymb}
\usepackage{cite}
\usepackage{hyperref}
\usepackage{braket}
\usepackage{booktabs}
\usepackage{caption}
\usepackage{amsfonts}
\usepackage{dsfont}

\makeatletter
\def\section{\@startsection {section}{1}{\z@}{-3.5ex plus -1ex minus -.2ex}{2.3ex plus .2ex}{\large\bf}}
\def\subsection{\@startsection{subsection}{2}{\z@}{-3.25ex plus -1ex
minus -.2ex}{1.5ex plus .2ex}{\normalsize\bf}}
\makeatother
\newcommand{\captionfonts}{\small}
\makeatletter  
\long\def\@makecaption#1#2{%
  \vskip\abovecaptionskip
  \sbox\@tempboxa{{\captionfonts #1: #2}}%
  \ifdim \wd\@tempboxa >\hsize
    {\captionfonts #1: #2\par}
  \else
    \hbox to\hsize{\hfil\box\@tempboxa\hfil}%
  \fi
  \vskip\belowcaptionskip}
\makeatother   

\catcode`@=11
\def\marginnote#1{}
\newcount\hour
\newcount\minute
\newtoks\amorpm
\hour=\time\divide\hour
by60
\minute=\time{\multiply\hour by60 \global\advance\minute
by-\hour}
\edef\standardtime{{\ifnum\hour<12 \global\amorpm={am}
\else\global\amorpm={pm}\advance\hour by-12 \fi
 \ifnum\hour=0
\hour=12 \fi
 \number\hour:\ifnum\minute<10
0\fi\number\minute\the\amorpm}}
\edef\militarytime{\number\hour:\ifnum\minute<10
0\fi\number\minute}
\def\draftlabel#1{{\@bsphack\if@filesw
{\let\thepage\relax
 \xdef\@gtempa{\write\@auxout{\string
\newlabel{#1}{{\@currentlabel}{\thepage}}}}}\@gtempa
 \if@nobreak
\ifvmode\nobreak\fi\fi\fi\@esphack}
\gdef\@eqnlabel{#1}}
\def\@eqnlabel{}
\def\@vacuum{}
\def\draftmarginnote#1{\marginpar{\raggedright\scriptsize\tt#1}}
\def\draft{\oddsidemargin
0.0truein
 \def\@oddfoot{\sl preliminary draft \hfil
\rm\thepage\hfil\sl\today\quad\militarytime}
 \let\@evenfoot\@oddfoot
\overfullrule 3pt
 \let\label=\draftlabel
\let\marginnote=\draftmarginnote
\def\@eqnnum{(\theequation)\rlap{\kern\marginparsep\tt\@eqnlabel}
\global\let\@eqnlabel\@vacuum}
}
\catcode`@=12

\newcommand{\beq}{\begin{eqnarray}}
\newcommand{\eeq}{\end{eqnarray}}

\newcommand{\gsim}{\raisebox{-0.13cm}{~\shortstack{$>$ \\[-0.07cm]
     $\sim$}}~}
\newcommand{\lsim}{\raisebox{-0.13cm}{~\shortstack{$<$ \\[-0.07cm]
     $\sim$}}~}
\newcommand{\s}{\newline \vspace*{-3.5mm}}

\newcommand{\treelevel}{tree-level}

\begin{document}

\thispagestyle{empty}

\begin{center}
\hfill KA-TP-41-2016 \\

\begin{center}

\vspace{1.7cm}

{\LARGE\bf Strong First Order Electroweak Phase Transition \\[0.2cm]
  in the CP-Conserving 2HDM Revisited}
\end{center}

\vspace{1.4cm}

\renewcommand{\thefootnote}{\fnsymbol{footnote}}
{\bf P.~Basler$^{\,1\,}$}\footnote{E-mail: \texttt{philipp.basler@kit.edu}}, {\bf M.~Krause$^{\,1\,}$}\footnote{E-mail: \texttt{marcel.krause@kit.edu}},{\bf M.~M\"uhlleitner$^{\,1}\,$}\footnote{E-mail: \texttt{milada.muehlleitner@kit.edu}}, {\bf J.~Wittbrodt$^{\,1,2\,}$}\footnote{E-mail: \texttt{jonas.wittbrodt@desy.de}} and {\bf A.~Wlotzka$^{\,1\,}$}\footnote{E-mail: \texttt{alexander.wlotzka@kit.edu}}\\

\vspace{1.2cm}

${}^1\!\!$
{\em Institute for Theoretical Physics, Karlsruhe Institute of
    Technology,} \\
{\em Wolfgang-Gaede-Str.~1, 76131 Karlsruhe, Germany}\\[3mm]
${}^2\!\!$
{\em Deutsches Elektronen-Synchrotron DESY, Notkestra{\ss}e 85, D-22607
Hamburg, Germany}
\\

\end{center}

\vspace{1.8cm}
\centerline{\bf Abstract}
\vspace{2 mm}
\begin{quote}
\small The discovery of the Higgs boson by the LHC experiments ATLAS
and CMS has marked a milestone for particle physics. Yet, there are
still many open questions that cannot be answered within the Standard
Model (SM). For example, the generation of the observed matter-antimatter
asymmetry in the universe through baryogenesis can only be explained
qualitatively in the SM. A simple extension of the SM compatible with the current
theoretical and experimental constraints is given by the 2-Higgs-Doublet
Model (2HDM) where a second Higgs doublet is added to the Higgs
sector. We investigate the possibility of a strong first order
electroweak phase transition in the CP-conserving 2HDM type I and type
II where either of the CP-even Higgs bosons is identified with the SM-like Higgs
boson. The renormalisation that we apply on the loop-corrected Higgs
potential allows us to efficiently scan the 2HDM parameter space and
simultaneously take into account all relevant theoretical and
up-to-date experimental constraints. The 2HDM parameter
regions found to be compatible with the applied constraints and a
strong electroweak phase transition are analysed systematically. Our
results show that there is a strong interplay between the
requirement of a strong phase transition and collider phenomenology
with testable implications for searches at the LHC.
\end{quote}

\newpage
\setcounter{page}{1}
\setcounter{footnote}{0}

\section{Introduction}
In 2012 the LHC experiments ATLAS and CMS announced the discovery of
the long-sought Higgs boson \cite{:2012gk,:2012gu}. Although it looks
very SM-like
\cite{Aad:2015mxa,Khachatryan:2014kca,Aad:2015gba,Khachatryan:2014jba}
it is quite possible that it is the scalar particle of a Higgs sector
beyond the SM (BSM). Despite the success of the SM, which
has been tested to highest precision at previous and current
colliders, there are still a lot of open questions that cannot be
answered within the SM and call for new physics (NP) extensions. One of the unanswered
problems is the origin of the observed matter-antimatter
asymmetry of the universe \cite{Bennett:2012zja}.
Electroweak (EW) baryogenesis is an
elegant mechanism to explain this asymmetry
\cite{Kuzmin:1985mm,Cohen:1990it,Cohen:1993nk,Quiros:1994dr,Rubakov:1996vz,Funakubo:1996dw,Trodden:1998ym,Bernreuther:2002uj,Morrissey:2012db},
which is related to physics at the weak scale, establishing a link
between collider phenomenology and cosmology. The asymmetry can be
generated provided the EW phase transition (PT) taking place in the early universe
is of strong first order \cite{Trodden:1998ym,Morrissey:2012db} and that
all three Sakharov conditions \cite{sakharov} are fulfilled, namely
baryon number violation, C and CP violation and departure from the
thermal equilibrium. The strong first order PT,
proceeding through bubble formation, suppresses
the baryon number violating sphaleron transitions in the false vacuum
\cite{Manton:1983nd,Klinkhamer:1984di}. CP-violating reflections of
top quarks from the bubble wall produce a hypercharge asymmetry which
is converted into a baryon asymmetry in the false vacuum. This
asymmetry  is transferred to the true vacuum when it passes the bubble wall
\cite{Konstandin:2013caa}, provided there is departure from the
thermal equilibrium.  Although in the SM all three Sakharov conditions
are fulfilled, the electroweak PT is not of first order
\cite{smnot}. Not only the Higgs boson mass
is too large \cite{notsm2}, but in addition the CP violation of the SM
from the Cabibbo-Kobayashi-Maskawa matrix is too small
\cite{Morrissey:2012db,Konstandin:2013caa,amountcpviol}. This calls
for physics BSM. Among the plethora of NP extensions the
2HDM \cite{Lee:1973iz,Branco:2011iw} belongs
to the simplest models that are in
accordance with present experimental constraints. Its Higgs sector features five
physical Higgs bosons, three neutral and two charged
ones. Their contributions to the effective Higgs
potential can strengthen the PT and in addition introduce new sources
of CP violation. Previous studies have shown that
2HDMs provide a good framework for successful baryogenesis
\cite{baryo2HDM,Dorsch:2013wja,Dorsch:2014qja} (see
\cite{Cline:1996mga,huberwithcp,Cline:2011mm,Dorsch:2016nrg,Haarr:2016qzq}
for studies in the CP-violating 2HDM). \s

In this work we will investigate the implications of a strong first
order PT required by baryogenesis on the LHC Higgs phenomenology in
the framework of the CP-conserving 2HDM. For this purpose we compute
the one-loop
corrected effective potential at finite temperature
\cite{ColemanWeinberg,Quiros:1999jp,Dolan:1973qd} including daisy
resummations for the bosonic masses \cite{Carrington:1991hz}
in two different approximations for the treatment of the thermal
masses \cite{Arnold:1992rz,Parwani:1991gq}. The renormalisation of the
loop-corrected potential is chosen such that not only the vacuum expectation
value (VEV) and all physical Higgs boson masses, but, for the first
time, also all mixing matrix elements remain at their tree-level
values. This allows to efficiently scan the whole 2HDM
parameter space with the tree-level masses and mixing angles as input
and at the same time test the compatibility of the model with the
theoretical and experimental constraints. The points
passing these tests will be investigated with respect to a first order
PT. The loop-corrected Higgs potential will be
minimised at increasing non-zero temperature to find the vacuum expectation value
$v_c$ at the critical temperature $T_c$, defined as the temperature
where two degenerate global minima exist. A value of
$v_c/T_c$\footnote{For discussions on the gauge dependence of
  $v_c/T_c$, see
  \cite{Dolan:1973qd,Patel:2011th,Wainwright:2011qy,Garny:2012cg}.}
larger than one is indicative for a strong first order PT
\cite{Quiros:1994dr,Moore:1998swa}.
In our analysis we will discard points leading to a
2-stage PT \cite{Land:1992sm,Hammerschmitt:1994fn}.
We will perform a systematic and comprehensive investigation of the
2HDM in four configurations, given by the 2HDM type I and type II where
either the lighter or the heavier of the two CP-even Higgs bosons is
identified with the SM-like Higgs boson. We will test the
compatibility of the model with both the experimental constraints and a strong
EW phase transition. The thus delineated regions in the
parameter space will be further investigated with respect to their
implications for collider phenomenology. We find that the link between cosmology
and high-energy collider constraints provides a powerful tool to
further constrain the allowed parameter regions of the 2HDM. At the same
time, the demand for a strong first order PT leads to testable
consequences at the collider experiments. \s

The outline of the paper is as follows: In section
\ref{sec:model_and_effective_pot} we introduce our notation and
provide the loop-corrected effective potential at non-vanishing temperature. In the
subsequent section \ref{sec:renorm} we describe in detail the
renormalisation procedure, which is chosen such that at zero
temperature the tree-level position of the minimum and the masses and
mixing matrix elements of the scalar
particles are preserved by the one-loop potential. Using the Higgs
boson masses and mixing angles as input parameters, this simplifies the
verification of the compatibility of the model with the Higgs data. In section
\ref{sec:numerical} the basic elements of our numerical analysis are
described, namely the minimisation procedure of the effective
potential in \ref{sec:minimization} and, in \ref{sec:scan}, the details
of the scan in the 2HDM parameter space together with the applied
theoretical and experimental constraints. Section \ref{sec:results}  is
devoted to our results. We present the parameter
regions compatible with the applied constraints and a strong first
order PT, and we then analyse the implications for collider
phenomenology. We end in section \ref{sec:conclusions}
with our conclusions. The paper is accompanied by an appendix
containing the formulae for the masses of the relevant
particles and, where appropriate, for the daisy resummed mass
corrections.

\section{The Effective Potential}\label{sec:model_and_effective_pot}
In this section we provide the loop-corrected effective potential of
the CP-conserving 2HDM for non-vanishing temperature. First, we set
our notation by introducing the model under investigation.

\subsection{The CP-conserving 2-Higgs-Doublet Model\label{sec:2hdm}}
In terms of the two $SU(2)_L$ Higgs doublets $\Phi_1$ and $\Phi_2$,
\beq
\Phi_1 = \left( \begin{array}{c} \phi_1^+ \\ \phi_1^0 \end{array} \right)
\quad \mbox{and} \quad
\Phi_1 = \left( \begin{array}{c} \phi_2^+ \\ \phi_2^0 \end{array} \right) \;,
\eeq
the tree-level potential of the 2HDM with a softly broken
$\mathbb{Z}_2$ symmetry, under which the doublets transform as
$\Phi_1\rightarrow \Phi_1,\ \Phi_2 \rightarrow -\Phi_2$, reads
\begin{align}
\begin{split}
V_{\text{tree}} &= m_{11}^2 \Phi_1^\dagger \Phi_1 + m_{22}^2
\Phi_2^\dagger \Phi_2 - \left[m_{12}^2 \Phi_1^\dagger \Phi_2 +
  \mathrm{h.c.} \right] + \frac{1}{2} \lambda_1 ( \Phi_1^\dagger
\Phi_1)^2 +\frac{1}{2} \lambda_2 (\Phi_2^\dagger \Phi_2)^2 \\
&\quad + \lambda_3 (\Phi_1^\dagger \Phi_1)(\Phi_2^\dagger\Phi_2) +
\lambda_4 (\Phi_1^\dagger \Phi_2)(\Phi_2^\dagger \Phi_1)
+ \left[ \frac{1}{2} \lambda_5 (\Phi_1^\dagger\Phi_2)^2  + \mathrm{h.c.} \right] \ .
\end{split}\label{eq:model_1}
\end{align}
The mass parameters $m_{11}^2$ and $m_{22}^2$ and the couplings
$\lambda_{1..4}$ are real parameters of the model. The mass and coupling
parameters $m_{12}^2$ and $\lambda_5$ can in general be complex, thereby
offering new sources of explicit CP violation in the Higgs sector. We
take them to be real as we work in the CP-conserving 2HDM.
After EW symmetry breaking the
two Higgs doublets acquire VEVs $\bar{\omega}_i
\in \mathbb{R}$ ($i=1,2,3$), about which the Higgs fields can be expanded
in terms of the charged CP-even and CP-odd fields $\rho_i$ and
$\eta_i$, and the neutral CP-even and CP-odd fields $\zeta_i$
and $\psi_i$, $i=1,2$,
\begin{align}
\Phi_1 & = \frac{1}{\sqrt{2}} \begin{pmatrix}
\rho_1 + \mathrm{i}\eta_1 \\ \bar{\omega}_1 + \zeta_1 +
\mathrm{i}\psi_1 \label{eq:model_3}
\end{pmatrix} \\
\Phi_2 & = \frac{1}{\sqrt{2}} \begin{pmatrix}
\rho_2 + \mathrm{i} \eta_2 \\ \bar{\omega}_2 +\mathrm{i} \bar{\omega}_3
+ \zeta_2 + \mathrm{i} \psi_2
\end{pmatrix} \,, \label{eq:model_4}
\end{align}
where, without loss of generality, the complex part of the VEVs has
been rotated to the second doublet
exclusively. Denoting the VEVs of our present vacuum at zero
temperature by\footnote{Strictly speaking, $T=2.7$~K. Setting $T=0$
  does not make a difference numerically.}
\beq
v_i \equiv \bar{\omega}_i|_{T=0} \;,
\eeq
we set
\beq
v_3 = 0 \;,
\eeq
whereas the remaining two VEVs are related to the SM VEV by
\beq
v_1^2 + v_2^2 \equiv v^2 \;.
\eeq
Introducing the angle $\beta$ through
\beq
\tan\beta = \frac{v_2}{v_1} \;,
\eeq
we have
\beq
v_1 = v \cos \beta \qquad \mbox{and} \qquad
v_2 = v \sin\beta\ . \label{eq:model_6}
\eeq
Through the complex part of the VEV, $\bar{\omega}_3$, we include the
possibility of generating at one-loop and/or non-zero temperature a
global minimum that is CP-violating.\footnote{In the 2HDM we can have
  three different types of minima: the normal EW breaking one, a
CP-breaking minimum, and a charge-breaking (CB) vacuum. It has been shown
that, at tree level, minima which break different symmetries cannot
coexist \cite{Ferreira:2004yd,Barroso:2007rr,Ivanov:2007de}. This means that,
if a normal minimum exists,
all CP or CB stationary points are proven to be saddle points. Recent
studies of the Inert 2HDM at one-loop level \cite{Ferreira:2015pfi},
which apply the effective potential
approach, indicate that these statements may not be true any more once
higher order corrections are included. We therefore allow for the
possibility of a CP-breaking vacuum.
Including the possibility of a charge breaking Higgs VEV makes the
present analysis considerably more complex.} 
The angle $\beta$ coincides with the angle of the rotation matrix
\begin{equation} \label{eq:model_alex1}
 R(\beta) =
 \begin{pmatrix}
  \cos\beta & \sin\beta \\
  -\sin\beta & \cos\beta
 \end{pmatrix}
\end{equation}
from
the gauge to the mass eigenstates of the charged Higgs sector, and also of
the neutral CP-odd sector. The physical states of the charged sector
are given by the charged Higgs
bosons $H^{\pm}$ with mass $m_{H^\pm}$ and the charged Goldstone
bosons $G^{\pm}$ which are massless at zero temperature,
\begin{equation}\label{eq:model_alex2}
\begin{pmatrix}
 G^\pm \\ H^\pm
\end{pmatrix}
= R(\beta)
\begin{pmatrix}
 \phi_1^\pm \\
 \phi_2^\pm
\end{pmatrix}\ .
\end{equation}
For the neutral CP-odd fields $\psi_1$ and $\psi_2$ the same rotation yields
the physical states $A$ with mass
$m_A$ and the neutral Goldstone boson $G^0$, massless at zero
temperature,
\begin{equation}\label{eq:model_alex3}
\begin{pmatrix}
 G^0 \\ A
\end{pmatrix}
= R(\beta)
\begin{pmatrix}
 \psi_1 \\
 \psi_2
\end{pmatrix}\ .
\end{equation}
Finally, in the neutral CP-even sector the rotation with the angle $\alpha$
transforms the fields $\zeta_1$ and $\zeta_2$ into the two physical
CP-even Higgs bosons $H$ and $h$ with masses $m_H$ and $m_h$, respectively,
\begin{equation}\label{eq:model_alex_4}
\begin{pmatrix}
 H \\ h
\end{pmatrix}
= R(\alpha)
\begin{pmatrix}
 \zeta_1 \\
 \zeta_2
\end{pmatrix}\ .
\end{equation}
In the minimum of the potential Eq.~\eqref{eq:model_1} the following minimum conditions have to be fulfilled,
\begin{align}
\left. \frac{\partial V_{\text{tree}}}{\partial \Phi_a^\dagger}\right|_{\Phi_i =\langle\Phi_i\rangle } &\overset{!}{=} 0 \qquad a,i\in\{1,2\}\ ,  \label{eq:model_12}
\end{align}
with the brackets denoting the Higgs field values in the minimum, {\it
i.e.}~$\langle\Phi_i\rangle = (0, v_i/\sqrt{2})$ at $T=0$.
This results in two equations
\begin{subequations}\label{eq:model_13}
\begin{align}
		m_{11}^2 & =  m_{12}^2 \frac{v_2}{v_1} - \frac{v_1^2}{2} \lambda_1 - \frac{v_2^2}{2} \left( \lambda_3 + \lambda_4 + \lambda_5 \right)  \label{eq:model_13_1}\\
		m_{22}^2 & =  m_{12}^2 \frac{v_1}{v_2} - \frac{v_2^2}{2} \lambda_2 - \frac{v_1^2}{2} \left( \lambda_3+\lambda_4+\lambda_5 \right)\ . \label{eq:model_13_2}
\end{align}
\end{subequations}
Exploiting the minimum conditions of the potential at zero
temperature, we use the following set of independent parameters of the
model,
\begin{equation}
 m_{h},\ m_{H},\ m_{A},\ m_{H^{\pm}},\ m_{12}^2,\ \alpha,\ \tan\beta, \ v \ .\label{eq:model_16}
\end{equation}
Due to the imposed $\mathbb{Z}_2$ symmetry each of the up-type quarks,
down-type quarks and charged leptons can only couple to one of the
Higgs doublets so that  flavour-changing neutral currents at
tree level are avoided. The possible combinations of Yukawa couplings
of the Higgs bosons to up-type quarks, down-type quarks or charged leptons
are classified as type I, type II, lepton-specific and flipped and
are defined in Table~\ref{tab:model1}.
\begin{table}[b!]
 \begin{center}
 \begin{tabular}{lcccc}
     \toprule
    & Type I & Type II & Lepton-Specific & Flipped \\
   \midrule
Up-type quarks & $\Phi_2$ & $\Phi_2$ & $\Phi_2$ & $\Phi_2$ \\
Down-type quarks & $\Phi_2$ & $\Phi_1$ & $\Phi_2$ & $\Phi_1$ \\
Leptons & $\Phi_2$ & $\Phi_1$ & $\Phi_1$ & $\Phi_2$ \\
   \bottomrule
  \end{tabular}
\caption{Classification of the Yukawa sector in the 2HDM according to the couplings of the fermions to the Higgs doublets.\label{tab:model1}}
   \end{center}
 \end{table}
The resulting couplings of the fermions normalised to the SM couplings
can be found in \cite{Fontes:2015mea}.
In this work we focus on real 2HDMs of type I and type II.

\subsection{One-Loop Effective Potential at Finite Temperature}\label{sec:one_loop_pot}
The one-loop contributions $V_1$ to the effective potential consist of two
parts: the Coleman-Weinberg (CW) contribution $V_{\text{CW}}$
\cite{ColemanWeinberg} which is already present at zero temperature,
and the contribution $V_{T}$ accounting for the thermal
corrections at finite temperature $T$. The one-loop corrected
effective potential then reads
\begin{equation}
 V = V_{\text{tree}} + V_1
\equiv V_{\text{tree}} + V_{\text{CW}} + V_{T} \ . \label{eq:looppot_11}
\end{equation}
The tree-level potential is given in Eq.~(\ref{eq:model_1}) with the
doublet $\Phi_1$ replaced by the classical constant field
configuration $\Phi_1^c = (0, \omega_1/\sqrt{2})$ and $\Phi_2$ by
$\Phi_2^c = (0,(\omega_2 + i\omega_3)/\sqrt{2})$. The Coleman-Weinberg
potential in the $\overline{\text{MS}}$ scheme is
given by \cite{Quiros:1999jp}
\begin{equation}
 V_{\text{CW}} (\{\omega\})= \sum_i \frac{n_i}{64\pi^2} (-1)^{2s_i} \,
 m_i^4(\{\omega\}) \,
 \left[\log\left(\frac{m_i^2(\{\omega\})}{\mu^2}\right)-c_i\right]\, ,
\label{eq:looppot_1}
\end{equation}
where the sum extends over the Higgs and Goldstone bosons, the massive
gauge bosons, the longitudinal photon  
and the fermions, $i=h, H, A, H^\pm, G^0, G^\pm, W^\pm, Z, \gamma_L,
f$ ($f=e,\mu,\tau, u,c,t,d,s,b$)\footnote{Note, that we
  assume the neutrinos to be massless.}. The $m_i^2$
is the respective eigenvalue for the particle $i$ of the mass matrix squared
expressed through the tree-level relations in terms of
$\omega_i$ ($i=1,2,3$). The explicit formulae can be found in
App.~\ref{sec:app_debye_corrections}.
The sum also includes the Goldstone bosons. Although we work in the
Landau gauge, where they are massless at $T=0$, they can acquire a
mass if the mass eigenvalues are determined at field configurations
other than the tree-level VEVs at $T=0$, which is required in the
minimisation procedure. Moreover, due to temperature corrections specified below,
the masses of the Goldstones and the 
longitudinal photon can be non-zero, which enforces also the inclusion
of $\gamma_L$ in the sum.
Note, that due to the choice of the Landau gauge there are no ghost
contributions. The variable $s_i$ denotes the
spin of the particle, $n_i$ represents the number of degrees of
freedom. Also for later use, we define the degrees of freedom of all
particles involved in the model. These are the neutral scalars $\Phi^0\equiv
h,H,A, G^0$, the charged scalars $\Phi^\pm \equiv H^\pm, G^\pm$,
the leptons $l$, the quarks $q$ and the longitudinal and transversal
gauge bosons, $V_L \equiv Z_L, W_L, \gamma_L$ and $V_T \equiv Z_T,
W_T, \gamma_T$, with the respective $n_i$,
\beq
\begin{array}{llll}
n_{\Phi^0} = 1\;,& \quad n_{\Phi^\pm} = 2\;, & \quad n_l = 4\;, &\quad
n_q = 12\;, \\
n_{W_T} = 4 \;, & \quad n_{W_L} = 2\;,  & \quad n_{Z_T} = 2 \;, &
\quad n_{Z_L} = 1 \;, \\
n_{\gamma_T} = 2 \;, & \quad n_{\gamma_L} = 1 \;.
\end{array}
\eeq
In the $\overline{\text{MS}}$ scheme employed here, the constants $c_i$ read
\begin{equation}\label{eq:looppot_3}
 c_i= \begin{cases}
       \frac{5}{6}\ , & i=W^\pm,Z,\gamma\ \\
       \frac{3}{2}\ , & \text{otherwise}\ .
      \end{cases}
\end{equation}
We fix the renormalisation scale $\mu$ by $\mu=v= 246.22\
\text{GeV}$. \s

In the thermal corrections $V_T$ we include the daisy resummation
\cite{Carrington:1991hz}
of the $n=0$ Matsubara modes of the longitudinal components of the
gauge bosons $W_L, Z_L, \gamma_L$ and the bosons $\Phi^0, \Phi^\pm$,
which adds to their masses at non-zero temperature the Debye corrections given in
App.~\ref{sec:app_debye_corrections}.
The thermal contributions $V_T$ to the potential can be written as
\cite{Dolan:1973qd,Quiros:1999jp}
\beq
V^T = \sum_k n_k \, \frac{T^4}{2\pi^2} \, J^{(k)}_{\pm} \;.
\label{eq:looppot_4}
\eeq
The sum extends over $k = W_L, Z_L,
\gamma_L,W_T,Z_T,\Phi^0,\Phi^\pm,f$. Note, that the Goldstone
bosons and the longitudinal part of the photon, which are massless at $T=0$,
acquire a mass at finite temperature and are included in the
sum. Denoting the mass eigenvalue including the thermal corrections
for the particle $k$ by $\overline{m}_k$, $J_\pm^{(k)}$ is given by (see {\it
  e.g.}~\cite{Carena:2008vj})
\beq
J^{(k)}_{\pm} = \left\{
\begin{array}{ll}
J_- \left( \frac{m_k^2}{T^2} \right) - \frac{\pi}{6} \left(\frac{\overline{m}_k^3}{T^3} -
\frac{m_k^3}{T^3}\right) & k = W_L, Z_L, \gamma_L, \Phi^0, \Phi^\pm \\
J_- \left( \frac{m_k^2}{T^2} \right) & k = W_T, Z_T \\
J_+ \left( \frac{m_k^2}{T^2} \right) & k=f
\end{array}
\right.
\label{eq:thermalj}
\eeq
with the thermal integrals
\begin{equation}
 J_{\pm}\left(\frac{m_k^2}{T^2}\right) = \mp \int_0^\infty \text{d}x \, x^2 \log\left[1\pm e^{-\sqrt{x^2+m_k^2/T^2}}\right]\ ,\label{eq:looppot_5}
\end{equation}
where $J_+$ ($J_-$) applies for $k$ being a fermion (boson).
For each temperature $T$ we
determine the VEVs $\bar{\omega}_i$, {\it i.e.} the field
configurations $\{ \bar{\omega} \} \equiv \{\bar{\omega}_1,
\bar{\omega}_2, \bar{\omega}_3 \}$, that minimise the loop-corrected potential
$V$, Eq.~(\ref{eq:looppot_11}).
These enter the tree-level mass matrices such that the masses $m_i$
depend  implicitly on the temperature 
$T$ through $\bar{\omega}_i = \bar{\omega}_i (T)$. The
$\overline{m}_k$ furthermore depend explicitly on $T$ through the thermal
corrections. The definition of $J^{(k)}_{\pm}$ Eq.~(\ref{eq:thermalj})
is the approach chosen in \cite{Arnold:1992rz}. A different prescription for
implementing the thermal corrections is proposed by
\cite{Parwani:1991gq} where the
Debye corrections are included for all the bosonic thermal loop
contributions\footnote{For a discussion and comparison, see also \cite{Cline:1996mga,Cline:2011mm}.}, so that we have
\beq
J^{(k)}_{\pm} = \left \{
\begin{array}{ll}
J_- \left( \frac{\overline{m}_k^2}{T^2} \right) & k = W_L, Z_L, \gamma_L,
\Phi^0, \Phi^\pm \\
J_- \left( \frac{m_k^2}{T^2} \right) & k = W_T, Z_T \\
J_+ \left( \frac{m_k^2}{T^2} \right) & k=f\ .
\end{array}
\right.
\label{eq:thermalj2}
\eeq
In this case, the Debye corrected masses are also used in the CW
potential Eq.~\eqref{eq:looppot_1} \cite{Cline:2011mm}. 
We refer to the first approach,
Eq.(\ref{eq:thermalj}), as 'Arnold-Espinosa' and to the 
second one, {\it i.e.}~Eq.(\ref{eq:thermalj2})
together with $V_{\text{CW}}$ including the thermal
  corrections in the bosonic masses, as 'Parwani' method.
The two approaches differ in the organisation of the
  perturbative series and hence by higher order terms. The
  'Arnold-Espinosa' method consistently implements the
  thermal masses at one-loop level in the high-temperature expansion,
  leading to Eq.~(\ref{eq:thermalj}). The 'Parwani' method admixes
  higher-order contributions, which at one-loop level could lead to
  dangerous artefacts. 
  Therefore, in the discussion of our results we
  will apply the 'Arnold-Espinosa' method. The 'Parwani' method will be
  used only to make contact to previous results in the
  literature.
\s

Since in the minimisation procedure the
numerical evaluation of the integral Eq.~\eqref{eq:looppot_5} at each
configuration in $\{\omega\}$ and $T$ is very time consuming, the
integrals $J_{\pm}$ are approximated by a series in $x^2 \equiv m^2/T^2$. For
small $x^2$ we use \cite{Cline:1996mga}
 \begin{align}
 \begin{split} \label{eq:looppot_6}
J_{+,\text{s}}(x^2,n) & = -\frac{7\pi^4}{360} +
\frac{\pi^2}{24}x^2 +\frac{1}{32}x^4
\left(\log x^2-c_{+}\right) \\
& \quad - \pi^2x^2 \sum_{l=2}^{n}
\left(-\frac{1}{4\pi^2}x^2\right)^l \frac{(2l-3)!!
  \zeta(2l-1)}{(2l)!! (l+1)} \left(2^{2l-1}-1\right)
\end{split} \\
\begin{split}\label{eq:looppot_7}
J_{-,\text{s}}(x^2,n) & =  - \frac{\pi^4}{45} +
\frac{\pi^2}{12}x^2 -
\frac{\pi}{6}\left(x^2\right)^{3/2} -
\frac{1}{32}x^4
\left(\log x^2-c_{-} \right) \\
& \quad + \pi^2 x^2 \sum_{l=2}^{n}
\left(-\frac{1}{4\pi^2}x^2 \right)^l \frac{(2l-3)!!
  \zeta(2l-1)}{(2l)!! (l+1)}  \ ,
\end{split}
\end{align}
with
\beq
c_+= 3/2+2\log\pi -2\gamma_E \quad \mbox{and} \quad c_-=c_+ +2\log 4 \;,
\eeq
where $\gamma_E$ denotes the Euler-Mascheroni constant, $\zeta(x)$ the
Riemann $\zeta$-function and $(x)!!$ the double factorial. For large
$x^2$ the expansion for both fermions and bosons reads
\cite{Cline:1996mga}
\begin{equation}\label{eq:looppot_8}
 J_{\pm,\text{l}}(x^2,n) = -\exp\left(-\left(x^2\right)^{1/2}\right)
 \left(\frac{\pi}{2}\left(x^2\right)^{3/2} \right)^{1/2}
 \sum_{l=0}^{n} \frac{1}{2^l l!} \frac{\Gamma(5/2+l)}{\Gamma(5/2-l)}
 \left(x^2\right)^{-l/2} \ ,
\end{equation}
with $\Gamma(x)$ denoting the Euler Gamma function. In order to
interpolate between the two approximations, first the point is determined where the
derivatives of the low- and high-temperature expansions can be
connected continuously. We then add a small finite shift to the small
$x^2$ expansion such that also the two expansions themselves are
connected continuously. We denote the values of $x^2$ where the connection
  is performed by $x^2_+$ and $x^2_-$ and the corresponding shifts by
  $\delta_\pm$ for the fermionic and bosonic contributions,
  respectively. They are given by
\beq
\begin{array}{lcllcl}
x^2_+ & = & 2.2161\;, & \qquad \delta_+ & = & -0.015603 \;,
\\
x^2_- & = & 9.4692 \;, & \qquad \delta_- & = & 0.0063109
\;.
\end{array}
\label{eq:looppot_10}
\eeq
We find that for small $x^2$ the expansion $J_{+,\text{s}}$ for
fermions approximates the exact result well by including terms of up
to order $n=4$, while for bosons this is the case for $n=3$ in
$J_{-,\text{s}}$. For large $x^2$, the integral is well
approximated by $n=3$ in both the fermion and the boson case,
$J_{\pm,\text{l}}$. This way, the deviation of the approximate results from
the numerical evaluation of the integrals is less than two
percent. The above approximations
Eqs.~\eqref{eq:looppot_6}-\eqref{eq:looppot_8} are only valid for
$m^2 \geq 0$. For bosons this is not necessarily the case as the
eigenvalues of the mass matrix of the neutral Higgs bosons can become
negative depending on the configuration $\{\omega\}$
and the temperature in the minimisation procedure. If this happens,
the value of the integral $J_-$, given by Eq.~\eqref{eq:looppot_5}, is
set to the real part of its numerical evaluation which is the relevant
contribution when extracting the global minimum
\cite{WeinbergWu}\footnote{Note, that negative masses squared
  correspond to a negative curvature of the potential, implying a local maximum and
  not a minimum.}. In practice, we evaluated the integral numerically
at several equidistant points in $m^2/T^2 < 0$, and in the
minimisation procedure we use the result obtained from the linear
interpolation between these points, which leads to a significant
speed-up. We explicitly verified that the difference between the exact
and the interpolated result is negligible for a
sufficiently large range of $m^2/T^2$.

\section{Renormalisation \label{sec:renorm}}
The Coleman-Weinberg potential, Eq.~(\ref{eq:looppot_1}), in the
one-loop effective potential Eq.~(\ref{eq:looppot_11}) contributes
already at $T=0$, so that the masses and mixing angles obtained from
the one-loop effective potential differ from those extracted from the
tree-level potential Eq.~\eqref{eq:model_1}. The loop-corrected masses
obtained in this way correspond to the full one-loop corrected masses in
the  approximation of vanishing external momenta. When we test for
compatibility of the model with the experimental constraints the
loop-corrected masses and the loop-corrected mixing angles,
which enter the couplings, have to be taken into account. For an
efficient scan over the parameter space of the model in terms of the
input parameters Eq.~\eqref{eq:model_16}, however, it is more
convenient to have the one-loop masses and angles directly as inputs, {\it i.e.}~they
should be the same as the tree-level ones. This can be achieved by an
appropriate renormalisation prescription, which will be described in the
following. \s

The Coleman-Weinberg potential given in Eq.~\eqref{eq:looppot_1} has
already been renormalised in the $\overline{\text{MS}}$ scheme. We modify
this scheme by including finite terms in the counterterm potential
that ensure the one-loop corrected masses and, for the first time,
also the mixing matrix elements to be equal to the tree-level
ones.\footnote{Previous works included only the
VEVs and (subsets of) the masses in the renormalisation conditions and required
them to be equal to their tree-level values \cite{Cline:2011mm,Dorsch:2013wja,Dorsch:2014qja,Cline:1996mga,huberwithcp}. In models with extended
Higgs sectors, the mixing angles, which
enter all Higgs boson observables through the Higgs couplings, are
crucial for the interpretation of the results. They are determined
from the diagonalisation of the loop-corrected mass matrix. The
renormalisation of the mixing matrix elements to their tree-level
values guarantees that the relevant quantities and observables
constraining the model can be tested with the tree-level input parameters.}
Introducing counterterms for each of the parameters of the tree-level
potential Eq.~\eqref{eq:model_1}, the counterterm potential
$V_{\text{CT}}$ added to the one-loop effective potential
Eq.~(\ref{eq:looppot_11}),
\beq
\tilde{V}= V + V_{\text{CT}} = V_{\text{tree}} + V_{\text{CW}} + V_{T} + V_{\text{CT}} \;, \label{eq:totalpot}
\eeq
reads
\begin{align}
\begin{split}
	V_{\text{CT}} &= \delta m_{11}^2 \frac{\omega_1^2}{2} + \delta
        m_{22}^2 \frac{\omega_2^2+\omega_3^2}{2} - \delta m_{12}^2
        \, \omega_1\omega_2 +
        \frac{\delta \lambda_1}{8} \omega_1^4 + \frac{\delta
          \lambda_2}{8} \left(\omega_2^2+\omega_3^2\right)^2 \\ &\quad
	+
        \left(\delta\lambda_3+\delta\lambda_4\right)\frac{\omega_1^2
          \left(\omega_2^2+\omega_3^2\right)}{4}
        + \delta\lambda_5
        \frac{\omega_1^2\left(\omega_2^2-\omega_3^2\right)}{4} \label{eq:VCT}\ .
	\end{split}
 \end{align}
The complete potential of Eq.~\eqref{eq:totalpot} will be minimised to find the global minimum
at a given temperature $T$.
As stated above, the counterterms $\delta p$ for the parameters $p$ of
the tree-level potential contain only the finite pieces, as the
divergent ones have already been absorbed by the
$\overline{\mbox{MS}}$ renormalised
$V_{\text{CW}}$.
We renormalise the effective potential such that at $T=0$ the tree-level
position of the minimum yields a local minimum, which is checked to be
the global one numerically. Furthermore, through our renormalisation
the masses and mixing angles of the scalar
particles are preserved at their tree-level values by the one-loop
potential. The corresponding
renormalisation conditions are imposed at $T=0$, which is where we
test for the compatibility with the experimental constraints.
The position of the minimum is determined by the first derivative of the
potential, whereas the masses and angles result from the second derivative, namely
the mass matrix. Formulae for both the first and the second
derivatives of the CW potential in the Landau gauge have been derived in
\cite{Camargo-Molina:2016moz}. We employ these formulae
in the gauge basis to calculate the
required derivatives. Consequently, for the renormalisation we also
express the counterterm potential and the tree-level potential in the gauge basis.
The renormalisation conditions for the first derivatives are then given by
($i=1,...,8$)
\beq
\partial_{\phi_i} \left.V_{\text{CT}} (\phi)\right|_{\phi= \langle
  \phi^c \rangle_{T=0}} =
- \partial_{\phi_i} \left.V_{\text{CW}} (\phi) \right|_{\phi=
  \langle\phi^c \rangle_{T=0}} \label{eq:rencond1}
\eeq
with
\beq
\phi_i \equiv \{ \rho_1, \eta_1, \rho_2, \eta_2, \zeta_1, \psi_1,
\zeta_2, \psi_2 \} \;,
\eeq
and $\langle \phi^c \rangle_{T=0}$ denoting the field configuration in
the minimum at $T=0$,
\beq
\langle \phi^c \rangle_{T=0} = (0,0,0,0,v_1,0,v_2,0) \;.
\eeq
This results in two non-trivial conditions
for the tree-level minimum at $T=0$ to be a CP-conserving extremum also at the
one-loop level. In order to ensure that both the masses and the mixing
angles remain at their tree-level values the complete $8\times 8$
mass matrix of the scalar sector should be preserved at its tree-level
value by the renormalised one-loop potential. This is achieved by
demanding ($i,j=1,...,8$)
\beq
\partial_{\phi_i} \partial_{\phi_j}\left.V_{\text{CT}}
  (\phi)\right|_{\phi= \langle\phi^c\rangle_{T=0}} &=&
- \partial_{\phi_i} \partial_{\phi_j}\left.V_{\text{CW}} (\phi)\right|_{\phi=
  \langle\phi^c\rangle_{T=0}} \label{eq:rencond2} \;.
\eeq
However, since we have only eight counterterms and after imposing
Eq.~\eqref{eq:rencond1} we are left with six to be set, the
resulting  system of equations is overconstrained and cannot in
general be solved. This means that we 
cannot renormalise all masses and mixing angles to exactly match
their tree-level values. We therefore pursue the following approach:
Both the tree-level and the one-loop mass matrix are rotated to the
mass basis with the tree-level rotation matrix. From the
resulting $8\times 8$ matrix we extract only the $2\times 2$ submatrix,
that corresponds to the physical charged Higgs bosons, and the $3\times
3$ submatrix for the neutral Higgs bosons. In the
CP-conserving case treated here, the latter decomposes into a
$2\times 2$ matrix for the CP-even Higgs bosons $h$ and $H$, and the
entry for the pseudoscalar $A$.
On these submatrices the renormalisation conditions are imposed, so that we have
\begin{equation}
 \partial_{\phi_i} \partial_{\phi_j}\left.V_{\text{CT}} (\phi)\right|_{\phi= \langle\phi^c\rangle_{T=0}} \big|_{\text{mass}}^{H^\pm} =
- \partial_{\phi_i} \partial_{\phi_j}\left.V_{\text{CW}} (\phi)\right|_{\phi=  \langle\phi^c\rangle_{T=0}}\big|_{\text{mass}}^{H^\pm}  \label{eq:rencond3}
\end{equation}
and
\begin{equation}
 \partial_{\phi_i} \partial_{\phi_j}\left.V_{\text{CT}} (\phi)\right|_{\phi= \langle\phi^c\rangle_{T=0}} \big|_{\text{mass}}^{h,H,A} =
- \partial_{\phi_i} \partial_{\phi_j}\left.V_{\text{CW}}
  (\phi)\right|_{\phi=
  \langle\phi^c\rangle_{T=0}}\big|_{\text{mass}}^{h,H,A}  \ . \label{eq:rencond4}
\end{equation}
The subscript 'mass' indicates that the mass matrix in the gauge basis
is rotated into the mass eigenbasis by means of the
rotation matrix that diagonalises the tree-level mass matrix. The
superscripts $H^\pm$ and $h,H,A$ indicate that from the resulting
matrix only the $2\times 2$ block for the physical charged Higgs
bosons and the $3\times 3$ block for the physical neutral  Higgs
bosons is considered, respectively. Equations~(\ref{eq:rencond3}) and
(\ref{eq:rencond4}) provide five independent non-trivial
renormalisation conditions.\footnote{
After application of Eq.~(\ref{eq:rencond1}) some matrix elements of
the extracted submatrices are linear combinations of other matrix
elements so that we do not have further conditions.} Together with
the two renormalisation
conditions from Eq.~(\ref{eq:rencond1}) we have altogether seven
renormalisation conditions to fix eight renormalisation constants,
{\it cf.}~Eq.~(\ref{eq:VCT}), so that one renormalisation constant is
left for determination. Inspecting the counterterm potential
Eq.~\eqref{eq:VCT}, we observe that the counterterms $\delta\lambda_3$ and
$\delta\lambda_4$  only appear as sum. Hence, we choose to use only
one of them and set $\delta\lambda_4=0$. The remaining seven
renormalisation constants are fixed by the conditions Eqs.~\eqref{eq:rencond1},
\eqref{eq:rencond3} and \eqref{eq:rencond4}. \s

We find that these renormalisation conditions allow us to preserve the
minimum, the masses and the mixing angles of the Higgs sector at their
tree-level values up to a very good approximation. Taking into account
numerical uncertainties, the minimum at one-loop remains at $v\pm 2\
\text{GeV}$, and all masses and mixing angles are preserved up to tiny
numerical fluctuations. \s

Equations~\eqref{eq:rencond3} and \eqref{eq:rencond4} require the
second derivative of the CW potential. It is a well-known problem that
this derivative leads to infrared divergences for the Goldstone
bosons in the Landau gauge
\cite{Cline:1996mga,Cline:2011mm,Dorsch:2013wja,Camargo-Molina:2016moz,Martin:2014bca,Elias-Miro:2014pca}. In
order to circumvent this problem, in \cite{Cline:1996mga} the logarithm is
redefined to capture on-shell effects regularising
  the divergence while in \cite{Cline:2011mm,Dorsch:2013wja,Dorsch:2014qja}
a non-vanishing infrared mass for the Goldstones is employed to
regulate the divergence.
In the effective potential approach itself, which is the approximation of
the full theory at vanishing external momenta, it is not possible to
cancel these divergences. Building up the complete self-energy of the
Higgs bosons from the effective potential and the momentum-dependent
parts obtained by a diagrammatic calculation, however, it becomes
apparent that the infrared divergences from the Goldstone
contributions cancel between the CW part and the momentum-dependent
part \cite{Camargo-Molina:2016moz,Elias-Miro:2014pca,Casas:1994us}. Taking the limit of vanishing external momenta afterwards we arrive at a finite expression for the second derivative of the CW potential. This cancellation
was checked explicitly using the results from the diagrammatic
calculation performed in \cite{Krause:2016oke,Krause:2016xku}. In
practice, this result can be obtained directly from the effective
potential approach by regularising the logarithmic divergence with a regulator
mass and then discarding the terms proportional to this logarithm
\cite{Camargo-Molina:2016moz}. The obtained results are independent of the
regulator mass and reflect the correct contributions present in the
effective potential approach.\footnote{Note, that this problem does
  not occur for the higher-order contributions to the Higgs masses
  resulting from loops with a photon inside, as the only class of
  diagrams possibly leading to infrared divergences (diagrams with a
  scalar and a vector boson in the loop) vanishes for zero external momenta.}

\section{Numerical Analysis \label{sec:numerical}}
\subsection{Minimisation of the Effective Potential \label{sec:minimization}}
The electroweak PT is considered to be strong if the ratio between the VEV $v_c$
at the critical temperature $T_c$ and the critical temperature $T_c$
is larger than one \cite{Quiros:1994dr,Moore:1998swa},
\beq
\xi_c \equiv \frac{v_c}{T_c} \ge 1 \;.
\eeq
The value $v$ at a given temperature $T$ is obtained as
\beq
v(T) = \sqrt{\bar{\omega}_1^2 + \bar{\omega}_2^2 + \bar{\omega}_3^2} \;.
\eeq
Remind that $\bar{\omega}_i$ are the field configurations that minimise the
loop-corrected effective potential at non-zero temperature.
The critical temperature $T_c$ is defined as the temperature where the
potential has two degenerate minima. For the determination of $T_c$ the effective
potential together with the counterterm potential,
Eq.~(\ref{eq:totalpot}), is minimised numerically at a given
temperature $T$. In a first order electroweak PT the VEV jumps from $v=v_c$ at
the temperature $T_c$ to $v=0$ for $T>T_c$.
In order to double-check the results of the
minimisation procedure, we apply two different minimisation
algorithms. One is the active \texttt{CMA-ES} algorithm as
implemented in \texttt{libcmaes}~\cite{CMAES}. This algorithm finds
the global minimum of a given function. As termination criterion we
choose the relative tolerance of the value of the effective potential
between two iterations to be smaller than $10^{-5}$. The other
algorithm that has been used is the local Nelder-Mead-Simplex algorithm from the
\texttt{GNU Scientific Library}~\cite{GSL}
(\texttt{gsl\_multimin\_fminimizer\_nmsimplex2}), also with a tolerance of
$10^{-5}$. For a given temperature, we start with 500 randomly distributed points in
the interval $\omega_{1,2,3} \in [-500,500]\ \mathrm{GeV}$ for which we compute
the minimum of the potential. Note that we have included $\bar{\omega}_3$ in
Eq.~(\ref{eq:model_4}) for the sake of generality.
The candidates for the global minimum obtained with the two algorithms
are compared to each other and the one with the lower value of the
effective potential is chosen as the global
minimum. Although there may be local minima that are CP-violating 
we find that in the global minimum $\bar{\omega}_3$ always
vanishes up to numerical fluctuations at both $T=0$ and $T=T_c$. Hence
we will not comment on it any further.
In order to determine the critical temperature $T_c$ where the phase
transition takes place, we employ a bisection method in the
temperature $T$, starting with the determination of the minimum at the
temperatures $T_S=0$~GeV and ending at $T_E=300$~GeV. The minimisation
procedure is terminated when the interval containing $T_c$ is smaller
than $10^{-2}\ \text{GeV}$. The temperature $T_c$ is then set to the
lower bound of the final interval. We exclude parameter points that
do not satisfy $\vert v(T=0) - 246.22 \mbox{ GeV} \vert \leq 2 \,
\mathrm{GeV}$, and parameter points where no PT is found for $T\leq 300\
\text{GeV}$\footnote{For temperatures $T_c \ge 246$~GeV the VEV would
  have to be larger than 246~GeV in order to fulfill the criterion of
  a strong first order PT. By choosing 300~GeV we apply
  an additional safety margin.}.
Moreover, we only retain parameter points with $T_c > 10\
\text{GeV}$. \s

The complete calculation and implementation was checked against an
independent calculation in {\tt{Mathematica}}. Profiting from significant
speed-up, the implementation above was used for the results presented
in this work.

\subsection{Constraints and Parameter Scan \label{sec:scan}}
We determine the value of $\xi_c$ only for those
  points that are compatible with theoretical and experimental
constraints. In order to obtain viable data sets we use
\texttt{ScannerS}~\cite{Coimbra:2013qq,Ferreira:2014dya}
to perform  extensive scans in the 2HDM parameter space and check for
compatibility with the constraints.
The program verifies if the tree-level
potential is bounded from below by applying the conditions given
in~\cite{Klimenko:1984qx} and checks for tree-level
perturbative unitarity as described in~\cite{Ginzburg:2003fe}.
In the CP-conserving 2HDM investigated here, the requirement
that the neutral CP-even tree-level minimum is the global one is tested
through a simple condition~\cite{Barroso:2013awa}. The consistency with
the EW precision constraints has been checked through the
oblique parameters $S$, $T$ and $U$ \cite{Peskin:1991sw} by applying
the general procedure for extended Higgs sectors as described
in~\cite{Grimus:2007if,Grimus:2008nb} and demanding for compatibility
with the SM fit~\cite{Baak:2014ora} within 2$\sigma$, including
correlations.
Constraints applied to the charged sector of the 2HDM are based on results from
the measurement of $R_b$~\cite{Haber:1999zh,Deschamps:2009rh} and
$B\rightarrow X_s \gamma$
\cite{Deschamps:2009rh,Mahmoudi:2009zx,Hermann:2012fc}
including the recent calculation~\cite{Misiak:2015xwa} that enforces
\begin{align}
  m_{H\pm}>480\,\mathrm{GeV}\label{eq:bsg}
\end{align}
in type II models. In type I models the bound is much
  weaker and more strongly dependent on $\tan\beta$.
Note, that the results
from LEP \cite{Abbiendi:2013hk} and the LHC
\cite{Aad:2014kga,Khachatryan:2015qxa}\footnote{The recent ATLAS
  results \cite{Aad:2015typ}  have not been translated into
  bounds so far.} require the charged Higgs mass to be above ${\cal O}
(100 \mbox{ GeV})$ depending on the model type.
For the check of the compatibility with the Higgs data we need the
Higgs production cross sections normalised to the corresponding SM
values and the Higgs branching ratios. The latter have been
computed with {\tt HDECAY} version
6.51\cite{Djouadi:1997yw,Butterworth:2010ym,Harlander:2013qxa}. This program
includes the state-of-the-art higher order QCD corrections and
off-shell decays. The Higgs production cross sections through gluon
fusion and $b$-quark fusion at the LHC have been obtained at NNLO QCD
from an interface with {\tt SusHi}
\cite{Harlander:2012pb,Harlander:2013qxa} and normalised to the
corresponding SM value
at NNLO QCD. The cross section ratio for associated production with a
heavy quark pair has been taken at LO. In the ratio involving CP-even
Higgs bosons the QCD corrections drop out. This is not the case for
the pseudoscalar. For associated production with
top quarks the cross section is very small. The associated production
with bottom quarks can be important for large values of
$\tan\beta$. However, here the QCD corrections in the
associated production of the pseudoscalar with the bottom quark pair
almost cancel against those of the SM counterpart due to the nearly
realised chiral limit for the small $b$-quark masses.
The remaining processes through gauge boson fusion and Higgs radiation
off a $W^\pm$ or
$Z$ boson only apply for a CP-even Higgs boson so that here the QCD
corrections drop out when normalised to the SM cross section. Since not all EW
corrections have been provided for the 2HDM so far they are
consistently neglected in all production and decay processes.
Agreement with the exclusion bounds from LHC Higgs searches has been
tested with {\tt HiggsBounds} \cite{Bechtle:2013wla}.
Compatibility with the observed signal of the 125~GeV Higgs boson has been
verified by calculating the reduced signal strengths and checking
against the two times one sigma bounds in the six parameter fit
of~\cite{Khachatryan:2016vau}. Further details on the various checks
can be found in \cite{Ferreira:2014dya}.\footnote{The respective
  experimental values cited there have been replaced by the latest
  experimental results.} \s

For the minimisation procedure we only use parameter points that are
in agreement with the described theoretical and experimental constraints. In order
to find viable parameter points we perform a scan in the 2HDM
parameter space given by the input parameters
Eq.~(\ref{eq:model_16}). The SM VEV given by the Fermi constant
$G_F$ through $v= 1/\sqrt{\sqrt{2} G_F}$, has been fixed to
\beq
v = 246. 22 \; \mbox{GeV} \;.
\eeq
The mixing angle $\alpha$ is varied in the
theoretically allowed region, {\it i.e.}
\beq
- \frac{\pi}{2} \le \alpha < \frac{\pi}{2} \;.
\eeq
In all scans, one of the masses of the CP-even Higgs bosons has been
fixed to \cite{Aad:2015zhl}
\beq
m_{h_{125}} = 125.09 \; \mbox{GeV} \;.
\eeq
This is the Higgs boson we identify with the SM-like Higgs boson
discovered at the LHC, and we denote it by $h_{125}$. We
performed two separate scans for the cases
where the lighter or the heavier of the two CP-even Higgs bosons is
identified with the SM-like Higgs, {\it  i.e.}~$m_h = m_{h_{125}}$
and $m_H= m_{h_{125}}$, respectively. The scan ranges for
the remaining parameters are given in
Table~\ref{tab1:scanranges1} in case of type I and in
Table~\ref{tab1:scanranges2} for type II.
\begin{table}[t!]
\centering
\begin{tabular}{rcccccc}
\toprule
\begin{centering}\# points\end{centering} & $m_{h}$ & $m_{H}$ & $m_{A}$ & $m_{H^\pm}$ & $m_{12}^2$ & $\tan(\beta)$\\
  &\multicolumn{4}{c}{in GeV}&in $\text{GeV}^2$&\\
\cmidrule(r){1-1}\cmidrule(lr){2-5}\cmidrule(lr){6-6}\cmidrule(l){7-7}
$1\,000\,000$ & $m_{h_{125}}$ & $130-1000$ & $30-1000$ & $65-1000$ & $0-5\times10^5$ & $1-35$\\
$100\,000$ & $30-120$ & $m_{h_{125}}$ & $30-1000$ & $65-1000$ & $0-5\times10^5$ & $1-35$\\
\bottomrule
\end{tabular}
\caption{Parameter ranges for the scan performed in the 2HDM type I.
The first column specifies the number of points
generated. \label{tab1:scanranges1}}
\end{table}
In our scans we required the neighboring non-SM-like
Higgs masses to deviate by at least 5~GeV from 125.09~GeV, in order to
avoid degenerate Higgs signals. The input masses for the non-SM-like
neutral Higgs bosons were chosen
within $30\ \text{GeV}$ and $1000\ \text{GeV}$ and the input mass for
the charged Higgs boson within $65\ \text{GeV}$ and $1000\ \text{GeV}$ to
cover most of the parameter space which is potentially interesting for
phenomenology and accessible by experiments.
The parameter $m_{12}^2$ is constrained by the tree-level global
minimum condition to be positive. The upper limits on $\tan\beta$ and
$m_{12}^2$ have been set by choice, but as we observe later, most of the
points compatible with the constraints and a strong PT are found for
rather small $\tan\beta$ so  that the chosen upper limit does not pose
a strong constraint. Type-specific choices for the ranges are the
lower bound on $\tan\beta$ in type I and the  lower bound on
$m_{H^\pm}$ in type II. They have been chosen such
  that they already leave out part of the parameter space that is
excluded by the constraints from $B\rightarrow X_s\gamma$
measurements. Moreover, in type II the lower
bound on $m_A$ in the second set, where $H \equiv
  h_{125}$, is motivated by the fact that
fulfilling constraints on the oblique parameters requires one Higgs to
be in vicinity of the charged Higgs boson. In the second set this
can only be the pseudoscalar Higgs $A$.  \s

\begin{table}[t!]
\centering
\begin{tabular}{rcccccc}
\toprule
\begin{centering}\# points\end{centering} & $m_{h}$ & $m_{H}$ & $m_{A}$ & $m_{H^\pm}$ & $m_{12}^2$ & $\tan(\beta)$\\
  &\multicolumn{4}{c}{in GeV}&in $\text{GeV}^2$&\\
\cmidrule(r){1-1}\cmidrule(lr){2-5}\cmidrule(lr){6-6}\cmidrule(l){7-7}
$1\,000\,000$ & $m_{h_{125}}$ & $130-1000$ & $30-1000$ & $480-1000$ & $0-5\times10^5$ & $0.1-35$\\
$100\,000$ & $30-120$ & $m_{h_{125}}$ & $450-1000$ & $480-1000$ & $0-5\times10^5$ & $0.1-35$\\
\bottomrule
\end{tabular}
\caption{Parameter ranges for the scan in the 2HDM type II. The first
column specifies the number of points generated. \label{tab1:scanranges2}}
\end{table}
For the SM parameters we have chosen
the following values: Apart from the computation of the oblique
parameters, where we use the fine structure constant at zero momentum transfer,
\beq
\alpha^{-1}_{\text{EM}} (0) = 137.0359997 \,,
\eeq
the fine structure constant is taken at the $Z$ boson mass scale
\cite{Agashe:2014kda},
\beq
\alpha^{-1}_{\text{EM}} (M_Z^2) = 128.962 \;.
\eeq
The massive gauge boson masses are chosen as
\cite{Agashe:2014kda,Denner:2047636}
\beq
M_W = 80.385 \mbox{ GeV} \qquad \mbox{and} \qquad M_Z = 91.1876 \mbox{
GeV} \;,
\eeq
the lepton masses as \cite{Agashe:2014kda,Denner:2047636}
\beq
m_e = 0.510998928 \mbox{ MeV} \;, \quad
m_\mu = 105.6583715 \mbox{ MeV} \;, \quad
m_\tau = 1.77682 \mbox{ GeV} \;,
\eeq
and the light quark masses are set following \cite{LHCHXSWG} to
\beq
m_u = 100 \mbox{ MeV} \;, \quad m_d = 100 \mbox{ MeV} \;, \quad
m_s = 100 \mbox{ MeV} \;.
\eeq
For consistency with the ATLAS and CMS analyses the on-shell top quark mass
\beq
m_t = 172.5 \mbox{ GeV}  \label{eq:ferm1}
\eeq
has been taken as recommended by the LHC Higgs Cross Section Working Group
(HXSWG) \cite{Denner:2047636,Dittmaier:2011ti}. The charm and
bottom quark on-shell masses are \cite{Denner:2047636}
\beq
m_c = 1.51 \mbox{ GeV} \qquad \mbox{and} \qquad
m_b = 4.92 \mbox{ GeV} \;. \label{eq:ferm2}
\eeq
We take the CKM matrix to be real, with the
CKM matrix elements given by \cite{Agashe:2014kda} \footnote{In the
computation of the counterterms we
choose $V_{\text{CKM}}=\mathds{1}$ for simplicity. The
impact of this choice on the counterterms and thereby on the potential
and its minimisation is negligible.}
\beq
V_{\text{CKM}} = \left( \begin{array}{ccc} V_{ud} & V_{us} & V_{ub} \\
V_{cd} & V_{cs} & V_{cb} \\ V_{td} & V_{ts} & V_{tb} \end{array}
\right) = \left( \begin{array}{ccc} 0.97427 & 0.22536 & 0.00355 \\
    -0.22522 & 0.97343 & 0.0414 \\ 0.00886 & -0.0405 &
    0.99914 \end{array} \right) \;. \label{eq:parameterscan_alex1}
\eeq

\section{Results \label{sec:results}}
We now turn to the presentation of our results. We will discuss the
specific features of the 2HDM parameter space that is compatible with
the theoretical and experimental constraints and at the same time
provides a strong first order PT. We will show results
both for the type I and the type II 2HDM. For comparison
with results in the literature, we show one plot where
  we have applied the 'Parwani' method in the treatment of the thermal masses, {\it
  cf.}~subsection~\ref{sec:one_loop_pot}. In the remaining discussion,
  however, we apply the 'Arnold-Esinosa' method for reasons
  discussed in \cite{Arnold:1992rz} and alluded to in section
  \ref{sec:one_loop_pot}. 
We will discuss scenarios where
the lighter of the CP-even Higgs bosons is identified with the discovered
Higgs boson, {\it i.e.}~$h \equiv h_{125}$, and where $H
\equiv h_{125}$. \s

For the interpretation of our results some general considerations on
first order PTs are in order. The value of $\xi_c$
is proportional to the couplings of the light bosonic particles to the
SM-like Higgs boson, and it decreases with the Higgs boson mass
\cite{Carena:2008vj}. The additional Higgs bosons in the 2HDM spectrum
allow for large trilinear bosonic couplings, in contrast to the
SM, where bosonic couplings are only due to the weak gauge couplings
between the Higgs boson and the EW gauge bosons. In the 2HDM,
the second CP-even Higgs boson with a non-vanishing VEV contributes to
the PT and can reduce its strength if $H$ is not light
enough. A strong electroweak PT therefore requires
$H$ either to be light or to have a vanishing VEV. The latter
corresponds to the alignment limit where only one of the physical
Higgs bosons has a VEV \cite{Gunion:2002zf}. Previous investigations
suggest that a 
first-order PT prefers a scalar spectrum, which is not too heavy
\cite{Cline:2011mm,Dorsch:2013wja,Dorsch:2014qja},
or else a large mass splitting between the heavy scalars
\cite{Dorsch:2014qja,Dorsch:2016nrg}. In the type II 2HDM the
requirement of a light Higgs spectrum puts some tension on the model,
as compatibility with the EW precision tests requires one of
the non-SM-like neutral Higgs bosons to be close to
$m_{H^\pm}$. Charged Higgs masses below 480~GeV on the other hand are
already excluded by $B\rightarrow X_s \gamma$.

\subsection{Type I: Parameter Sets with $h \equiv h_{125}$ \label{sec:Ihh125}}
\begin{figure}[t!]
\begin{center}
\includegraphics[width=0.5\textwidth]{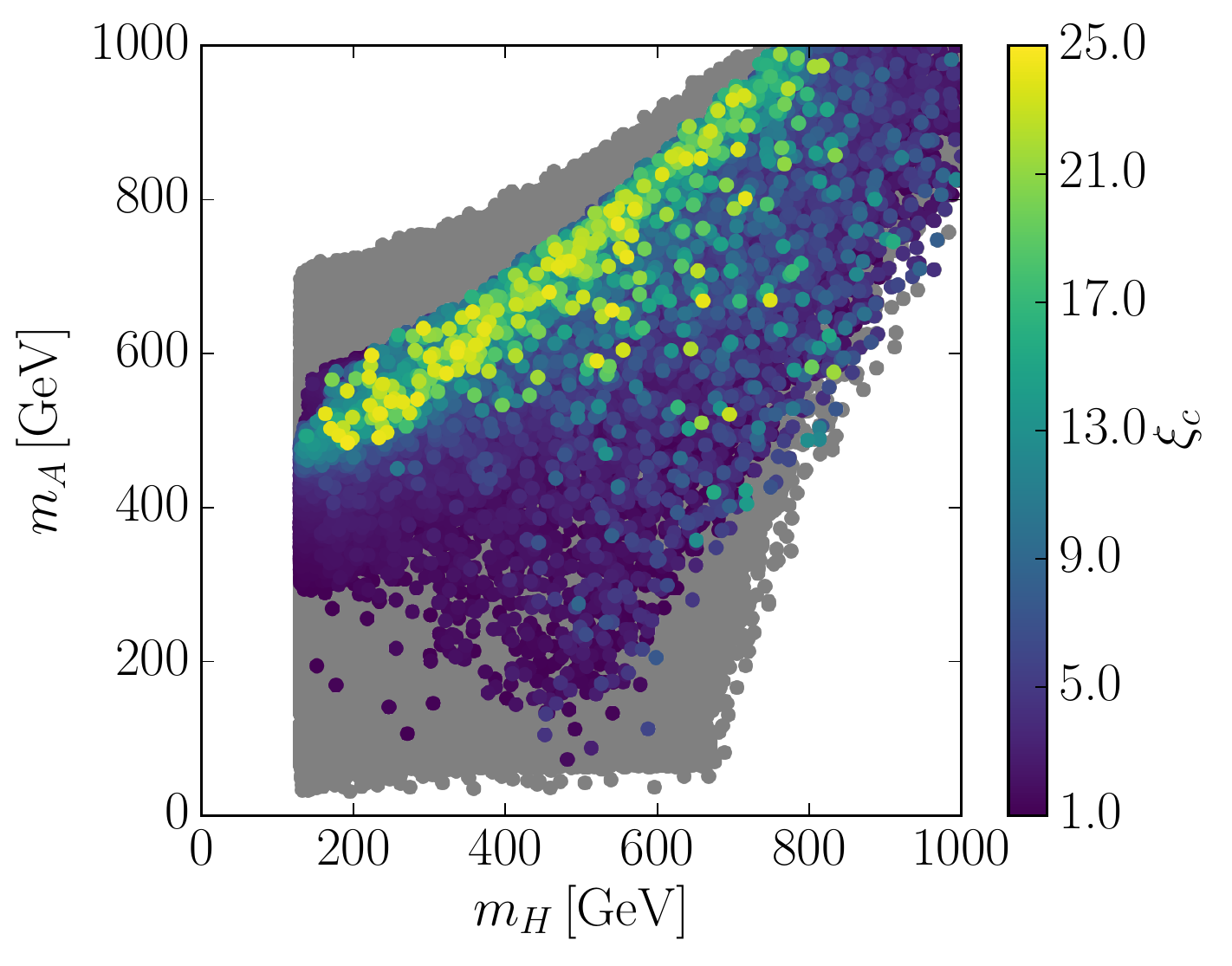}\hfill
\includegraphics[width=0.5\textwidth]{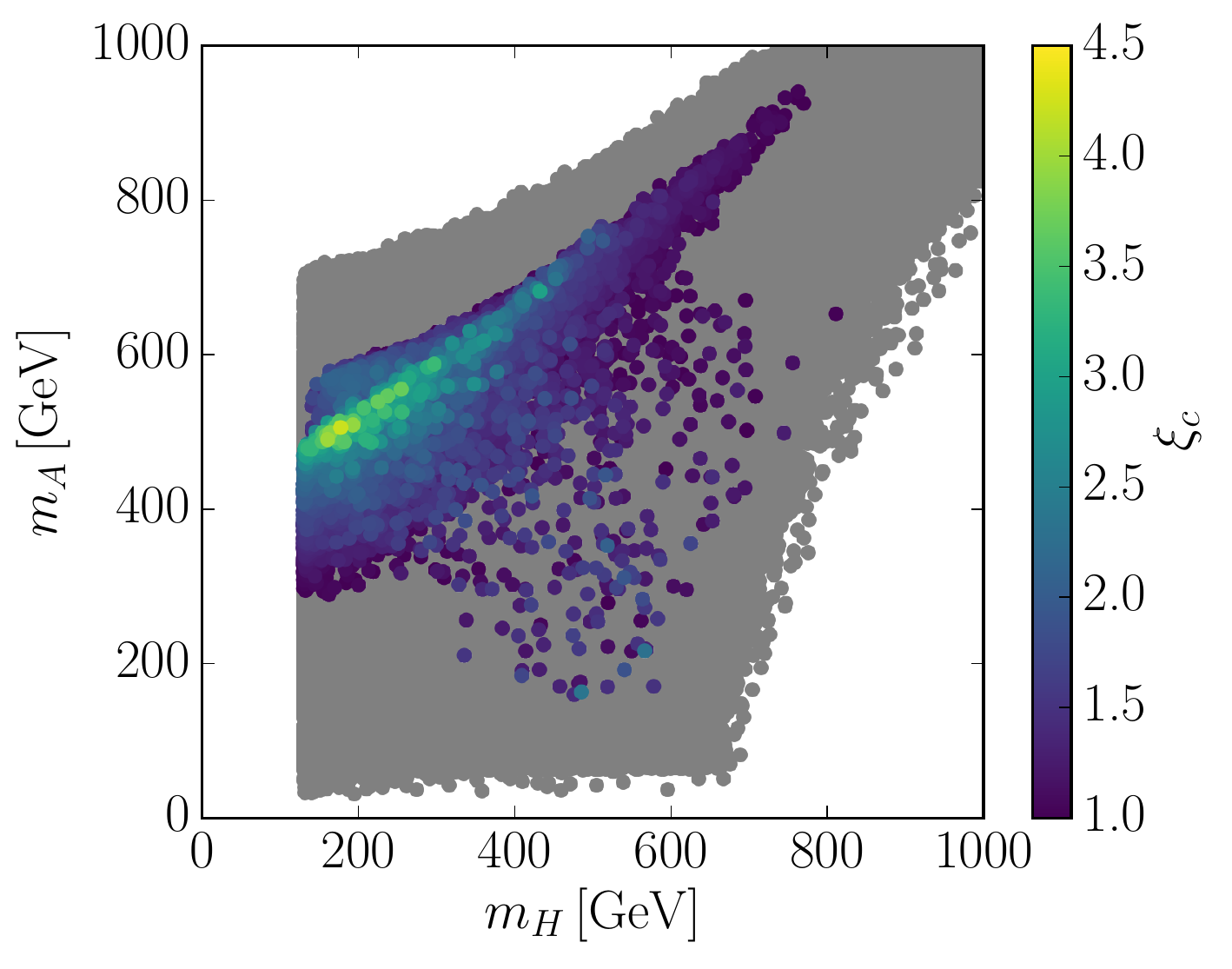}
\vspace*{-0.2cm}
\caption{Type I, $h\equiv h_{125}$:
           Results in the $m_A$ versus $m_H$-plane, showing in grey the
           parameter points passing all
           applied constraints. Points highlighted in color have a PT of strong first
           order, where the value of $\xi_c$ is indicated
           by the color code. Left: 'Parwani' method, right:
           'Arnold-Espinosa' method. \label{fig:t1mamh}}
\end{center}
\vspace*{-0.6cm}
\end{figure}
\begin{figure}[b!]
\begin{center}
\includegraphics[width=8.5cm]{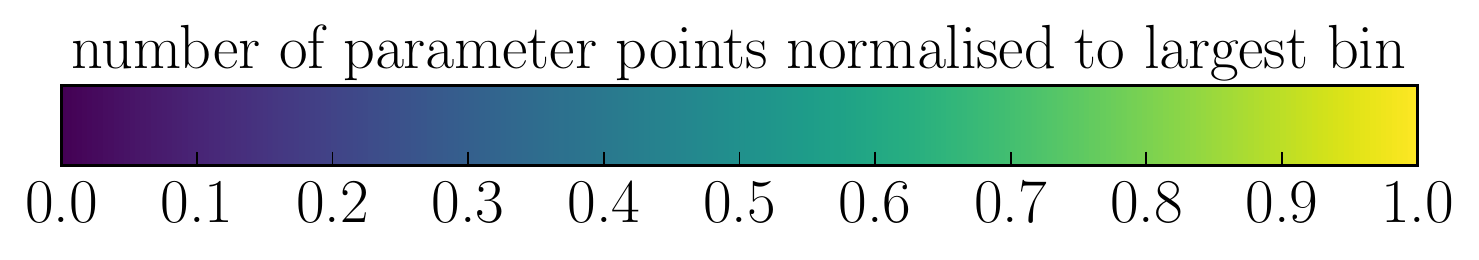} \\
\includegraphics[width=0.49\textwidth]{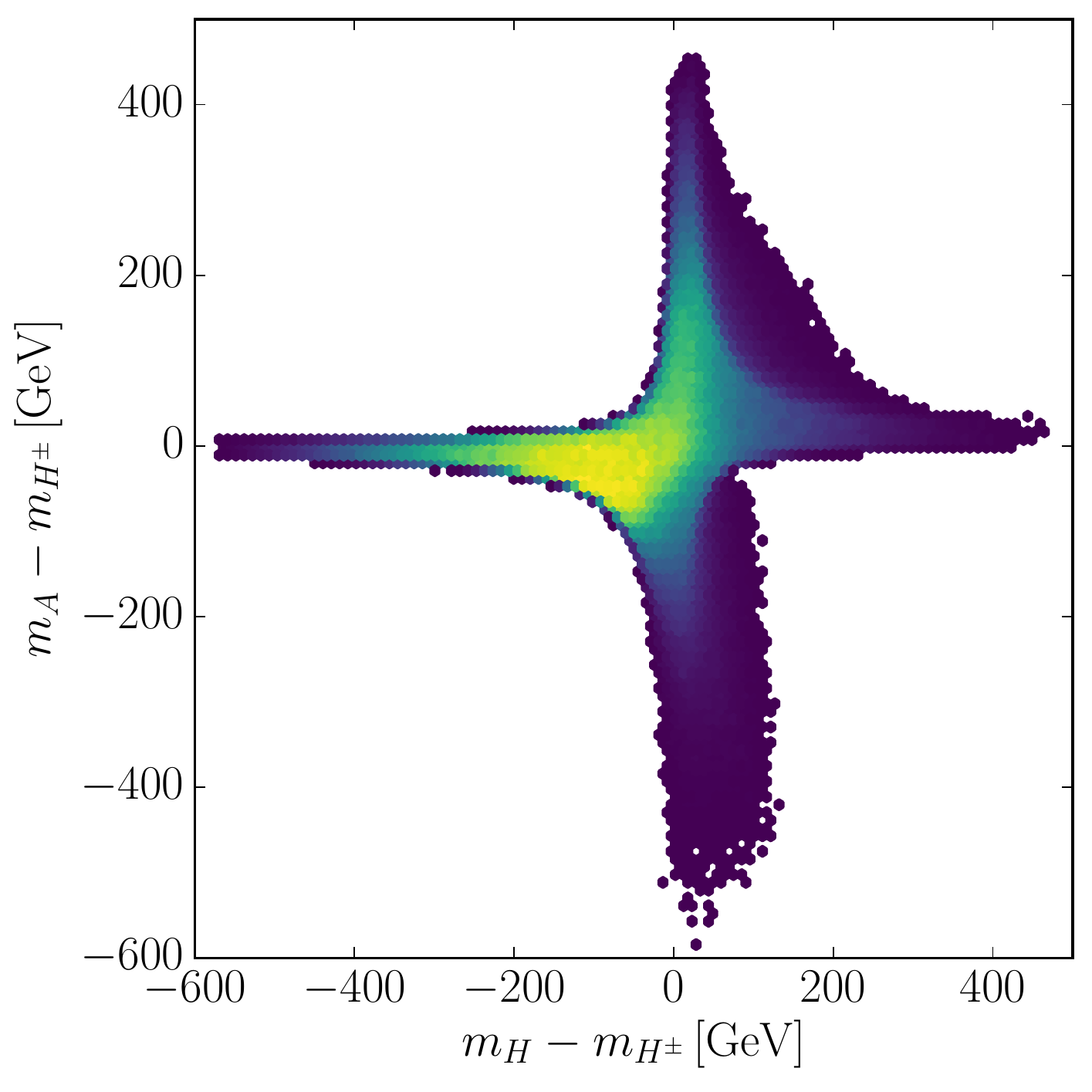}\hfill
\includegraphics[width=0.49\textwidth]{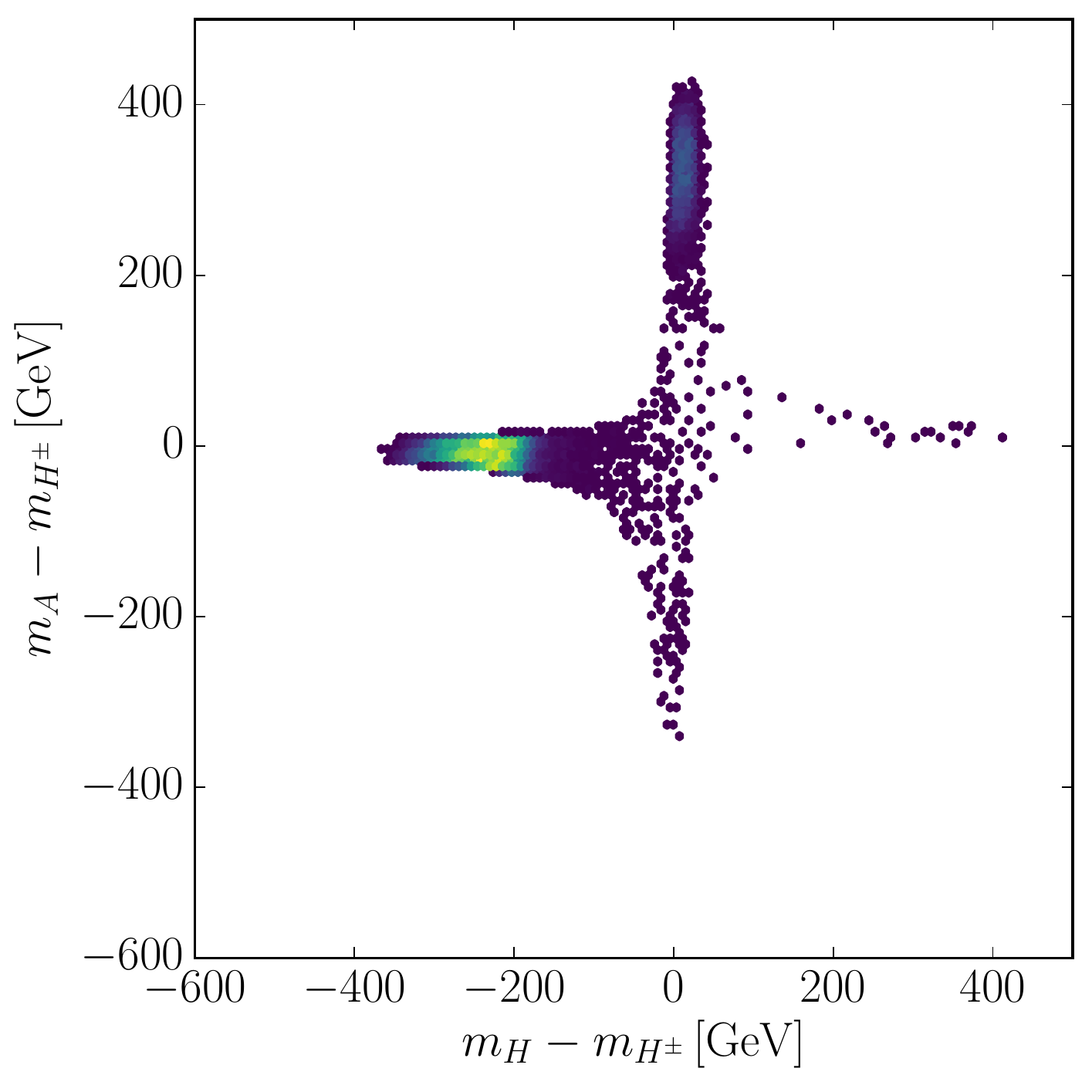} 
\vspace*{-0.2cm}
\caption{Type I, $h\equiv h_{125}$: The mass difference $m_A-m_{H^\pm}$ versus
  $m_H-m_{H^\pm}$. The colour code shows the
  relative frequency of left: all points passing the constraints; right:
  all points with additionally $\xi_c \ge 1$ ('Arnold-Espinosa'
  method). \label{fig:t1massdiffs}}
\end{center}
\vspace*{-0.6cm}
\end{figure}
We start with the analysis of the results in the 2HDM type I.
Figure~\ref{fig:t1mamh} shows in the $m_A$ versus $m_H$ plane all parameter
points that pass the applied constraints, for scenarios where $h\equiv
h_{125}$. The coloured points are those for which we obtain a
strong first order PT, {\it i.e.}~where $\xi_c \ge
1$. In the treatment of the thermal masses we have applied the
'Parwani' method (left plot) in order to compare to the results of
\cite{Dorsch:2014qja}, where the 'Parwani' method was applied.  In the
right plot we show the results for the 
'Arnold-Espinosa' method, which we will use in the remainder of the discussion.
As can be inferred from the plots, in the 2HDM type I first order
PTs are still possible taking into account the
up-to-date LHC Higgs data and all theoretical constraints on the 2HDM
Higgs potential. The comparison of the left and right plot, however,
also shows that the results obtained for $\xi_c$ are significantly
different when the two different approximations in the treatment of
the thermal masses are applied. Overall, the regions in the parameter
space compatible with $\xi_c \ge 1$ are smaller when the 'Arnold-Espinosa'
method is applied. Furthermore, the maximum values of $\xi_c$ that can be
obtained with the 'Parwani' method are by a factor five larger than
those obtained with the 'Arnold-Espinosa'
method. Working with a one-loop effective potential
  only, the 'Parwani' method cannot be applied consistently, which is
  reflected in the very different results for both
  methods. Note also that the unrealistically large values for $\xi_c$ obtained in the
  'Parwani' method imply very low critical temperatures $T_c$ where
  the phase transition takes place. This again questions the
  way the thermal masses are implemented so that the results of the
  'Parwani' method have to be taken with care.
In the following, we will only show plots for the 'Arnold-Espinosa'
method. \s

\begin{figure}[t!]
\begin{center}
\includegraphics[width=8.5cm]{Scale} \\
\includegraphics[width=0.49\textwidth]{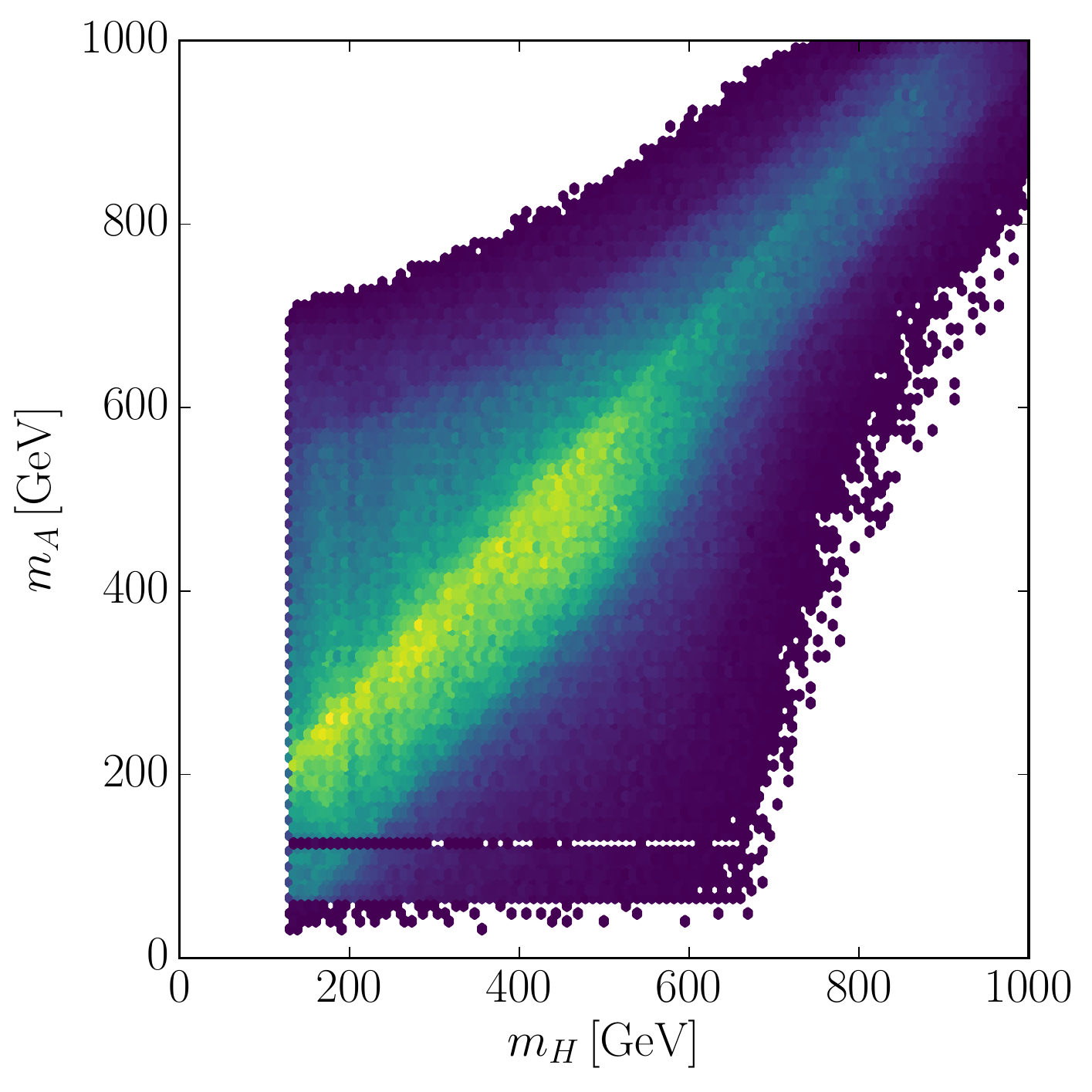}\hfill
\includegraphics[width=0.49\textwidth]{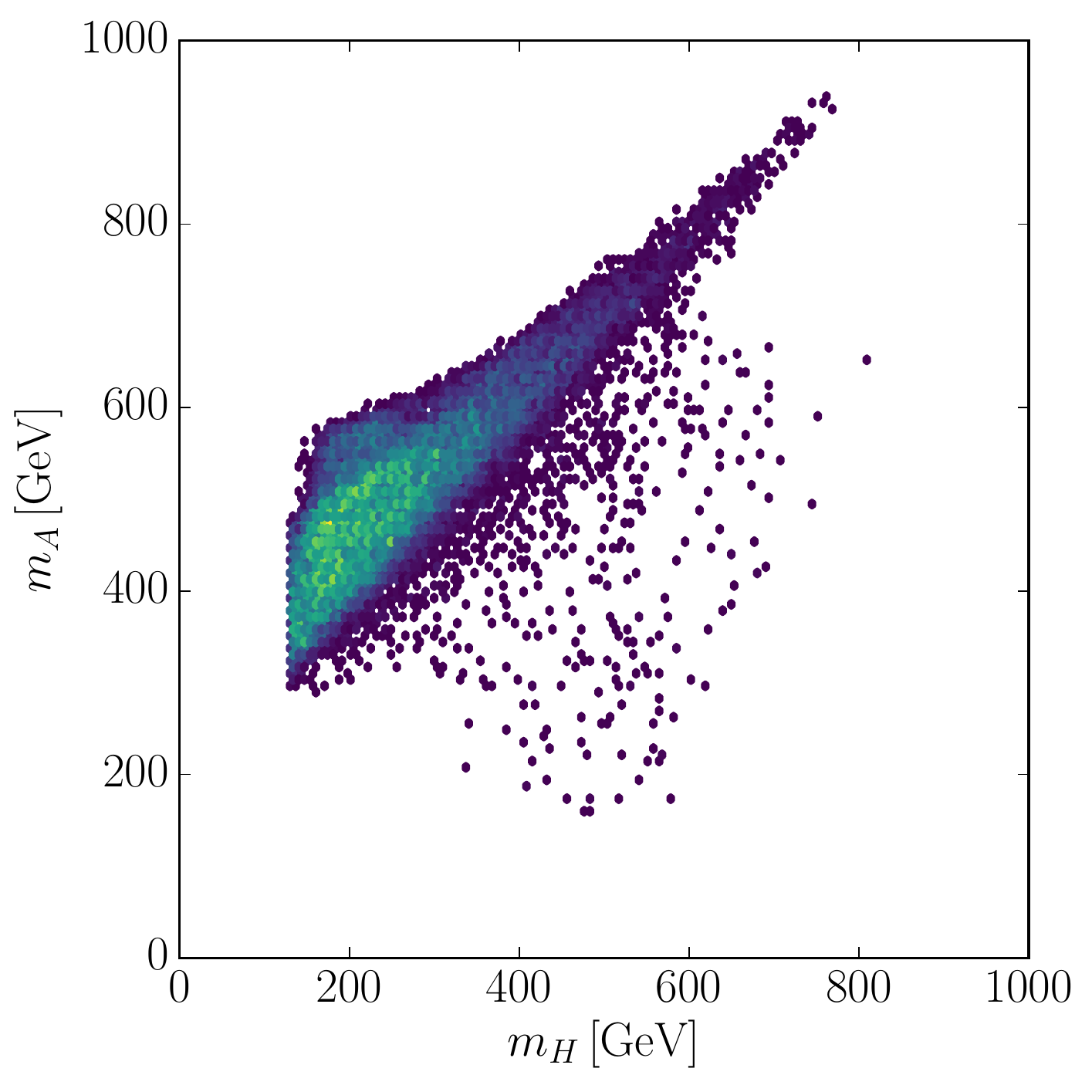} 
\vspace*{-0.2cm}
\caption{Type I, $h\equiv h_{125}$:
  The masses $m_A$ versus $m_H$. The colour code shows the
  relative frequency of left: all points passing the constraints; right:
  all points with additionally $\xi_c \ge 1$ ('Arnold-Espinosa'
  method). \label{fig:t1furtherplots}}
\end{center}
\vspace*{-0.6cm}
\end{figure}
In order to examine how the requirement of a strong first order phase
transition translates into LHC Higgs phenomenology we show in
Fig.~\ref{fig:t1massdiffs} the mass differences between the
non-SM-like Higgs bosons. The left plot shows the frequency of the
points that pass the constraints. The right
plot displays the frequency of the points when additionally a
strong EW phase transition is required. As can be inferred from the
left plot, the EW precision tests, namely the measurement of the $\rho$
parameter, force the mass differences between the charged Higgs boson
and at least one of the non-SM-like Higgs bosons to be small and
strongly favour mass spectra where all of the non-$h_{125}$ masses are
close to each other. The requirement of a strong EW phase transition,
however, favours scenarios where the pseudoscalar mass is close to
$m_{H^\pm}$ with a larger mass gap relative to a lighter $H$.
In Fig.~\ref{fig:t1furtherplots} we display the
relative frequencies in the $m_A$ versus $m_H$ plane for all point passing the
constraints (left) and for those points which additionally fulfill
$\xi_c \ge 1$ (right). The comparison of the two plots shows that
the requirement of a strong PT favours a mass spectrum where the
heaviest Higgs bosons $A$ and $H^\pm$ have masses around 400-500~GeV
and $m_H \approx 200$~GeV.
$H$, which acquires a VEV, should be light, so that the strength of
the PT is not reduced by a heavy $H$.
Consequently $m_{12}^2$ is small\footnote{The masses
of the heavy Higgs bosons $\Phi = H,A,H^\pm$ are given by $m_\phi^2 =
m_{12}^2 / (\sin\beta \cos\beta) \, c_\phi^2 + f(\lambda_i) v^2$, where
$f(\lambda_i)$ is a linear combination of $\lambda_1$-$\lambda_5$ and
$c_\phi= 1$ for $\phi=A,H^\pm$ and $\sin(\beta-\alpha)$ for $\phi= H$
\cite{Krause:2016xku}.}, which means that the strength of the phase
transition is governed by the quartic couplings $\lambda_4$ and
$\lambda_5$, {\it cf.}~also \cite{Dorsch:2013wja}. 
The next important mass configuration is given by scenarios where again the
mass gap between $A$ and $H$ is large, but now overall pushed to higher
mass values, {\it i.e.}~$m_H \approx
m_{H^\pm}$ and $m_A-m_{H} \approx 350$~GeV, {\it
  cf.}~Fig.~\ref{fig:t1massdiffs} (right). 
Since $h \equiv h_{125}$ and hence $\sin (\beta-\alpha) \approx 1$,
the coupling $g_{ZAH} \sim \sin(\beta- \alpha)$ between $A$, $Z$ and
$H$ is significant. The requirement of a strong PT prefers scenarios
where the decay $A \to ZH$ is kinematically allowed so that this decay
can become important. 
These scenarios can be searched for at the LHC, as has
been found earlier in \cite{Dorsch:2014qja} and proposed by the
authors as possible benchmark scenarios. Still,
Fig.~\ref{fig:t1massdiffs} demonstrates that also scenarios are
compatible with $\xi_c \ge 1$ where all
three non-SM-like Higgs bosons are close in mass or where the decay $H \to
AZ$ is possible, {\it i.e.}~where $m_H > m_A$ and
  either $m_H - m_{H^\pm} \approx 0$ or $m_A - m_{H^\pm} \approx
  0$. While our results confirm earlier 
results in the literature \cite{Dorsch:2014qja,Dorsch:2016nrg}, our
results also show that a decay $A \to ZH$ is not unique for a 2HDM type
I featuring a strong first order PT. \s
\begin{figure}[t!]
\begin{center}
\includegraphics[width=8.5cm]{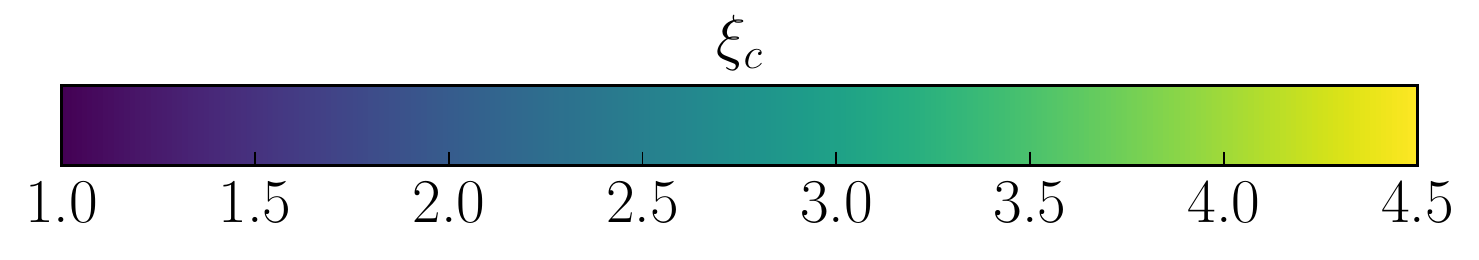} \\
\includegraphics[width=0.49\textwidth]{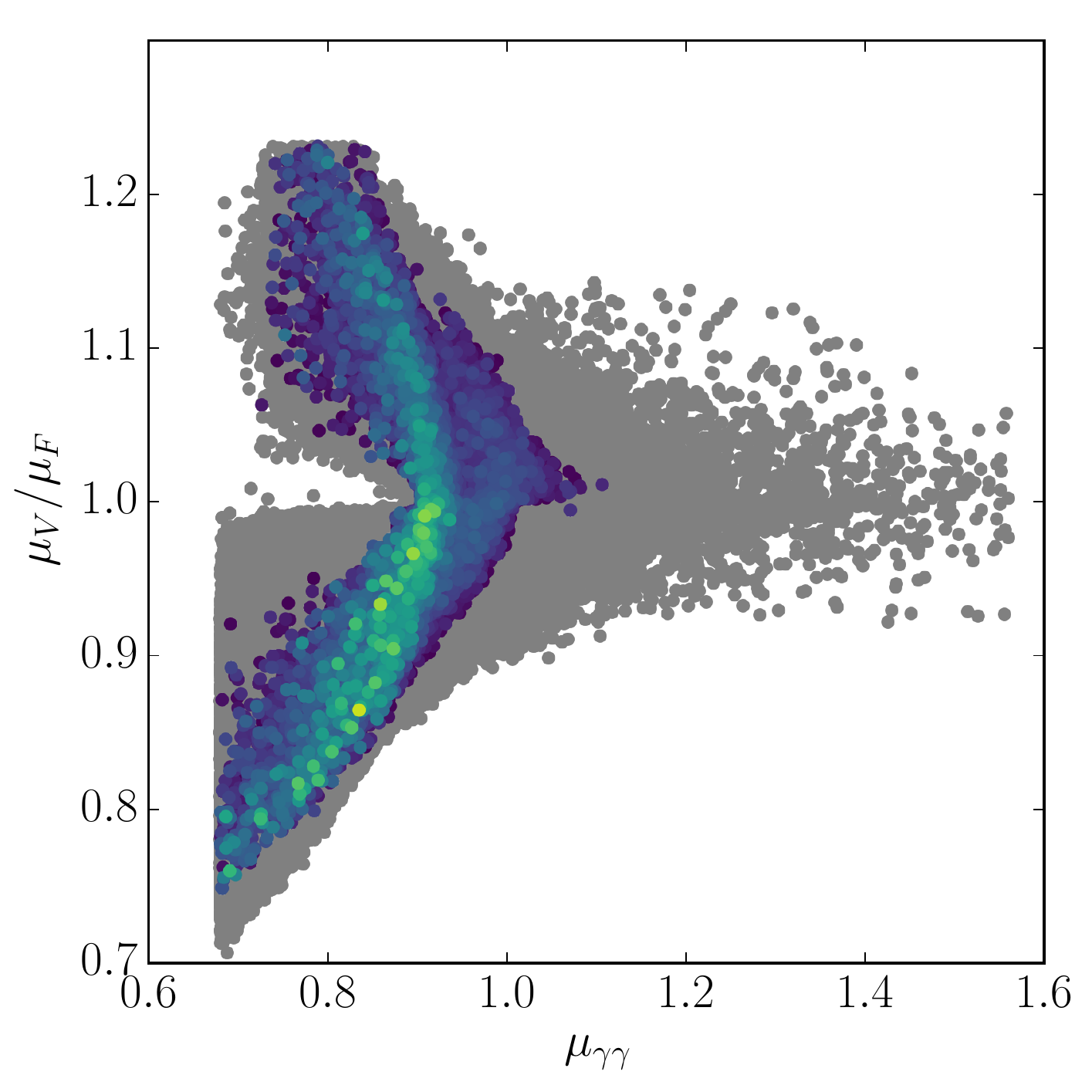}\hfill
\includegraphics[width=0.49\textwidth]{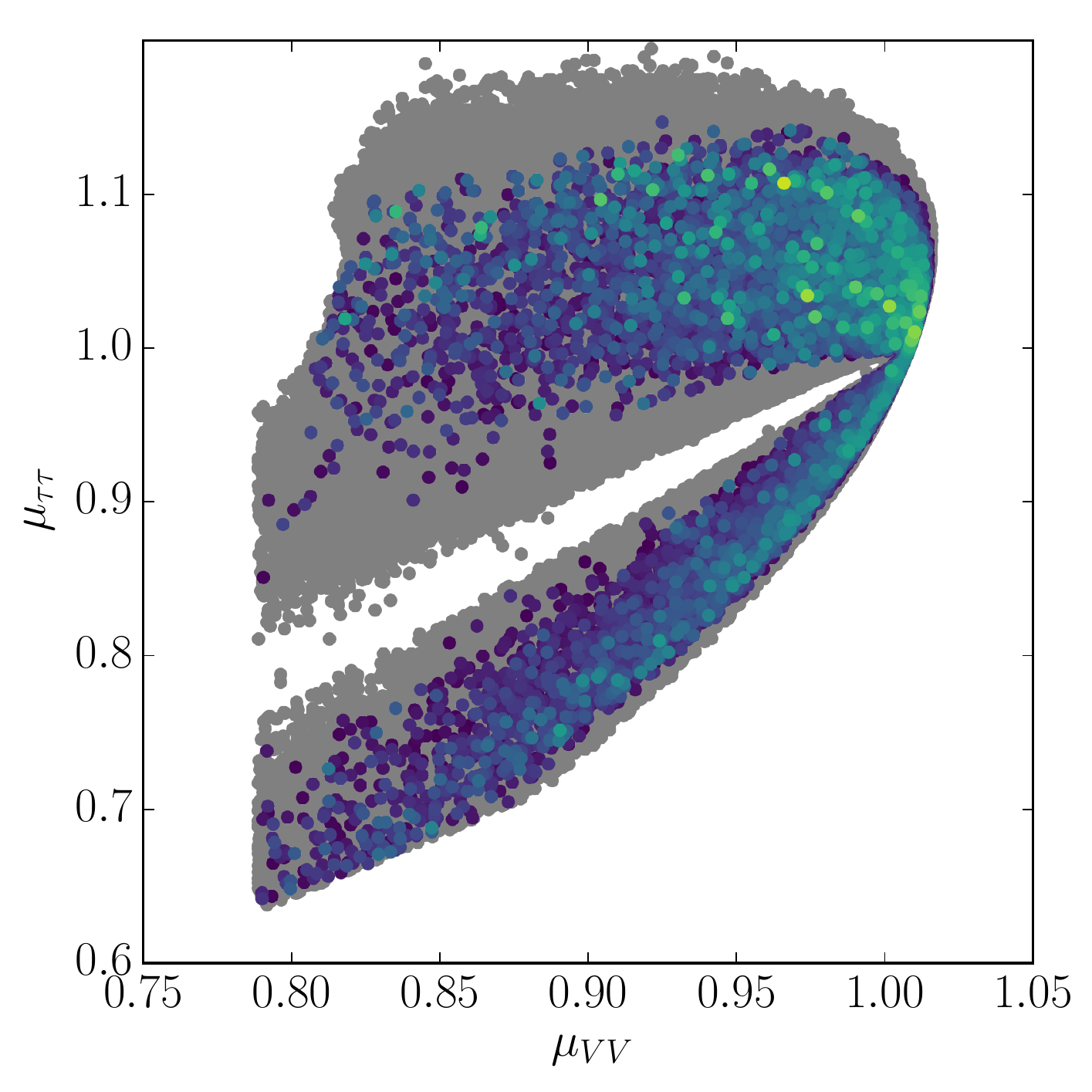} 
\vspace*{-0.2cm}
\caption{Type I, $h\equiv h_{125}$:
  $\mu_V / \mu_F$ versus $\mu_{\gamma\gamma}$ (left)
  and $\mu_{\tau\tau}$ versus $\mu_{VV}$ (right); grey: all points
    passing the applied constraints, colour:
  all points with additionally $\xi_c \ge 1$ ('Arnold-Espinosa'
  method). The colour code indicates the value of
  $\xi_c$. \label{fig:t1rates}}
\end{center}
\vspace*{-0.6cm}
\end{figure}

The majority of the scenarios we find is very close to the alignment
limit, {\it i.e.}~$\sin (\beta-\alpha) \approx 1$ with $\tan\beta$
close to its smallest possible value of about 2.5. While this is a
feature resulting already from the constraints applied, the
requirement of a strong PT overall pushes the Higgs rates towards SM
values, as can be inferred from Fig.~\ref{fig:t1rates}. It shows in
grey the distribution of the Higgs signal strengths for the scenarios
passing the constraints and in colour the scenarios that are
additionally compatible with a strong PT. The colour code indicates
the strength of the PT. The left plot shows $\mu_V/\mu_F$ versus
$\mu_{\gamma\gamma}$ and the right one $\mu_{\tau\tau}$ versus
$\mu_{VV}$.
Here $\mu_F$ denotes the fermion initiated cross section (gluon fusion
and associated production with a heavy quark pair) of the
SM-like Higgs boson ($h_{125}$) normalised to the SM, and $\mu_V$ the
normalised production cross section through massive gauge bosons (gauge boson
fusion and associated production with a vector boson).
The value $\mu_{xx}$ is defined as
\beq
\mu_{xx} = \mu_F \, \frac{\mbox{BR}_{\text{2HDM}} (h_{125} \to
  xx)}{\mbox{BR}_{\text{SM}} (H_{\text{SM}}\to xx)} \;,
\eeq
where $H_{\text{SM}}$ is the SM Higgs boson with mass 125~GeV. The left plot shows
that for $\mu_V/\mu_F$ close to 1, enhanced signal rates in the
photonic final states with $\mu_{\gamma\gamma}$ of up to about 1.5 are
still allowed. However, including the requirement for a strong first
order PT the possible range of an enhanced $\mu_{\gamma\gamma}$ is
strongly restricted down to $\mu_{\gamma\gamma} \approx 1.1$.
On the other hand, the limits on the $\tau$ or gauge
boson final states are not as significantly changed,
as can be inferred from Fig.~\ref{fig:t1rates} (right).

\subsection{Type II: Parameter Sets with $h \equiv
  h_{125}$ \label{sec:IIhh125}}
\begin{figure}[t!]
\begin{center}
\includegraphics[width=0.6\textwidth]{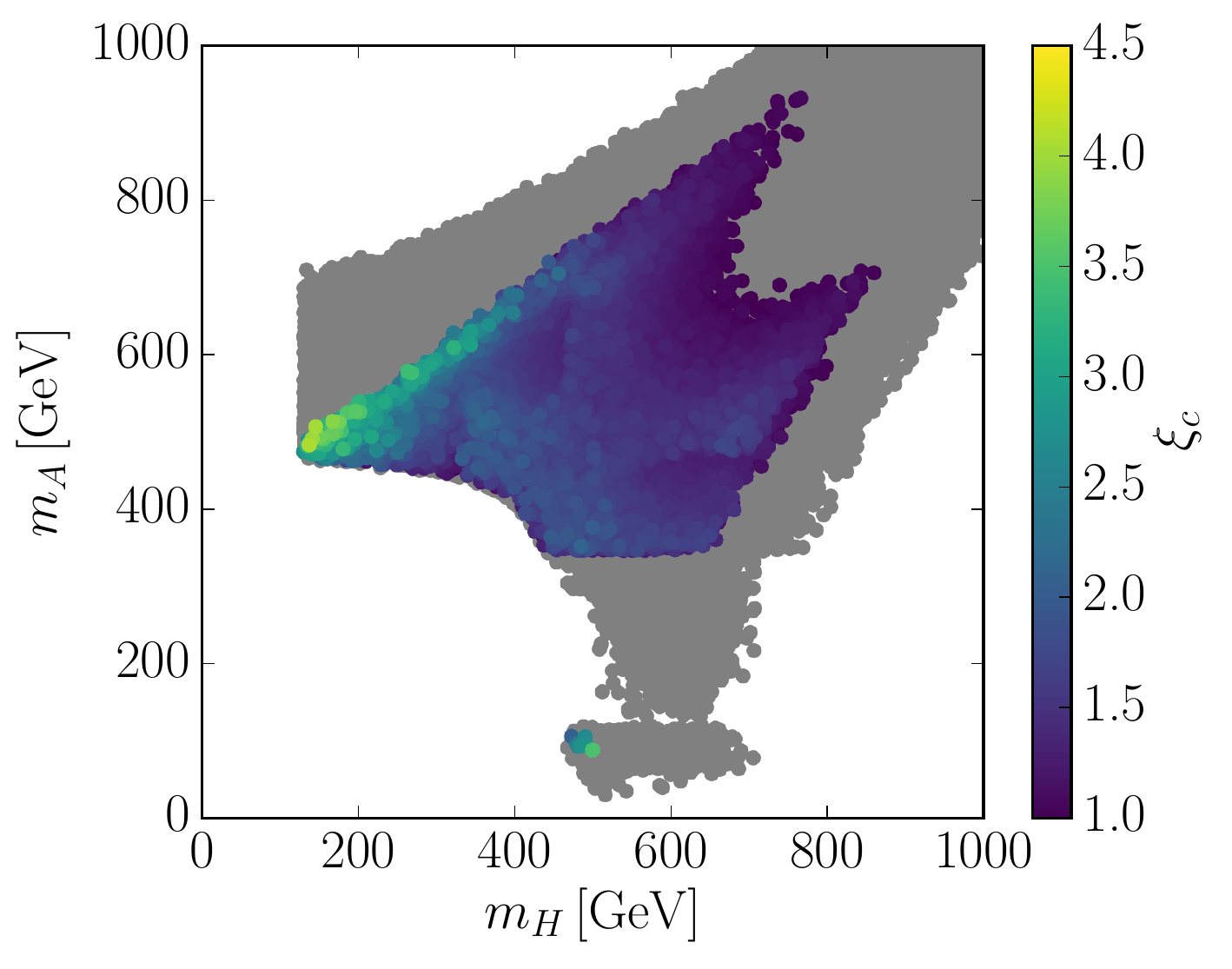}
\vspace*{-0.2cm}
\caption{Type II, $h\equiv h_{125}$: Results in the $m_A$ versus $m_H$-plane,
  showing in grey the
           parameter points passing all
           applied constraints. Points highlighted in color have a PT of strong first
           order, where the value of $\xi_c$ is indicated
           by the color code ('Arnold-Espinosa' method). \label{fig:t2mamh}}
\end{center}
\vspace*{-0.6cm}
\end{figure}
We now turn to the discussion of the compatibility of the 2HDM type II
with the requirement of $\xi_c \ge 1$ for scenarios with $h =
h_{125}$. Figure~\ref{fig:t2mamh}, which displays the values of
$\xi_c$ for all parameter points compatible with our
constraints, shows that also in the 2HDM type II there are scenarios
allowing for a strong first order PT. 
The constraints from $B$-physics observables and the EW precision tests raise
the mass scale for $m_{H^\pm}$ and at least one of the non-SM-like
Higgs bosons to higher values. For $m_A \lsim
  350$~GeV we only find few scenarios compatible with the experimental
  constraints. The pseudoscalar with $m_A \lsim 350$~GeV has a
  significant branching 
ratio into $Zh$ (up to 10\%). This final state has
been searched for by the LHC experiments. The resulting exclusion
limits severely constrain this parameter region so that there the
amount of points compatible with the experimental constraints is
substantially smaller than above the top quark pair
threshold where $A$ dominantly decays into $t\bar{t}$.\footnote{In type I, where
  also $H$ or $H^\pm$ can be 
  light and hence $A \to ZH$ or $A \to W^\pm H^\mp$ decays are
  possible, the LHC searches, which focus on the $A\to Zh_{125}$ decays, are less
restrictive.} 
When additionally a strong first order
PT is required, the mass region $130 \mbox{
    GeV} \lsim m_A \lsim 340$~GeV is completely
  excluded. As can be inferred from the plot, for these values of
  $m_A$ the heavy Higgs mass ranges between $\sim 450$ and $700\
  \text{GeV}$. In this range the occurrence of a strong first order PT is strongly limited
 by deviations from the exact alignment limit at the per mille
 level. The small portions of the VEV assigned to $H$ by these tiny
 deviations already suppress the strength of the PT strongly due to the large
 $m_H$. Once $m_H > 650\ \text{GeV}$, even for the parameter points
 extremely close to the alignment limit, the $H$ mass is finally too
 heavy to allow for a strong PT. 
The restrictions for $m_A \lsim 120$~GeV on the
other hand are less severe, as there are less experimental studies
in this mass region so that we have more points allowed
by the experimental constraints.  This increases the chances of
finding a strong first order
PT and explains why we have some coloured points for $m_A \lsim
120$~GeV. \s

\begin{figure}[h!]
\begin{center}
\includegraphics[width=8.5cm]{Scale} \\
\includegraphics[width=0.49\textwidth]{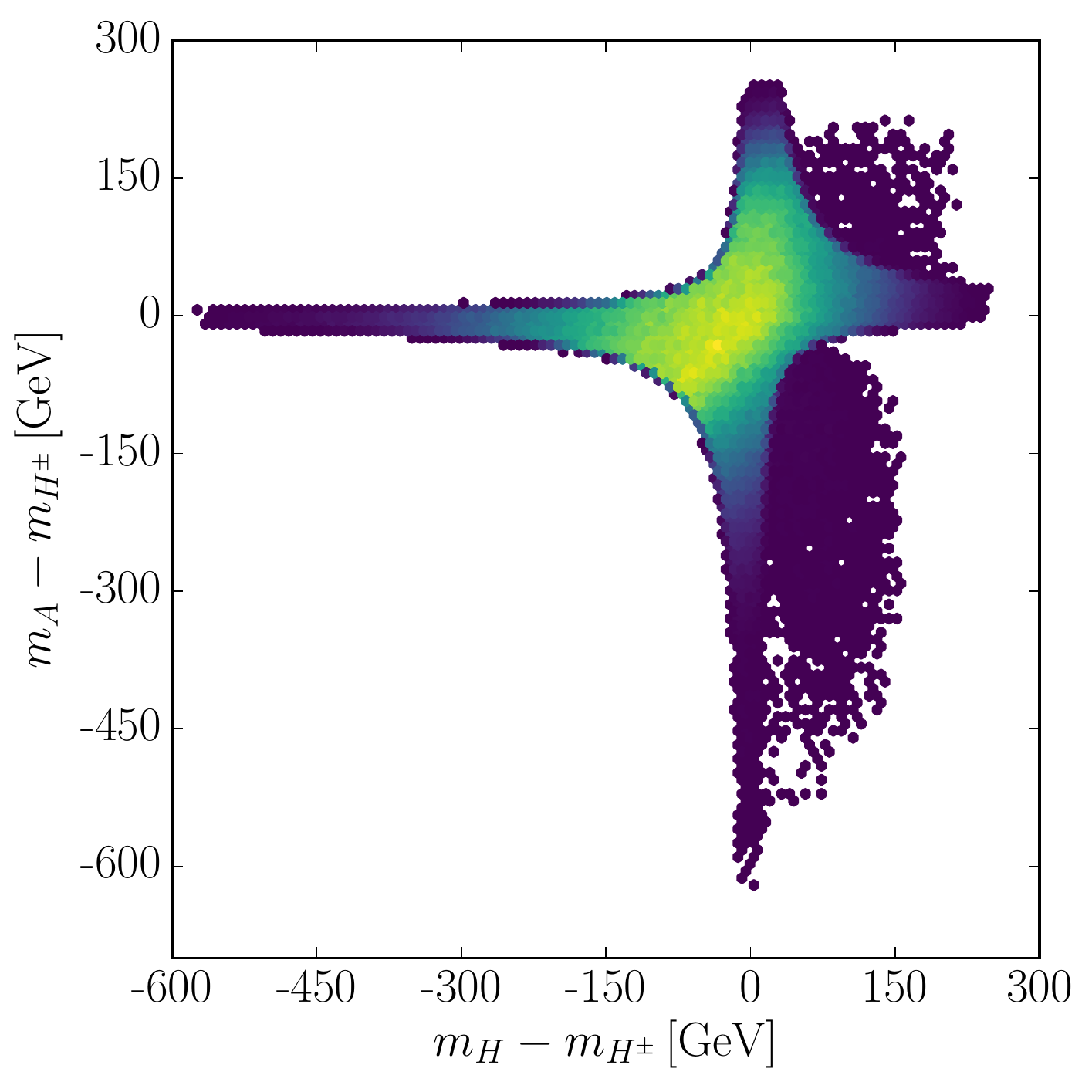}\hfill
\includegraphics[width=0.49\textwidth]{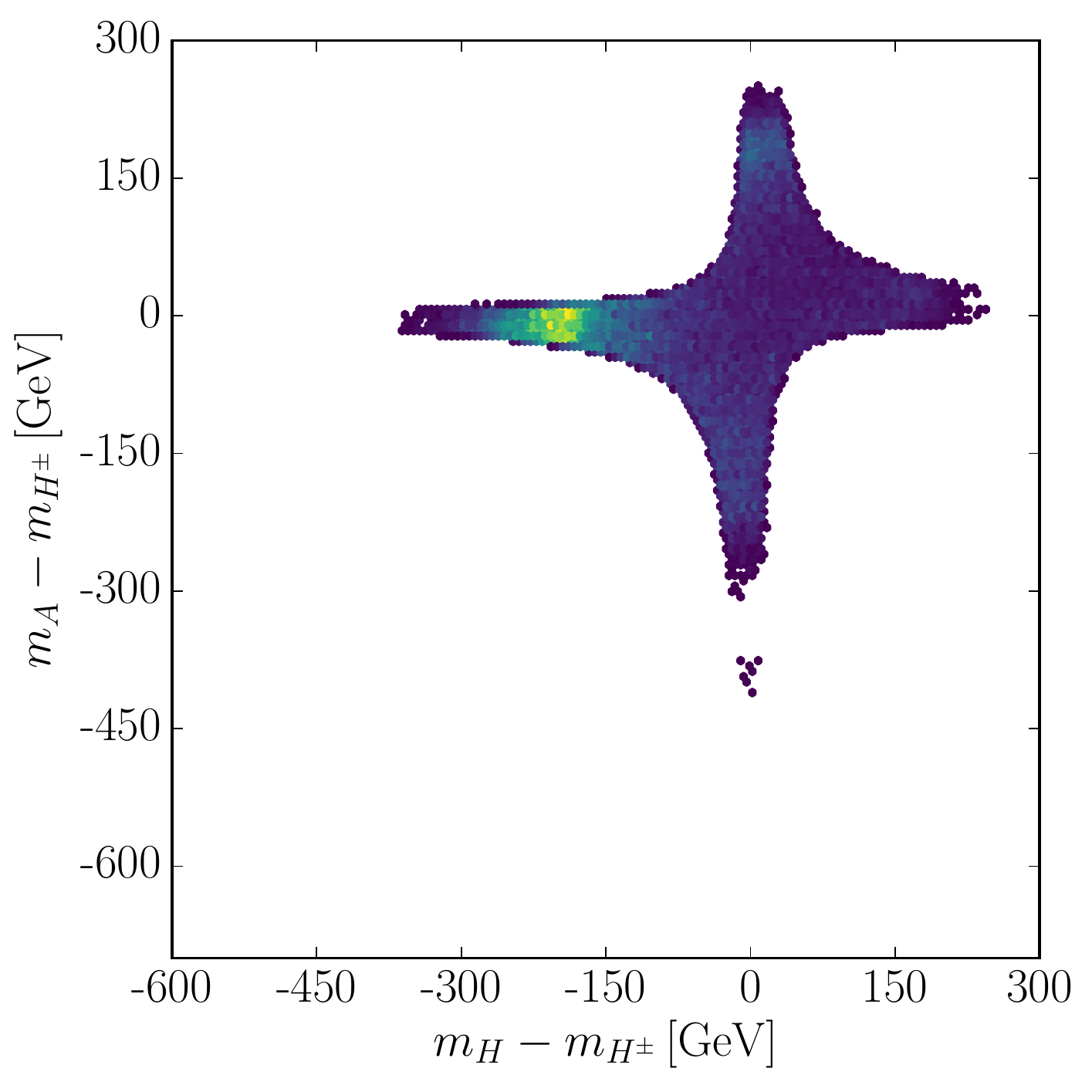} 
\vspace*{-0.2cm}
\caption{Type II, $h\equiv h_{125}$: The mass difference $m_A-m_{H^\pm}$ versus
  $m_H-m_{H^\pm}$. The colour code shows the
  relative frequency of left: all points passing the constraints; right:
  all points with additionally $\xi_c \ge 1$ ('Arnold-Espinosa'
  method). \label{fig:t2massdiffs}}
\end{center}
\vspace*{-0.6cm}
\end{figure}
The implications of the requirement of a strong first order phase
transition in the type II 2HDM for LHC phenomenology can be read off
Fig.~\ref{fig:t2massdiffs}. Scenarios with all non-SM-like Higgs
masses being close to each other are favoured
by the experimental constraints, {\it cf.}~Fig.~\ref{fig:t2massdiffs}
(left)\footnote{The dark blue points with non-zero mass gaps are
  points in the wrong-sign regime (see below). Due to the different
  coupling structure in the wrong-sign regime the constraints from the
  $\rho$ parameter have a different shape when projected into the
  plane of the plot. }. A
strong first order PT, however, prefers scenarios with $m_A
\approx m_{H^\pm}$ and with a large positive mass gap between $m_{H^\pm}$ and
$m_H$ and hence also $m_A - m_H \gsim 180$~GeV, {\it
  cf.}~Fig.~\ref{fig:t2massdiffs} (right). Scenarios where $m_H
\approx m_{H^\pm}$ and $|m_A-m_{H^{\pm}} (\approx m_H)| > 0$ and also
those where $m_A \approx m_{H^\pm}$ and $m_H - m_{H^\pm} > 0$
are rarer, as they would require a much heavier $H$, given that
$m_{H^\pm} \ge 480$~GeV in type II models. A heavy $H$ with
non-vanishing VEV tends to reduce the strength of the phase
transition. For the same reason scenarios where all non-SM-like Higgs
bosons have similar masses are not very
probable either. While again $A \to ZH$ is a typical
decay that is possibly realised for strong first order PTs, the
non-discovery of such a decay does not exclude $\xi_c \ge 1$ as other
scenarios can be realised as well. We find that scenarios with $m_A
\gsim 460$~GeV are preferred 
and namely those scenarios that are
located in the alignment limit with $\tan\beta
\approx 1$. This is, however, not due to the first order PT but already
found by only imposing the theoretical and experimental constraints.
\s

In the type II 2HDM there are parameter regions compatible with the
experimental constraints where the coupling of the $h_{125}$ to the
massive gauge bosons is of opposite sign with respect to the coupling
to down-type fermions. This wrong-sign regime
\cite{Ferreira:2014naa,Ferreira:2014dya,Fontes:2014tga} has
interesting phenomenological implications like the non-decoupling of
heavy particles \cite{Ferreira:2014naa,Krause:2016xku}. Future
precision measurements of the signal rates will allow to constrain or exclude
this parameter region
\cite{Ferreira:2014naa,Ferreira:2014dya,Muhlleitner:2016mzt}. The
question arises to which extent the requirement of a strong
PT is able to restrict the wrong-sign regime.
\begin{figure}[t!]
\begin{center}
\includegraphics[width=8.5cm]{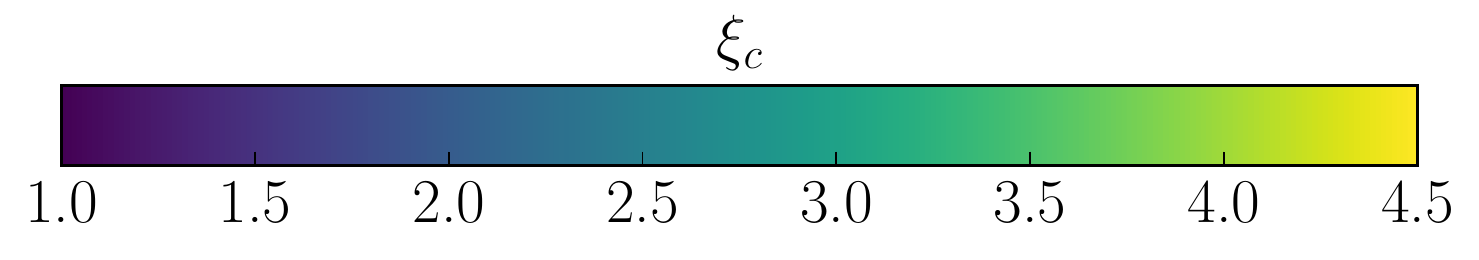} \\
\includegraphics[width=0.49\textwidth]{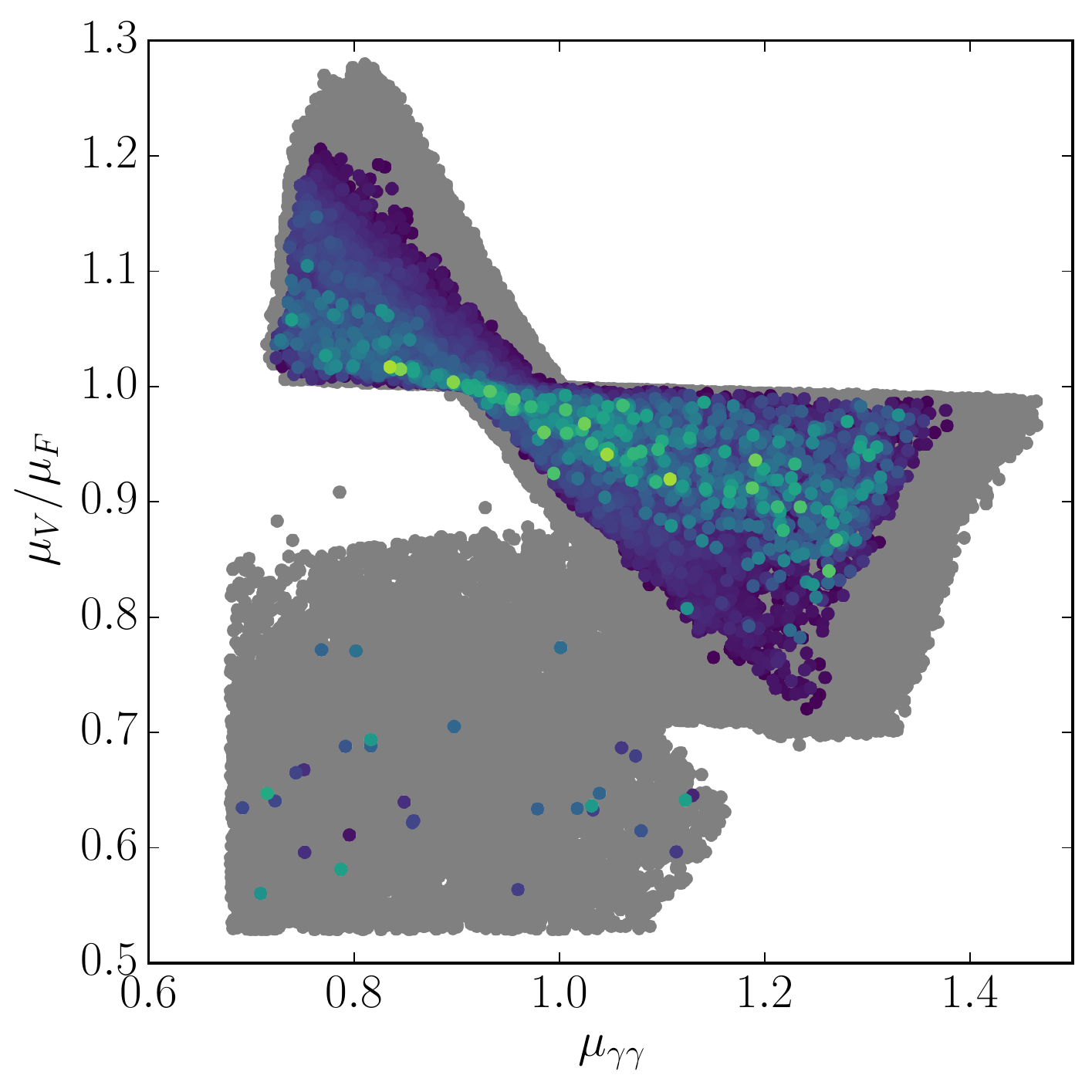}\hfill
\includegraphics[width=0.49\textwidth]{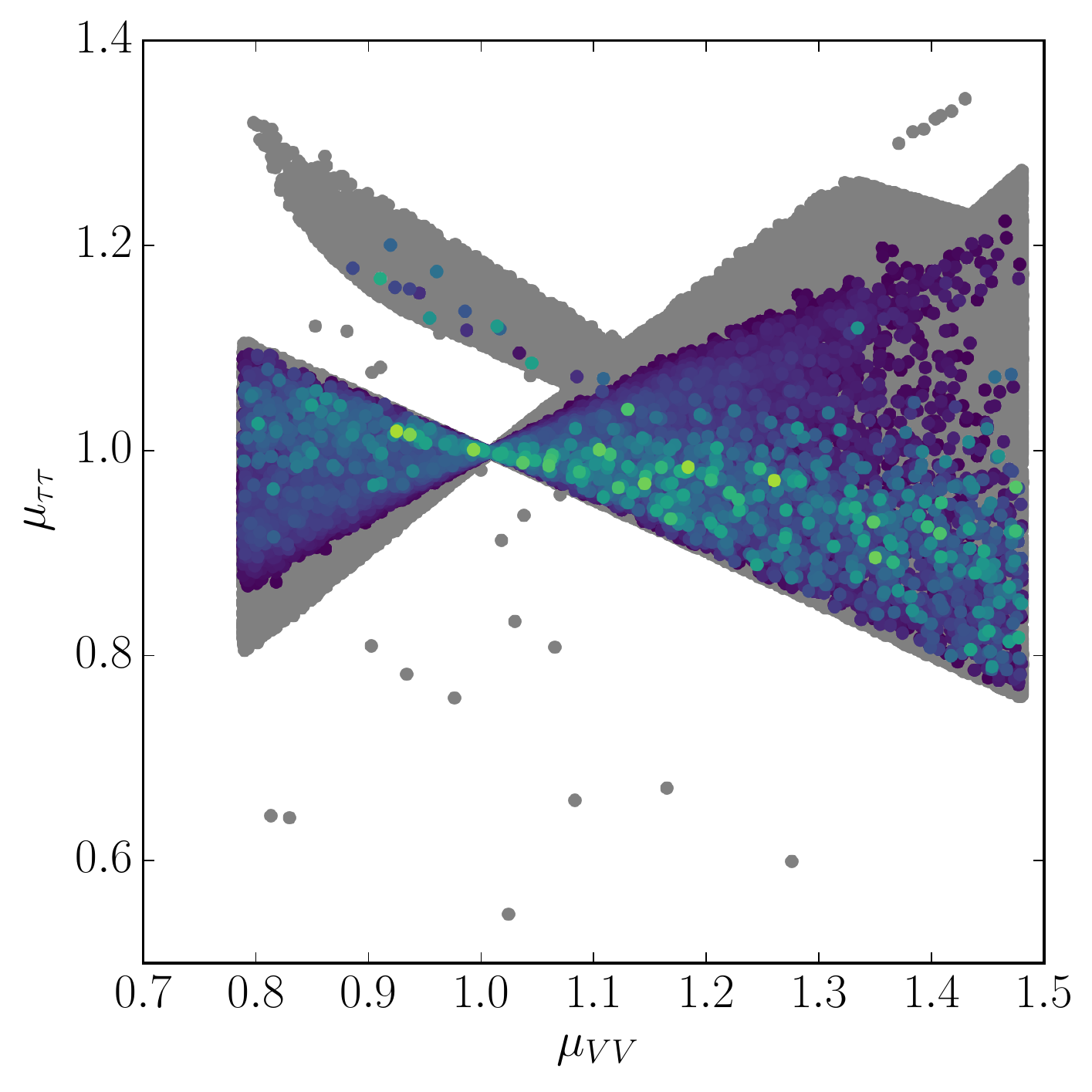} 
\vspace*{-0.2cm}
\caption{Type II, $h\equiv h_{125}$:
  $\mu_V / \mu_F$ versus $\mu_{\gamma\gamma}$ (left)
  and $\mu_{\tau\tau}$ versus $\mu_{VV}$ (right); grey: all points
    passing the applied constraints, colour:
  all points with additionally $\xi_c \ge 1$ ('Arnold-Espinosa'
  method). The colour code indicates the value of
  $\xi_c$. \label{fig:t2rates}}
\end{center}
\vspace*{-0.6cm}
\end{figure}
Figure~\ref{fig:t2rates} (left) displays $\mu_V/\mu_F$ versus
$\mu_{\gamma\gamma}$. Among the grey points, which show the scenarios
passing all constraints, the outliers in the left bottom corner of the
plot correspond to the wrong-sign regime. The coloured points fulfill
$\xi_c \ge 1$ and show that a strong PT strongly disfavours the
wrong-sign regime. This can also be observed in Fig.~\ref{fig:t2rates}
(right) where the distribution of $\mu_{\tau\tau}$ versus $\mu_{VV}$
is displayed. The wrong-sign regime scenarios are given by the
outliers in the upper left corner of the plot. This behaviour can be
understood by the fact that the VEV $\langle H \rangle$ of the heavy
CP-even Higgs normalised to the SM VEV for $h \equiv h_{125}$ is
given by
\beq
\frac{\langle H \rangle^2}{v^2} = \cos^2 (\beta-\alpha) \;.
\eeq
In the wrong-sign regime non-zero values of $\cos
(\beta-\alpha)$ are still compatible with the data. This means that
$H$ can take a significant fraction of the VEV and drive the PT. If
$H$ is not light enough, the PT is reduced to values $\xi_c < 1$.
We also observe that the maximum value of
  $\mu_{\gamma\gamma}$ is reduced from about 1.46 to about 1.38.

\subsection{Type I: Parameter Sets with $H \equiv
  h_{125}$ \label{sec:Ihhh125}}
\begin{figure}[t!]
 \begin{center}
\includegraphics[width=0.6\textwidth]{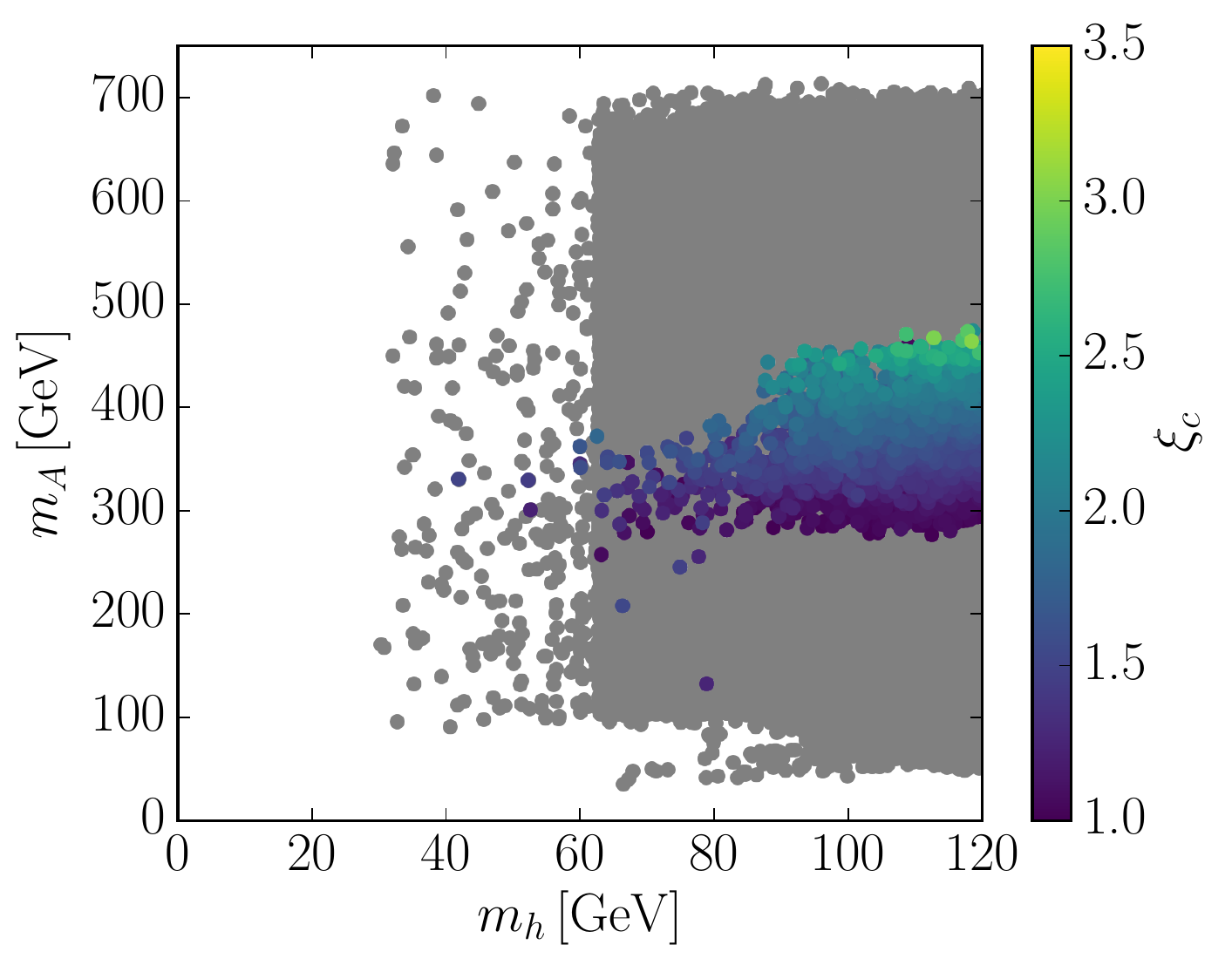}
\vspace*{-0.2cm}
\caption{Type I, $H \equiv h_{125}$: Results in the $m_A$ versus $m_h$-plane,
   showing in grey the
           parameter points passing all
           applied constraints. Points highlighted in color have a PT of strong first
           order, where the value of $\xi_c$ is indicated
           by the color code ('Arnold-Espinosa' method). \label{fig:t1mamh_inverted}}
\vspace*{-0.6cm}
 \end{center}
\end{figure}
We now investigate scenarios where the heavier of the
two CP-even Higgs bosons is the SM-like Higgs boson, {\it
  i.e.}~$H\equiv h_{125}$. Figure~\ref{fig:t1mamh_inverted} displays
in the $m_A$ versus $m_h$ plane in grey all points passing the constraints and
in colour all parameter points also compatible with $\xi_c \ge
1$. First, we observe that independent of the strength of the PT,
there are only few scenarios with $m_h \lsim 65$~GeV. This is due to
the fact that the decay $H \equiv h_{125} \to hh$
can change the total width of $h_{125}$ such that its branching ratios
into SM final states lead to signal rates not compatible with the LHC
data any more. In this mass region there are hardly any points with
$\xi_c \ge 1$. The requirement of $\xi_c \ge 1$ also restricts the
mass of the pseudocscalar 
to the region $280 \mbox{ GeV} \lsim m_A \lsim 480 \mbox{ GeV}$, with
the exception of a few outliers.
\begin{figure}[t!]
\begin{center}
\includegraphics[width=8.5cm]{Scale}
\includegraphics[width=0.49\textwidth]{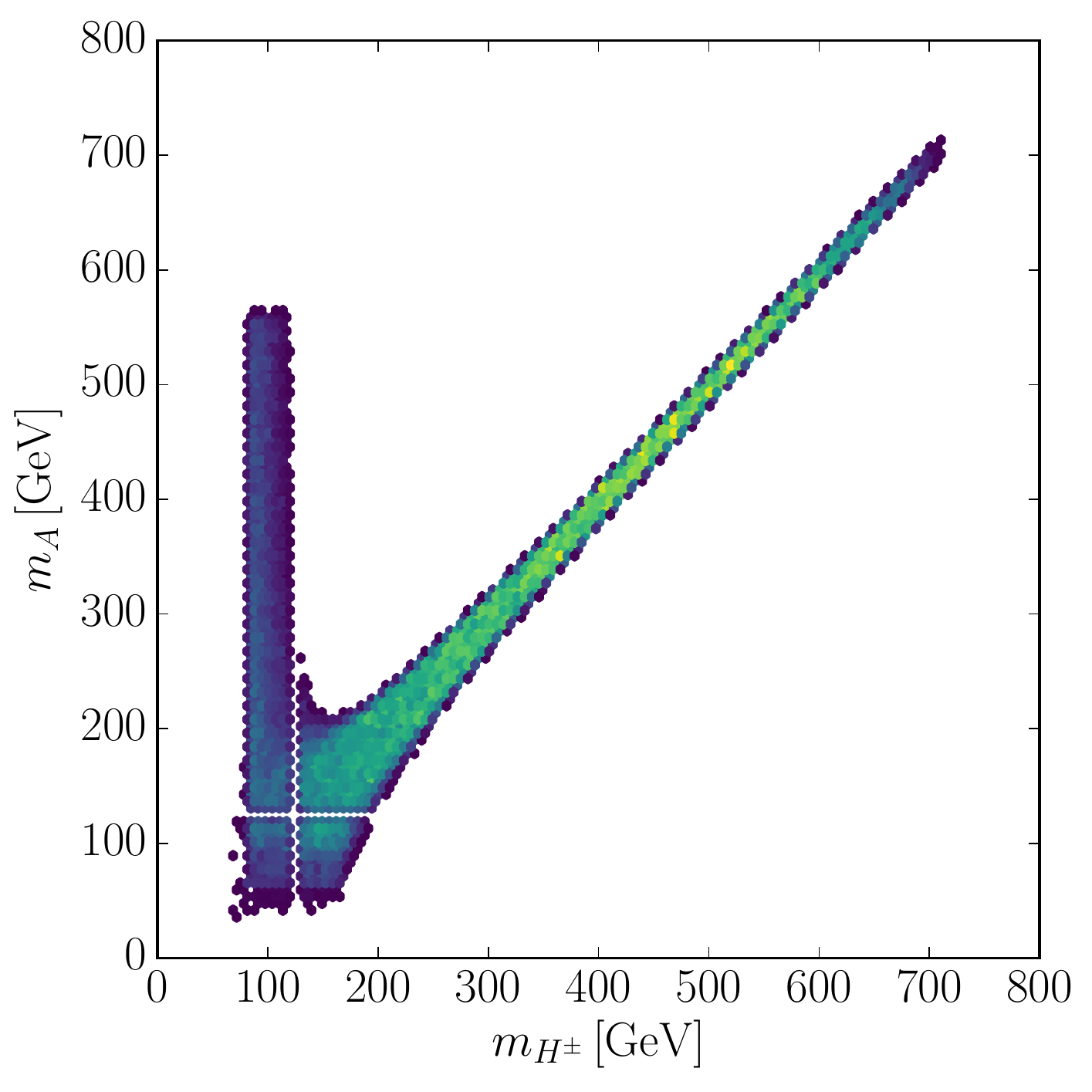}\hfill
\includegraphics[width=0.49\textwidth]{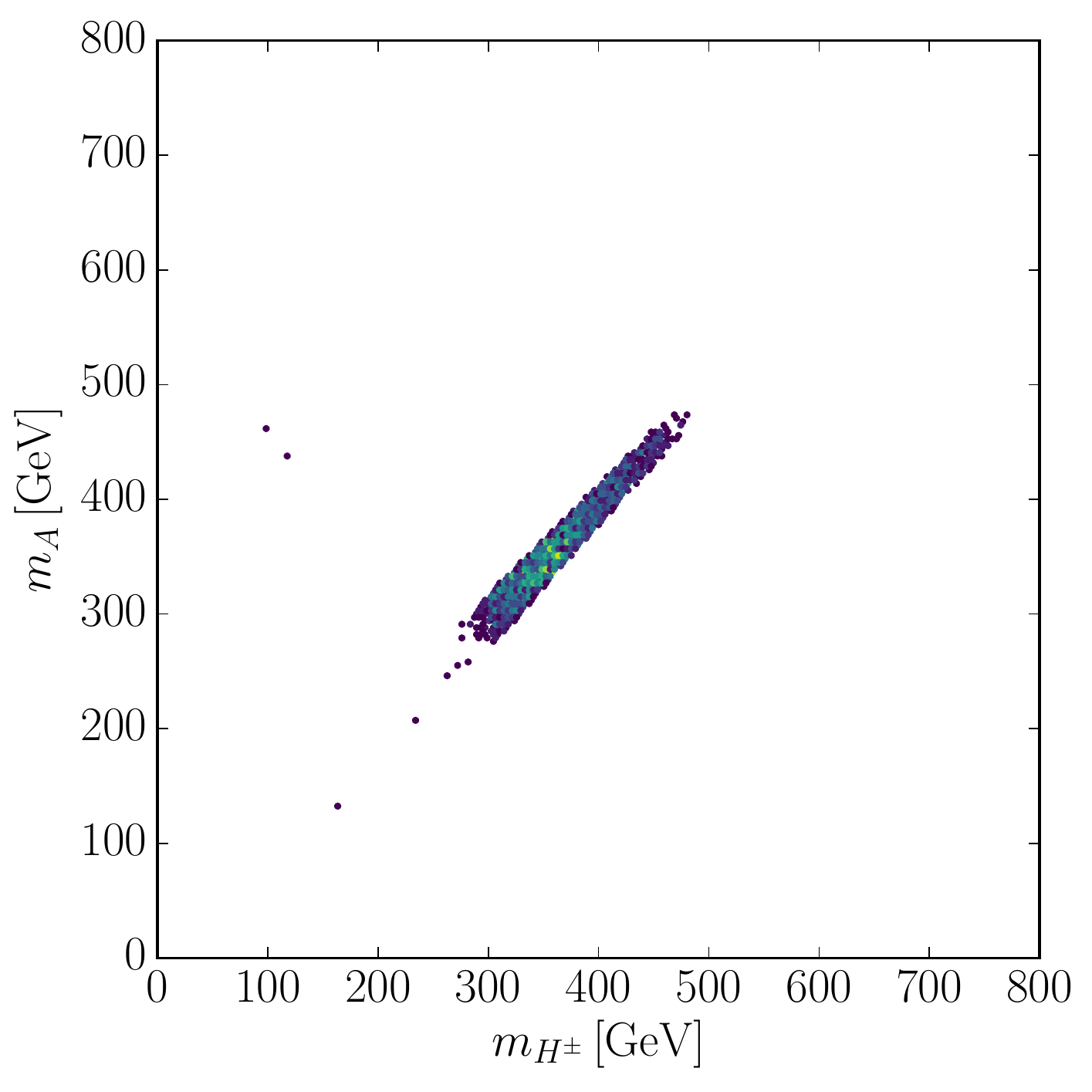} 
\vspace*{-0.2cm}
\caption{Type I, $H\equiv h_{125}$:
  The masses $m_A$ versus $m_{H^\pm}$. The colour code shows the
  relative frequency of left: all points passing the constraints; right:
  all points with additionally $\xi_c \ge 1$ ('Arnold-Espinosa'
  method). \label{fig:t1hfurtherplots}}
\end{center}
\vspace*{-0.6cm}
\end{figure}
%
The strongest PTs are reached for larger $m_A$,
close to 480~GeV. Figure~\ref{fig:t1hfurtherplots} displays the
distribution of the masses for $A$ and $H^\pm$ after applying all
constraints (left) and when in addition $\xi_c \ge 1$ is demanded
(right). With the exception of a few outliers, the strong PT
restricts the mass region of the charged Higgs
boson to $300\mbox{  GeV} \lsim m_{H^\pm} \lsim 480 \mbox{ GeV}$.
As we demand the heavier of the two CP-even Higgs
bosons to be light, the mass scale $m_{12}^2$, which determines its
mass, cannot be too large. For the PT to be strong we need large
quartic couplings. Since $\lambda_2$, which enters $m_H$, must
not be large, we are left with $\lambda_4$ and $\lambda_5$ driving the
PT, as can be inferred from the rather large mass values for $A$ and
$H^\pm$, namely the mass gap between $m_H$ and $m_{A, H^\pm}$. When,
on the other hand, the masses of the heavy Higgs bosons $A$ and
$H^\pm$ become larger by increasing the involved quartic couplings the
interplay of Higgs 
self-couplings and masses reduces $\xi_c$ again. We conclude that a
strong PT in the 2HDM type I with two light CP-even Higgs boson excludes
heavy Higgs bosons above about 500~GeV and enforces a mass gap between
$m_A \approx m_{H^\pm}$ and $m_H$. The decay $A \to HZ$, however, is
suppressed because of $\sin (\beta- \alpha) \sim 0$ for $H \equiv
h_{125}$. The decay $A \to hZ$ on the other hand, is allowed. For
pseudoscalar masses above the top pair threshold, it competes, however,
with the decay $A \to t\bar{t}$. \s
\begin{figure}[t!]
\begin{center}
\includegraphics[width=8.5cm]{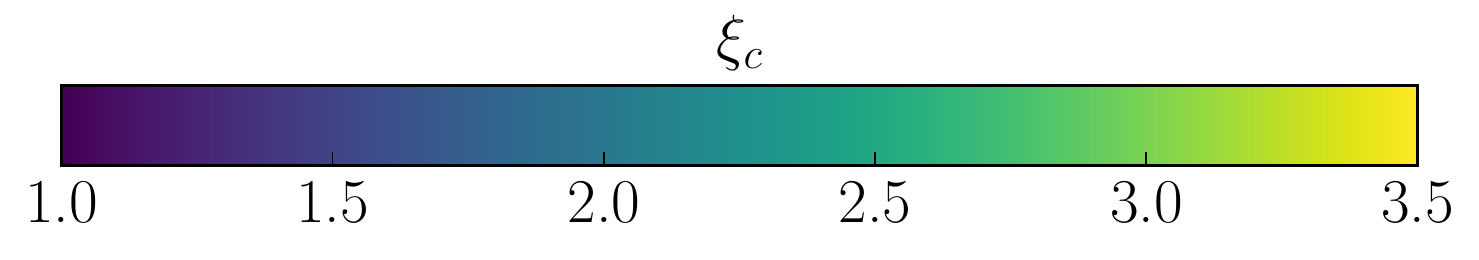} \\
\includegraphics[width=0.49\textwidth]{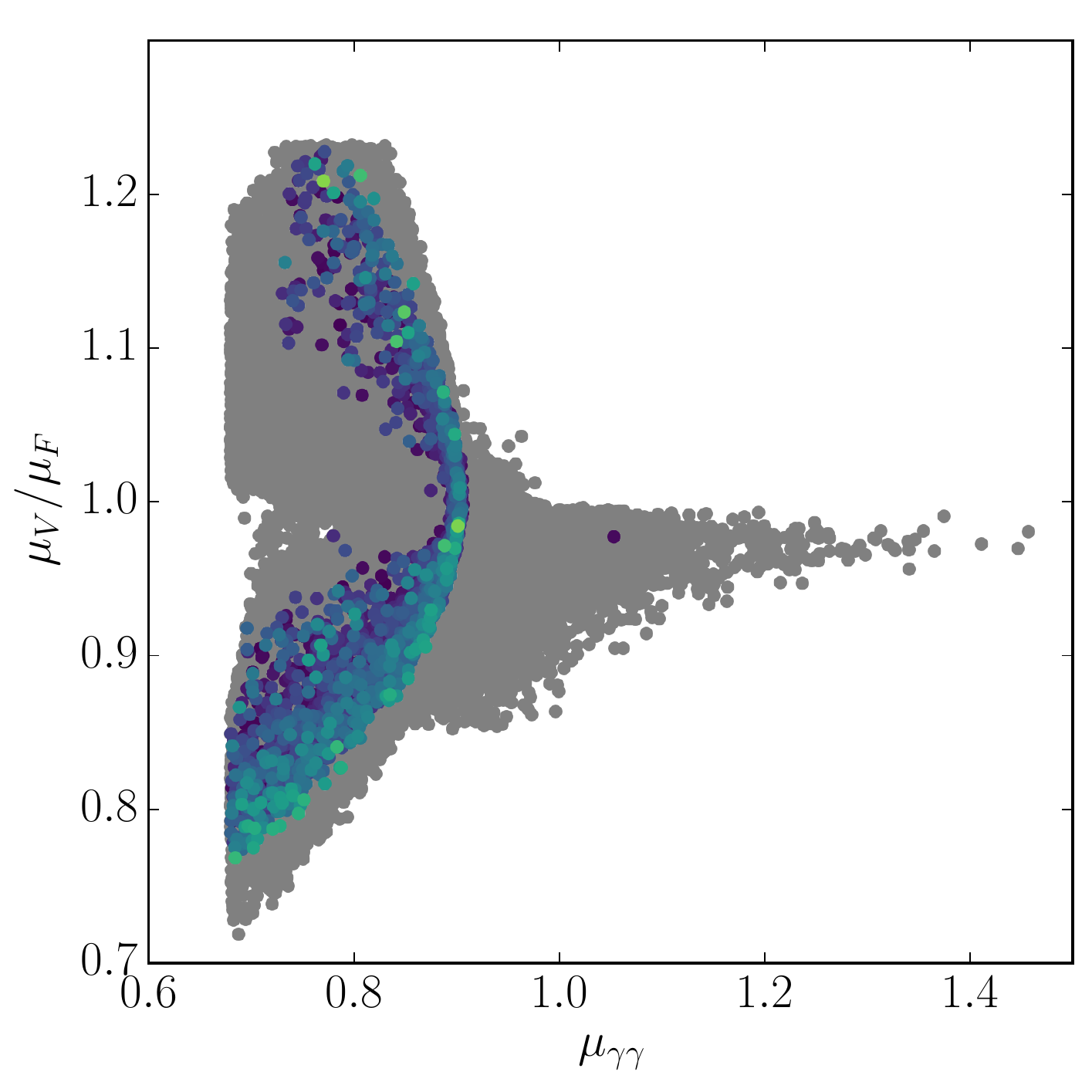}\hfill
\includegraphics[width=0.49\textwidth]{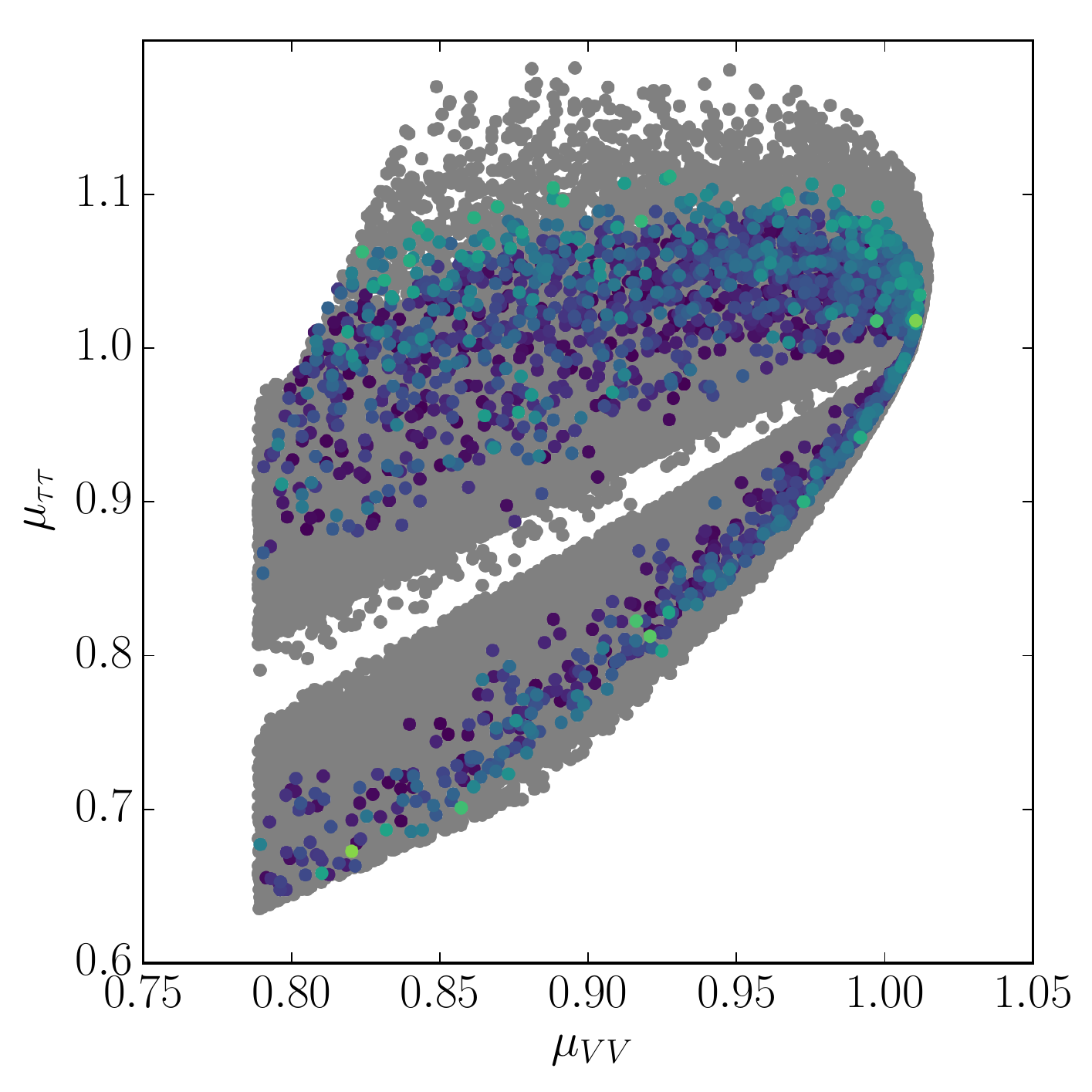} 
\vspace*{-0.2cm}
\caption{Type I, $H \equiv h_{125}$:
   $\mu_V / \mu_F$ versus $\mu_{\gamma\gamma}$ (left)
  and $\mu_{\tau\tau}$ versus $\mu_{VV}$ (right); grey: all points
    passing the applied constraints, colour:
  all points with additionally $\xi_c \ge 1$ ('Arnold-Espinosa'
  method). The colour code indicates the value of
  $\xi_c$. \label{fig:t1rates_inverted}}
\end{center}
\vspace*{-0.6cm}
\end{figure}

The implications of a strong PT for the Higgs data are shown in
Fig.~\ref{fig:t1rates_inverted}. There are practically
no points any more with values beyond 0.9 for the photonic rate,
although rates of up to about
1.46 are still compatible with the
Higgs data. Also the decays into $\tau$ final states cannot exceed
1.11 in case of $\xi_c \ge 1$. 

\subsection{Type II: Parameter Sets with $H \equiv
  h_{125}$ \label{sec:IIhhh125}}
\begin{figure}[t!]
 \begin{center}
\includegraphics[width=0.6\textwidth]{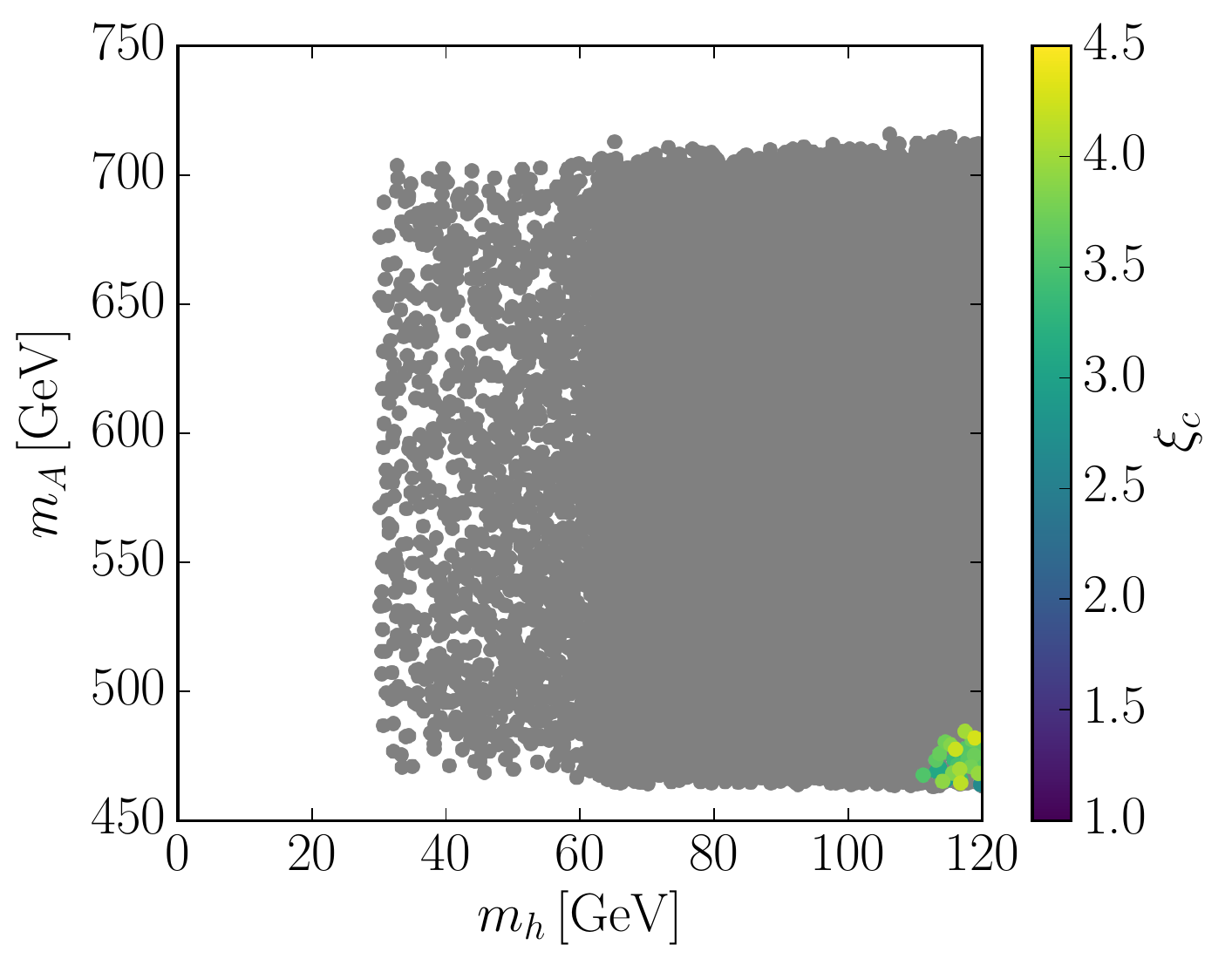}
\vspace*{-0.3cm}
\caption{Type II, $H \equiv h_{125}$: Results in the $m_A$ versus $m_H$-plane,
   showing in grey the
           parameter points passing all
           applied constraints. Points highlighted in color have a PT of strong first
           order, where the value of $\xi_c$ is indicated
           by the color code ('Arnold-Espinosa' method).
\label{fig:t2mamh_inverted}}
 \end{center}
\vspace*{-0.6cm}
\end{figure}

In the 2HDM type II with $H\equiv h_{125}$ the implications of a
strong PT on the mass pattern are very pronounced,
as can be inferred from Fig.~\ref{fig:t2mamh_inverted}.
The requirement of $\xi_c \ge 1$ excludes a large portion of the
parameter space, which is still compatible with the applied
constraints. Scenarios with
$m_A \approx m_{H^\pm} \gsim 480$~GeV
are forbidden if $\xi_c \ge 1$. Furthermore, very light scalars with
$m_h \lsim 110$~GeV are not  
compatible with a strong PT. The tension between the requirement of
light scalar masses and the wish to have a strong PT makes a strong
link between baryogenesis and collider phenomenology. \s

\begin{figure}[t!]
\begin{center}
\includegraphics[width=8.5cm]{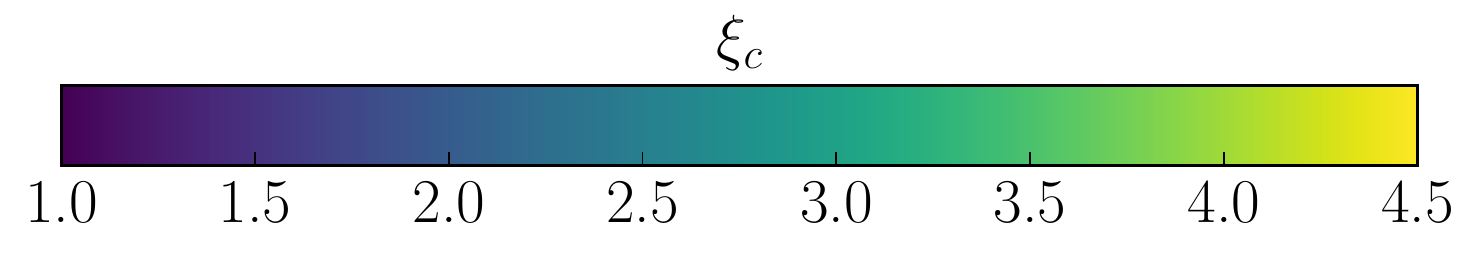} \\
\includegraphics[width=0.49\textwidth]{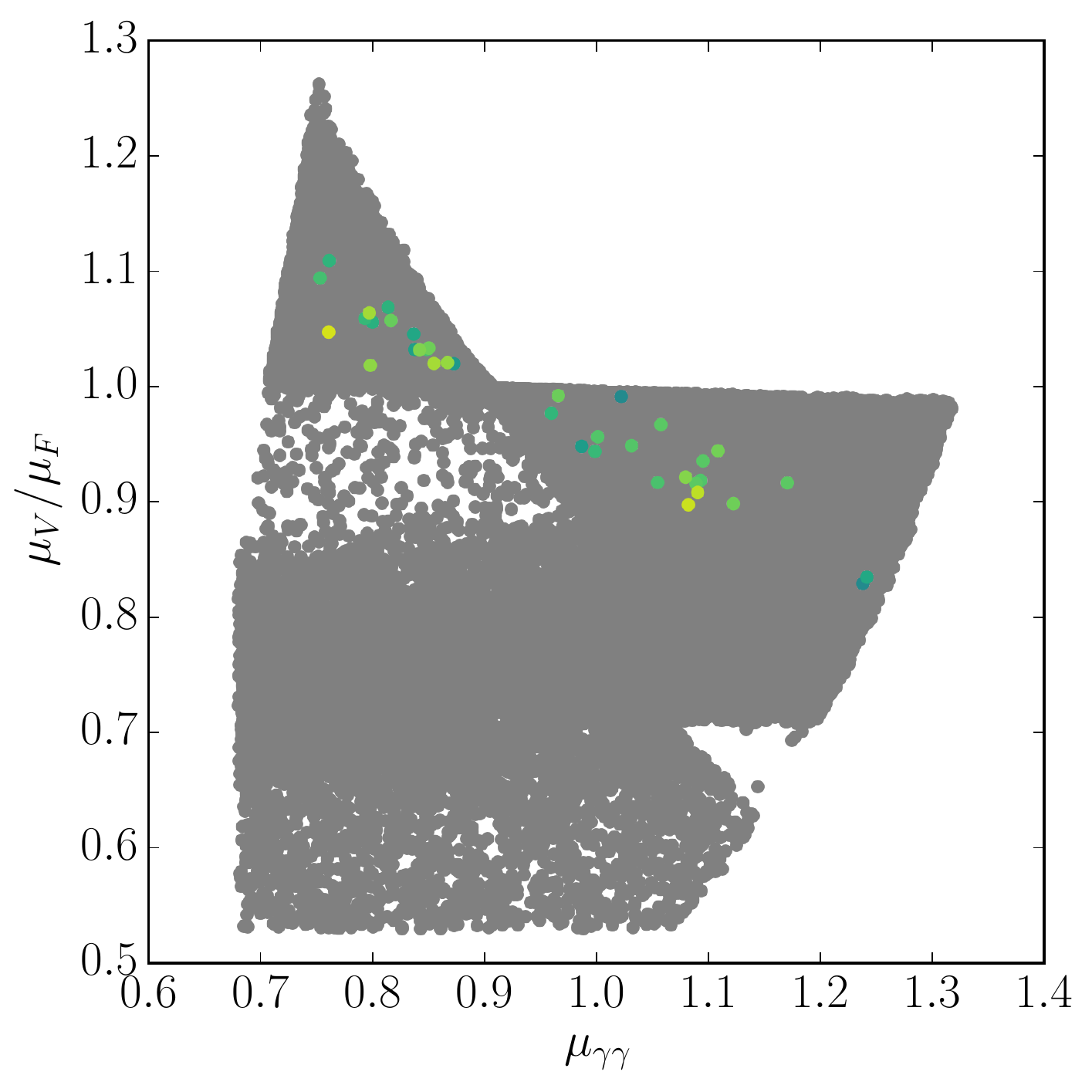}\hfill
\includegraphics[width=0.49\textwidth]{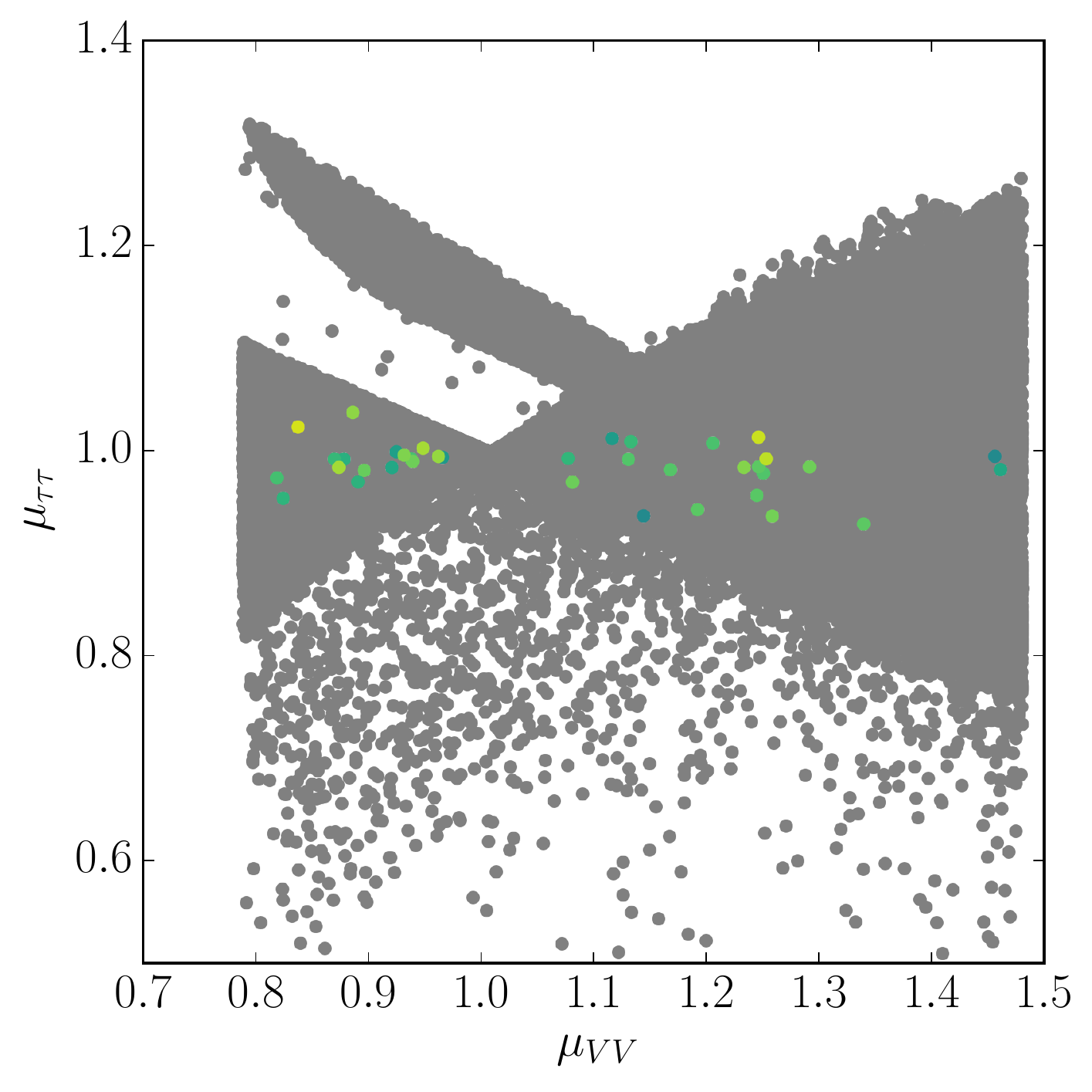} 
\vspace*{-0.2cm}
\caption{Type II, $H \equiv h_{125}$:
   $\mu_V / \mu_F$ versus $\mu_{\gamma\gamma}$ (left)
  and $\mu_{\tau\tau}$ versus $\mu_{VV}$ (right); grey: all points
    passing the applied constraints, colour:
  all points with additionally $\xi_c \ge 1$ ('Arnold-Espinosa' 
  method). The colour code indicates the value of
  $\xi_c$. \label{fig:t2rates_inverted}}
\end{center}
\vspace*{-0.6cm}
\end{figure}

Further implications for LHC phenomenology are shown in
Fig.~\ref{fig:t2rates_inverted} where the signal rates are displayed
before (grey) and after (coloured) imposing a strong PT.
All scenarios with $\xi_c \ge 1$ are located in the
  correct-sign regime (given by the triangle areas in the plots),
  whereas the wrong-sign regime (given by the outliers) is completely
  excluded by a strong PT. For the Higgs
measurements, this means that the observation of $\mu_V/\mu_F < 1$
together with $\mu_{\gamma\gamma} \lsim 0.9$ is excluded,
as well as the observation of $\mu_{\tau\tau} \gsim 1.04$. Furthermore, the
region where $\mu_{\tau\tau} \lsim 0.9$ because 
of possible decays $H \to hh$, is excluded by the demand
of a strong PT. 

\section{Conclusions \label{sec:conclusions}}
In this paper we investigated the strength of the EW phase transition
in the framework of the CP-conserving 2HDM. For this purpose we computed
the loop-corrected effective potential at non-zero temperature including the
resummation of the daisy graphs for the bosonic masses
following the 'Arnold-Espinosa' method.
We applied a renormalisation scheme that preserves the position of the
minimum and where both the loop-corrected
masses of the Higgs bosons and the mixing angles are renormalised to
their tree-level values. This is in contrast to earlier works which
focus solely on the Higgs boson masses. Our renormalisation allows us
to efficiently scan the whole 2HDM parameter space and
test the compatibility of the model with the theoretical and
experimental constraints. This is possible since our renormalisation fixes not only
the Higgs mass values but, through the mixing angles, also the Higgs
couplings to their tree-level values. 

We performed an extensive scan in the parameter space of the 2HDM and
retained only those points that are compatible with the state-of-art
theoretical and experimental constraints. For these parameter points
we determined the value of $\xi_c$. Subsequently, we performed a
comprehensive and systematic 
analysis in four 2HDM configurations: For the 2HDM type I and II, with
either $h$ or $H$ identified with the SM-like Higgs boson, we
investigated the implications of the requirement of a strong PT, {\it
  i.e.}~$\xi_c \ge 1$, for LHC phenomenology. Our results can be
summarised as follows: Both the 2HDM type I and type II, with either of the CP-even
Higgs bosons being the SM-like Higgs boson, are found to be compatible with the
theoretical and experimental constraints on the model and a strong
first order PT. The strong PT, however, strongly
  constrains the enhanced rates into photonic final states for the
  2HDM type I, and to some extent also for type II.
Furthermore, in the 2HDM type II with $h=h_{125}$ the wrong-sign
regime is strongly restricted by the requirement of $\xi_c \ge 1$. In
case $H=h_{125}$ the wrong-sign regime is even excluded for a strong
PT. In more detail, our results for the four different realisations
of the 2HDM are:
\begin{itemize}
\item For the 2HDM type I with $h \equiv h_{125}$, we confirm earlier
  results which find that  a large mass splitting between the heavy
  scalars is favourable for a strong PT. The preferred scenarios are
  the ones with $m_A \approx m_{H^\pm} \approx 400-500$~GeV and $m_A -
  m_H \gsim 200$~GeV.
 However, we also find that scenarios
  with different hierarchies among the heavy Higgs bosons (but at
  least one of $H$ and $A$ nearly mass degenerate with $H^\pm$) or
  with degenerate heavy Higgs bosons $H$, $A$ and $H^\pm$ are allowed,
  though much less frequent. The maximally allowed photonic rate is
  reduced from 1.5 to 1.1 in case of $\xi_c \ge 1$.
\item We find in the 2HDM type II with $h=h_{125}$ that scenarios
  with $130 \mbox{ GeV} \lsim m_A \lsim 340 \mbox{
  GeV}$, which are already strongly constrained by LHC searches in $A
\to Zh_{125}$, are completely excluded by the requirement of a strong PT.
This requirement also restricts the wrong-sign regime considerably.
The maximum value of $\mu_{\gamma\gamma}$ is reduced
  from about 1.46 to about 1.38.
\item In the 2HDM type I with two light CP-even Higgs bosons, namely $H\equiv
h_{125}$, the heavy Higgs masses cannot exceed 480~GeV, although
experimentally still allowed, if $\xi_c \ge 1$. Furthermore, this
enforces a mass gap of about 155~GeV
between the heavier and the lighter Higgs bosons.
A strong PT is found to exclude
almost completely scenarios with $m_h \lsim 65$~GeV. It strongly
reduces $\mu_{\gamma\gamma}$ from 1.46 down to 0.9
(with very few exceptions) and limits the
$\tau$ final state rates to values below 1.11.
\item In the 2HDM type II with $H \equiv h_{125}$ the
    tension between
  a light CP-even Higgs mass spectrum and a strong PT excludes large portions
  of the parameter space.  The observation of heavy
Higgs bosons with masses above 480 GeV or of a light
Higgs boson with a mass below 110 GeV is excluded by the requirement of a strong
PT. Furthermore, simultaneously reduced values of $\mu_V/\mu_F <1$ and 
$\mu_{\gamma\gamma} \lsim 0.9$ are not compatible with $\xi_c \ge 1$,
nor values of $\mu_{\tau\tau} \gsim 1.04$. The reason is that the
wrong-sign regime is excluded by a strong PT.  The
requirement of a strong PT also excludes parameter
regions with reduced $\mu_{\tau\tau} \lsim 0.9$, resulting from
Higgs-to-Higgs decays $H \to hh$, which contribute to the total width
of the SM-like Higgs boson. 
\end{itemize}

Our results show that there is a strong interplay between the
requirement of successful baryogenesis and LHC phenomenology. The
realisation of a strong EW phase transition leads to testable
consequences for collider phenomenology. The
systematic investigations performed in this work serve as basis for
further analyses of the LHC phenomenology of 2HDM models featuring 
a strong EW phase transition.

\vspace*{0.5cm}

\section*{Acknowledgements}
MM wants to thank Prof. Dr. Werner Bernreuther for suggesting the
study of the 2HDM phase transition in her diploma
thesis. This triggered a long time ago the realisation of this paper.
We are particularly grateful to Marco Sampaio for many enlightening discussions.
We also want to thank David Lopez-Val and
Florian Staub for very helpful discussions. We are indebted to Rui
Santos for comments on our draft. 
We furthermore thank Jos\'e Eliel Camargo-Molina, Pedro Ferreira,
Luminita Mihaila, Ben O'Leary, Michael
Spira and Hanna Ziesche for useful comments.
MK acknowledges financial support by the Graduiertenkolleg ``GRK
1694: Elementarteilchenphysik bei h\"ochster Energie und h\"ochster
Pr\"azision''. 
AW acknowledges financial support by the ``Karlsruhe School of Elementary Particle
and Astroparticle Physics: Science and Technology 
(KSETA)''.

\vspace*{0.5cm}
\section*{Appendix}
\appendix
\section{Masses with thermal
  corrections}\label{sec:app_debye_corrections}
In the following we give the mass formulae for the SM particles in
terms of the field configurations $\omega_k$ ($k=1,2,3$). The masses
of the physical particles are obtained from these formulae when the
$\omega_k$ take the values that minimise the loop-corrected
effective potential $\tilde{V}$, Eq.~(\ref{eq:totalpot}). At non-zero
temperature these are the $\bar{\omega}_k$, which we obtain from the
numerical determination of the global minimum of
$\tilde{V}$ at fixed $T$. For $T=0$ they are given by
the VEVs $v_1$ and $v_2$.
We only need the tree-level
relations for the masses. At non-zero temperature we furthermore
include the Debye corrections to the masses of the scalars and the
longitudinal gauge bosons stemming from the resummation of the
daisy graphs. The mass formulae will be specified in the following.

\subsection{Fermion Masses}
The fermion masses do not get a Debye correction, and therefore the
mass squared of a fermion $f$ at temperature $T$ is given by
\begin{align}
	m^2_f(T) &= \frac{1}{2} y_f^2 \vert \phi_k^{c,0} \vert^2 =
        m^2_f(T=0) \, \frac{\vert \phi_k^{c,0} \vert^2}{v_k^2} \,,
\end{align}
where $y_f$ is the \treelevel{} Yukawa coupling and $k=1,2$ denotes the
classical constant field configuration doublet $\Phi^c_k$ to
which the fermion
couples. This depends on the
type of the 2HDM, {\it cf.}~Table~\ref{tab:model1}. For the
neutral components of the doublets we have
\begin{align}
	\vert \phi_1^{c,0} \vert^2 &= \omega_1^2 \\
	\vert \phi_2^{c,0} \vert^2 &= \omega_2^2+\omega_3^2 \,.
\end{align}
The fermion mass at $T=0$ is given by the tree-level VEV $v_k$ of the
doublet $\Phi^c_k$ as
\beq
m_f (T=0) = \frac{y_f}{\sqrt{2}} v_k \;. \label{eq:fermt0}
\eeq

\subsection{Gauge Boson Masses}
The longitudinal gauge bosons get a Debye correction to their mass matrix. The
masses including the thermal corrections, denoted in
section~\ref{sec:one_loop_pot} by $\overline{m}$, in terms of the field
configurations $\omega_k$ are given by
\begin{align}
\overline{m}_W^2 &= \frac{g^2}{4} \omega^2 + 2g^2 T^2 \\
\overline{m}_\gamma^2 &= \left(g^2 + g'^2\right) \left(T^2+\frac{\omega^2}{8}\right) - \frac{1}{8} \sqrt{\left(g^2-g'^2\right)^2\left(64T^4+16T^2\omega^2\right) + \left(g^2+g'^2\right)^2 \omega^4} \\
\overline{m}_Z^2 &= \left(g^2 + g'^2\right) \left(T^2+\frac{\omega^2}{8}\right) + \frac{1}{8} \sqrt{\left(g^2-g'^2\right)^2\left(64T^4+16T^2\omega^2\right) + \left(g^2+g'^2\right)^2 \omega^4} \,,
\end{align}
where $g$ and $g'$ denote the $SU(2)_L$ and $U(1)_Y$ gauge couplings,
respectively, and
\beq
\omega^2 = \sum_{i=1,2,3} \omega_i^2 \;.
\eeq
Again, the physical masses are obtained for $\omega_i \equiv
\bar{\omega}_i$, and at $T=0$ we recover the well-known relations for
the physical gauge boson masses ($v^2 = v_1^2 + v_2^2 = \sum_{i=1,2,3}
\left.\bar{\omega}_i^2\right|_{T=0}$)
\beq
m_W^2 = \frac{g^2}{4} v^2 \;, \quad
m_Z^2 = \frac{g^2+g'^2}{4} v^2 \quad \mbox{and} \quad
m_\gamma^2 = 0 \;.
\eeq

\subsection{Masses of the Higgs Bosons}
The tree-level relations for the mass matrices of the Higgs bosons in
the interaction basis in terms of the $\omega_k$ are obtained by
differentiating the tree-level Higgs potential $V_{\text{tree}}$
Eq.~(\ref{eq:model_1}) twice with respect to the real interaction
fields
\beq
\phi_i \equiv \{ \rho_1, \eta_1, \rho_2,\eta_2,\zeta_1,
\psi_1, \zeta_2, \psi_2 \} \label{eq:phi}
\eeq
and replacing the fields with their classical constant field
configurations
\beq
\phi_i^c \equiv \{ 0, 0, 0, 0, \omega_1, 0, \omega_2,
\omega_3 \} \;, \label{eq:phic}
\eeq
leading to the mass matrix
\beq
({\cal M})_{ij} = \frac{1}{2} \left.\frac{\partial^2 V_{\text{tree}}}{\partial \phi_i \partial
    \phi_j}\right|_{\phi = \phi^c} \;. \label{eq:massmat}
\eeq
The physical masses are given by the field values in the global
minimum of the potential where $\omega_k \equiv \bar{\omega}_k$, which
at $T=0$ reduces to $\bar{\omega}_{1,2}|_{T=0} = v_{1,2}$ and
$\bar{\omega}_3|_{T=0} = 0$. Because of charge conservation the mass matrix of
Eq.~(\ref{eq:massmat}) decomposes into a $4\times 4$ matrix ${\cal M}^C$ for the
charged fields $\rho_1,\eta_1, \rho_2, \eta_2$ and a
$4\times 4$ matrix ${\cal M}^N$ for the neutral states $\zeta_1,
\psi_1, \zeta_2, \psi_2$. In the CP-conserving 2HDM the neutral CP-even and
CP-odd fields do not mix so that the latter matrix further decomposes into two
$2\times 2$ matrices, one for the CP-even Higgs states $\zeta_{1,2}$
and one for the pseudoscalar states $\psi_{1,2}$.
\\We introduce the following definitions
\begin{align}
y_t^2 &= \frac{2}{v_2^2} \, m_t^2(T=0) \\
y_b^2 &= \begin{cases} \frac{2}{v_2^2} \, m_b^2(T=0) & \text{Type I \&
    Lepton Specific} \\ \frac{2}{v_1^2} \, m_b^2(T=0) & \text{Type II
    \& Flipped} \end{cases}  \\
d_1 &= \frac{1}{48}
\left[12\lambda_1+8\lambda_3+4\lambda_4+3\left(3g^2+g'^2\right)\right]
\\
d_2 &= \frac{1}{48}
\left[12\lambda_2+8\lambda_3+4\lambda_4+3\left(3g^2+g'^2\right)+12y_t^2
\right]  \\
c_1 &= \begin{cases} d_1 & \text{Type I \& Lepton Specific} \\ d_1 + \frac{1}{4}y_b^2 & \text{Type II \& Flipped} \end{cases} \\
c_2 &= \begin{cases} d_2+\frac{1}{4}y_b^2 & \text{Type I \& Lepton Specific} \\ d_2 & \text{Type II \& Flipped} \end{cases} \,,
\end{align}
where we take for the top and bottom quark masses at
  zero temperature,
$m_{t,b} (T=0)$, the input values given in Eqs.~(\ref{eq:ferm1}) and
(\ref{eq:ferm2}).
The masses of the charged Higgs boson and the charged Goldstone boson
including the thermal corrections are then given by
\begin{align}
\overline{m}_{H^\pm}^2 &= \frac{1}{2} \left(\mathcal{M}^C_{11} +
  \mathcal{M}^C_{22} + \left(c_1+c_2\right)T^2\right) + \frac{1}{2}
\sqrt{\left(\mathcal{M}^C_{11}-\mathcal{M}^C_{22}+\left(c_1-c_2\right)T^2\right)^2
  + 4\left((\mathcal{M}^C_{12})^2 + (\mathcal{M}^C_{13})^2\right)}  \\
\overline{m}_{G^\pm}^2 &= \frac{1}{2} \left(\mathcal{M}^C_{11} +
  \mathcal{M}^C_{22} + \left(c_1+c_2\right)T^2\right) - \frac{1}{2}
\sqrt{\left(\mathcal{M}^C_{11}-\mathcal{M}^C_{22}+\left(c_1-c_2\right)T^2\right)^2
  + 4\left(({\mathcal{M}^C_{12}})^2 + ({\mathcal{M}^C_{13}})^2\right)} \,,
\end{align}
with
\begin{align}
\mathcal{M}^C_{11} &= m_{11}^2 + \lambda_1 \frac{\omega_1^2}{2} +
\lambda_3 \frac{\omega_2^2+\omega_3^2}{2} \\
\mathcal{M}^C_{22} &= m_{22}^2 + \lambda_2
\frac{\omega_2^2+\omega_3^2}{2} + \lambda_3 \frac{\omega_1^2}{2} \\
\mathcal{M}^C_{12} &= \frac{\omega_1\omega_2}{2} \left(\lambda_4 +
  \lambda_5 \right) -m_{12}^2 \\
\mathcal{M}^C_{13} &=\frac{\omega_1\omega_3}{2} \left(\lambda_4-\lambda_5\right) \,.
\end{align}
The thermal masses of the neutral Higgs bosons are given as the eigenvalues of
\begin{align}
	\overline{\mathcal{M}}^N &= ({\cal M}^N)^2+T^2 \mathrm{diag}(c_1,c_1,c_2,c_2)
\end{align}
in the basis $(\zeta_1,\psi_1,\zeta_2,\psi_2)$ with
\begin{align}
\mathcal{M}^N_{11} &= m_{11}^2 + \frac{3}{2} \lambda_1\omega_1^2 +
\frac{\lambda_3+\lambda_4}{2} \left(\omega_2^2+\omega_3^2\right) +
\frac{1}{2} \lambda_5 \left(\omega_2^2-\omega_3^2\right) \\
\mathcal{M}^N_{22} &= m_{11} + \frac{\lambda_1}{2} \omega_1^2 +
\frac{\lambda_3+\lambda_4}{2} \left(\omega_2^2+\omega_3^2\right) -
\frac{1}{2} \lambda_5 \left(\omega_2^2-\omega_3^2\right) 
\end{align}
\begin{align}
\mathcal{M}^N_{33} &= m_{22}^2 + \frac{1}{2} \lambda_2
\left(3\omega_2^2+\omega_3^2\right) + \frac{1}{2}
\left(\lambda_3+\lambda_4 + \lambda_5\right) \omega_1^2 \\
\mathcal{M}^N_{44} &= m_{22}^2 + \frac{\lambda_2}{2}
\left(\omega_2^2+3\omega_3^2\right) + \frac{1}{2}
\left(\lambda_3+\lambda_4 - \lambda_5\right)\omega_1^2  \\
\mathcal{M}^N_{12} &= \lambda_5 \omega_2 \omega_3 \\
\mathcal{M}^N_{13} &= -m_{12}^2 +
\left(\lambda_3+\lambda_4+\lambda_5\right)\omega_1\omega_2 \\
\mathcal{M}^N_{14} &= \left(\lambda_3+\lambda_4-\lambda_5\right) \omega_1\omega_3 \\
\mathcal{M}^N_{23} &= \lambda_5\omega_1\omega_3 \\
\mathcal{M}^N_{24} &= -m_{12}^2 + \lambda_5 \omega_1\omega_2 \\
\mathcal{M}^N_{34} &= \lambda_2 \omega_2\omega_3 \,.
\end{align}
The physical masses at $T=0$ are recovered after replacing the
$\omega_k$ with the VEVs at $T=0$. In particular, the Goldstone masses
become zero in the Landau gauge, in which we are working.

\vspace*{0.5cm}


\begin{thebibliography}{10}

\bibitem{:2012gk}
 G.~Aad {\it et al.}  [ATLAS Collaboration],
   Phys.\ Lett.\ B {\bf 716} (2012) 1
   [arXiv:1207.7214 [hep-ex]];
 G.~Aad {\it et al.}  [ATLAS Collaboration], ATLAS-CONF-2012-162.

 \bibitem{:2012gu}
 S.~Chatrchyan {\it et al.}  [CMS Collaboration],
   Phys.\ Lett.\ B {\bf 716} (2012) 30
   [arXiv:1207.7235 [hep-ex]];
 S.~Chatrchyan {\it et al.}  [CMS Collaboration], CMS-PAS-HIG-12-045.

\bibitem{Aad:2015mxa}
  G.~Aad {\it et al.} [ATLAS Collaboration],
  Eur.\ Phys.\ J.\ C {\bf 75} (2015) 10,  476
  [arXiv:1506.05669 [hep-ex]].

\bibitem{Khachatryan:2014kca}
  V.~Khachatryan {\it et al.} [CMS Collaboration],
  Phys.\ Rev.\ D {\bf 92} (2015) 1,  012004
  [arXiv:1411.3441 [hep-ex]].

\bibitem{Aad:2015gba}
  G.~Aad {\it et al.} [ATLAS Collaboration],
  Eur.\ Phys.\ J.\ C {\bf 76} (2016) no.1,  6
  [arXiv:1507.04548 [hep-ex]].

\bibitem{Khachatryan:2014jba}
  V.~Khachatryan {\it et al.} [CMS Collaboration],
  Eur.\ Phys.\ J.\ C {\bf 75} (2015) 5,  212
  [arXiv:1412.8662 [hep-ex]].

\bibitem{Bennett:2012zja}
  C.~L.~Bennett {\it et al.} [WMAP Collaboration],
  Astrophys.\ J.\ Suppl.\  {\bf 208} (2013) 20
  [arXiv:1212.5225 [astro-ph.CO]].

\bibitem{Kuzmin:1985mm}
  V.~A.~Kuzmin, V.~A.~Rubakov and M.~E.~Shaposhnikov,
  Phys.\ Lett.\ B {\bf 155} (1985) 36.

\bibitem{Cohen:1990it}
  A.~G.~Cohen, D.~B.~Kaplan and A.~E.~Nelson,
  Nucl.\ Phys.\ B {\bf 349} (1991) 727.

\bibitem{Cohen:1993nk}
  A.~G.~Cohen, D.~B.~Kaplan and A.~E.~Nelson,
  Ann.\ Rev.\ Nucl.\ Part.\ Sci.\  {\bf 43} (1993) 27
  [hep-ph/9302210].

\bibitem{Quiros:1994dr}
 M.~Quiros,
 Helv.\ Phys.\ Acta {\bf 67} (1994) 451.

\bibitem{Rubakov:1996vz}
  V.~A.~Rubakov and M.~E.~Shaposhnikov,
  Usp.\ Fiz.\ Nauk {\bf 166} (1996) 493
   [Phys.\ Usp.\  {\bf 39} (1996) 461]
  [hep-ph/9603208].

\bibitem{Funakubo:1996dw}
  K.~Funakubo,
  Prog.\ Theor.\ Phys.\  {\bf 96} (1996) 475
  [hep-ph/9608358].

\bibitem{Trodden:1998ym}
  M.~Trodden,
  Rev.\ Mod.\ Phys.\  {\bf 71} (1999) 1463
  [hep-ph/9803479].

\bibitem{Bernreuther:2002uj}
  W.~Bernreuther,
  Lect.\ Notes Phys.\  {\bf 591} (2002) 237
  [hep-ph/0205279].

\bibitem{Morrissey:2012db}
D.~E.~Morrissey and M.~J.~Ramsey-Musolf,
  New J.\ Phys.\  {\bf 14} (2012) 125003
  [arXiv:1206.2942 [hep-ph]].

\bibitem{sakharov}
A.D.~Sakharov, ZhETF Pis'ma {\bf 5} (1967) 32 (JETP Letters {\bf 5}
(1967) 24).

\bibitem{Manton:1983nd}
  N.~S.~Manton,
  Phys.\ Rev.\ D {\bf 28} (1983) 2019.

\bibitem{Klinkhamer:1984di}
  F.~R.~Klinkhamer and N.~S.~Manton,
  Phys.\ Rev.\ D {\bf 30} (1984) 2212.

\bibitem{Konstandin:2013caa}
 T.~Konstandin,
 Phys.\ Usp.\  {\bf 56} (2013) 747
  [Usp.\ Fiz.\ Nauk {\bf 183} (2013) 785]
 [arXiv:1302.6713 [hep-ph]].

\bibitem{smnot}
 K.~Kajantie, K.~Rummukainen and M.~E.~Shaposhnikov,
  Nucl.\ Phys.\ B {\bf 407} (1993) 356
  [hep-ph/9305345];
   Z.~Fodor, J.~Hein, K.~Jansen, A.~Jaster and I.~Montvay,
  Nucl.\ Phys.\ B {\bf 439} (1995) 147
  [hep-lat/9409017];
  K.~Kajantie, M.~Laine, K.~Rummukainen and M.~E.~Shaposhnikov,
  Nucl.\ Phys.\ B {\bf 466} (1996) 189
  [hep-lat/9510020];
  K.~Jansen,
  Nucl.\ Phys.\ Proc.\ Suppl.\  {\bf 47} (1996) 196
  [hep-lat/9509018].

\bibitem{notsm2}
 K.~Kajantie, M.~Laine, K.~Rummukainen and M.~E.~Shaposhnikov,
  Phys.\ Rev.\ Lett.\  {\bf 77} (1996) 2887
  [hep-ph/9605288];
F.~Csikor, Z.~Fodor and J.~Heitger,
  Phys.\ Rev.\ Lett.\  {\bf 82} (1999) 21
  [hep-ph/9809291].

\bibitem{amountcpviol}
J.~M.~Cline,
  hep-ph/0609145.

\bibitem{Lee:1973iz}
  T.~D.~Lee,
  Phys.\ Rev.\ D {\bf 8} (1973) 1226.

\bibitem{Branco:2011iw}
  G.~C.~Branco, P.~M.~Ferreira, L.~Lavoura, M.~N.~Rebelo, M.~Sher and J.~P.~Silva,
  Phys.\ Rept.\  {\bf 516} (2012) 1
  [arXiv:1106.0034 [hep-ph]].

\bibitem{baryo2HDM}
A.~I.~Bochkarev, S.~V.~Kuzmin and M.~E.~Shaposhnikov,
  Phys.\ Lett.\ B {\bf 244} (1990) 275;
L.~D.~McLerran, M.~E.~Shaposhnikov, N.~Turok and M.~B.~Voloshin,
  Phys.\ Lett.\ B {\bf 256} (1991) 451;
A.~I.~Bochkarev, S.~V.~Kuzmin and M.~E.~Shaposhnikov,
  Phys.\ Rev.\ D {\bf 43} (1991) 369;
N.~Turok and J.~Zadrozny,
  Nucl.\ Phys.\ B {\bf 358}, 471 (1991);
A.~G.~Cohen, D.~B.~Kaplan and A.~E.~Nelson,
  Phys.\ Lett.\ B {\bf 263} (1991) 86;
N.~Turok and J.~Zadrozny,
  Nucl.\ Phys.\ B {\bf 369} (1992) 729;
A.~E.~Nelson, D.~B.~Kaplan and A.~G.~Cohen,
  Nucl.\ Phys.\ B {\bf 373} (1992) 453;
K.~Funakubo, A.~Kakuto and K.~Takenaga,
  Prog.\ Theor.\ Phys.\  {\bf 91} (1994) 341;
A.~T.~Davies, C.~D.~froggatt, G.~Jenkins and R.~G.~Moorhouse,
  Phys.\ Lett.\ B {\bf 336} (1994) 464;
K.~Funakubo, A.~Kakuto, S.~Otsuki, K.~Takenaga and F.~Toyoda,
  Prog.\ Theor.\ Phys.\  {\bf 94} (1995) 845;
K.~Funakubo, A.~Kakuto, S.~Otsuki and F.~Toyoda,
  Prog.\ Theor.\ Phys.\  {\bf 96} (1996) 771;
J.~M.~Cline, K.~Kainulainen and A.~P.~Vischer,
  Phys.\ Rev.\ D {\bf 54} (1996) 2451;
K.~Fuyuto and E.~Senaha,
  Phys.\ Lett.\ B {\bf 747}, 152 (2015);
C.~W.~Chiang, K.~Fuyuto and E.~Senaha,
  Phys.\ Lett.\ B {\bf 762}, 315 (2016)
  [arXiv:1607.07316 [hep-ph]].

\bibitem{Dorsch:2013wja}
  G.~C.~Dorsch, S.~J.~Huber and J.~M.~No,
  JHEP {\bf 1310} (2013) 029
  [arXiv:1305.6610 [hep-ph]].

\bibitem{Dorsch:2014qja}
G.~C.~Dorsch, S.~J.~Huber, K.~Mimasu and J.~M.~No,
  Phys.\ Rev.\ Lett.\  {\bf 113} (2014) no.21,  211802
  [arXiv:1405.5537 [hep-ph]].

\bibitem{Cline:1996mga}
 J.~M.~Cline and P.~A.~Lemieux,
 Phys.\ Rev.\ D {\bf 55} (1997) 3873
 [hep-ph/9609240].

\bibitem{huberwithcp}
L.~Fromme, S.~J.~Huber and M.~Seniuch,
  JHEP {\bf 0611} (2006) 038.

\bibitem{Cline:2011mm}
  J.~M.~Cline, K.~Kainulainen and M.~Trott,
  JHEP {\bf 1111} (2011) 089
  [arXiv:1107.3559 [hep-ph]]. 

\bibitem{Dorsch:2016nrg}
  G.~C.~Dorsch, S.~J.~Huber, T.~Konstandin and J.~M.~No,
  arXiv:1611.05874 [hep-ph].

\bibitem{Haarr:2016qzq}
  A.~Haarr, A.~Kvellestad and T.~C.~Petersen,
  arXiv:1611.05757 [hep-ph].

\bibitem{ColemanWeinberg}
  S.~Coleman and E.~Weinberg,
  Phys.~Rev.~D {\bf 7} (1973) 1888.

\bibitem{Dolan:1973qd}
  L.~Dolan and R.~Jackiw,
  Phys.\ Rev.\ D {\bf 9} (1974) 3320.

\bibitem{Quiros:1999jp}
  M.~Quiros,
  hep-ph/9901312.

\bibitem{Carrington:1991hz}
 M.~E.~Carrington,
 Phys.\ Rev.\ D {\bf 45} (1992) 2933.

\bibitem{Arnold:1992rz}
  P.~B.~Arnold and O.~Espinosa,
  Phys.\ Rev.\ D {\bf 47} (1993) 3546
   Erratum: [Phys.\ Rev.\ D {\bf 50} (1994) 6662]
  [hep-ph/9212235].

\bibitem{Parwani:1991gq}
  R.~R.~Parwani,
  Phys.\ Rev.\ D {\bf 45} (1992) 4695
   Erratum: [Phys.\ Rev.\ D {\bf 48} (1993) 5965].

\bibitem{Patel:2011th}
  H.~H.~Patel and M.~J.~Ramsey-Musolf,
  JHEP {\bf 1107} (2011) 029
  [arXiv:1101.4665 [hep-ph]].

\bibitem{Wainwright:2011qy}
  C.~Wainwright, S.~Profumo and M.~J.~Ramsey-Musolf,
  Phys.\ Rev.\ D {\bf 84} (2011) 023521
 [arXiv:1104.5487 [hep-ph]].

\bibitem{Garny:2012cg}
  M.~Garny and T.~Konstandin,
  JHEP {\bf 1207} (2012) 189
  [arXiv:1205.3392 [hep-ph]].

\bibitem{Moore:1998swa}
  G.~D.~Moore,
  Phys.\ Rev.\ D {\bf 59} (1999) 014503.

\bibitem{Land:1992sm}
  D.~Land and E.~D.~Carlson,
  Phys.\ Lett.\ B {\bf 292} (1992) 107
  [hep-ph/9208227].

\bibitem{Hammerschmitt:1994fn}
  A.~Hammerschmitt, J.~Kripfganz and M.~G.~Schmidt,
  Z.\ Phys.\ C {\bf 64} (1994) 105
  [hep-ph/9404272].


\bibitem{Ferreira:2004yd}
  P.~M.~Ferreira, R.~Santos and A.~Barroso,
  Phys.\ Lett.\ B {\bf 603} (2004) 219
   Erratum: [Phys.\ Lett.\ B {\bf 629} (2005) 114]
  [hep-ph/0406231].

\bibitem{Barroso:2007rr}
  A.~Barroso, P.~M.~Ferreira and R.~Santos,
  Phys.\ Lett.\ B {\bf 652} (2007) 181
  [hep-ph/0702098 [HEP-PH]].

\bibitem{Ivanov:2007de}
  I.~P.~Ivanov,
  Phys.\ Rev.\ D {\bf 77} (2008) 015017
  [arXiv:0710.3490 [hep-ph]].

\bibitem{Ferreira:2015pfi}
  P.~M.~Ferreira and B.~Swiezewska,
  JHEP {\bf 1604} (2016) 099
  [arXiv:1511.02879 [hep-ph]].

\bibitem{Fontes:2015mea}
  D.~Fontes, J.~C.~Romao, R.~Santos and J.~P.~Silva,
  JHEP {\bf 1506} (2015) 060
  [arXiv:1502.01720 [hep-ph]].

\bibitem{Carena:2008vj}
  M.~Carena, G.~Nardini, M.~Quiros and C.~E.~M.~Wagner,
  Nucl.\ Phys.\ B {\bf 812} (2009) 243
  [arXiv:0809.3760 [hep-ph]].

\bibitem{WeinbergWu}
  E.~J.~Weinberg and A.~Wu,
  Phys.~Rev.~D {\bf 36} (1987) 2474.

\bibitem{Camargo-Molina:2016moz}
  J.~E.~Camargo-Molina, A.~P.~Morais, R.~Pasechnik, M.~O.~P.~Sampaio
  and J.~Wess\'en,
  JHEP {\bf 1608} (2016) 073
  [arXiv:1606.07069 [hep-ph]].

\bibitem{Martin:2014bca}
  S.~P.~Martin,
  Phys.\ Rev.\ D {\bf 90} (2014) no.1,  016013
  [arXiv:1406.2355 [hep-ph]].

\bibitem{Elias-Miro:2014pca}
  J.~Elias-Miro, J.~R.~Espinosa and T.~Konstandin,
  JHEP {\bf 1408} (2014) 034
  [arXiv:1406.2652 [hep-ph]].

\bibitem{Casas:1994us}
  J.~A.~Casas, J.~R.~Espinosa, M.~Quiros and A.~Riotto,
  Nucl.\ Phys.\ B {\bf 436} (1995) 3
   Erratum: [Nucl.\ Phys.\ B {\bf 439} (1995) 466]
  [hep-ph/9407389].

\bibitem{Krause:2016oke}
  M.~Krause, R.~Lorenz, M.~Muhlleitner, R.~Santos and H.~Ziesche,
  JHEP {\bf 1609} (2016) 143
  [arXiv:1605.04853 [hep-ph]].

\bibitem{Krause:2016xku}
 M.~Krause, M.~Muhlleitner, R.~Santos and H.~Ziesche,
 arXiv:1609.04185 [hep-ph]

\bibitem{CMAES}
Emmanuel Benazera \& Nikolaus Hansen, libcmaes,
[\url{https://github.com/beniz/libcmaes}]

\bibitem{GSL}
M. Galassi {\it et al.},
GNU Scientific Library Reference Manual 3rd Edition
[\url{http://www.gnu.org/software/gsl/}]

\bibitem{Coimbra:2013qq}
 R.~Coimbra, M.~O.~P.~Sampaio and R.~Santos,
 Eur.\ Phys.\ J.\ C {\bf 73} (2013) 2428
[arXiv:1301.2599 [hep-ph]].

\bibitem{Ferreira:2014dya}
P.~M.~Ferreira, R.~Guedes, M.~O.~P.~Sampaio and R.~Santos,
JHEP {\bf 1412} (2014) 067
[arXiv:1409.6723 [hep-ph]].

\bibitem{Klimenko:1984qx}
K.~G.~Klimenko,
 Theor.\ Math.\ Phys.\  {\bf 62} (1985) 58
  [Teor.\ Mat.\ Fiz.\  {\bf 62} (1985) 87].

\bibitem{Ginzburg:2003fe}
I.~F.~Ginzburg and I.~P.~Ivanov,
hep-ph/0312374.

\bibitem{Barroso:2013awa}
 A.~Barroso, P.~M.~Ferreira, I.~P.~Ivanov and R.~Santos,
 JHEP {\bf 1306} (2013) 045
[arXiv:1303.5098 [hep-ph]].

\bibitem{Peskin:1991sw}
  M.~E.~Peskin and T.~Takeuchi,
  Phys.\ Rev.\ D {\bf 46} (1992) 381.

\bibitem{Grimus:2008nb}
  W.~Grimus, L.~Lavoura, O.~M.~Ogreid and P.~Osland,
  Nucl.\ Phys.\ B {\bf 801} (2008) 81
 [arXiv:0802.4353 [hep-ph]].

\bibitem{Grimus:2007if}
  W.~Grimus, L.~Lavoura, O.~M.~Ogreid and P.~Osland,
  J.\ Phys.\ G {\bf 35} (2008) 075001
 [arXiv:0711.4022 [hep-ph]].

\bibitem{Baak:2014ora}
  M.~Baak {\it et al.} [Gfitter Group Collaboration],
  Eur.\ Phys.\ J.\ C {\bf 74} (2014) 3046
 [arXiv:1407.3792 [hep-ph]].

\bibitem{Haber:1999zh}
 H.~E.~Haber and H.~E.~Logan,
 Phys.\ Rev.\ D {\bf 62} (2000) 015011
[hep-ph/9909335].

\bibitem{Deschamps:2009rh}
  O.~Deschamps, S.~Descotes-Genon, S.~Monteil, V.~Niess, S.~T'Jampens and V.~Tisserand,
  Phys.\ Rev.\ D {\bf 82} (2010) 073012
 [arXiv:0907.5135 [hep-ph]].

\bibitem{Mahmoudi:2009zx}
  F.~Mahmoudi and O.~Stal,
  Phys.\ Rev.\ D {\bf 81} (2010) 035016
 [arXiv:0907.1791 [hep-ph]].

\bibitem{Hermann:2012fc}
  T.~Hermann, M.~Misiak and M.~Steinhauser,
  JHEP {\bf 1211} (2012) 036
 [arXiv:1208.2788 [hep-ph]].

\bibitem{Misiak:2015xwa}
  M.~Misiak {\it et al.},
  Phys.\ Rev.\ Lett.\  {\bf 114} (2015) no.22,  221801
 [arXiv:1503.01789 [hep-ph]].

\bibitem{Abbiendi:2013hk}
  G.~Abbiendi {\it et al.} [ALEPH and DELPHI and L3 and OPAL and LEP Collaborations],
  Eur.\ Phys.\ J.\ C {\bf 73} (2013) 2463
 [arXiv:1301.6065 [hep-ex]].

\bibitem{Aad:2014kga}
  G.~Aad {\it et al.} [ATLAS Collaboration],
  JHEP {\bf 1503} (2015) 088
 [arXiv:1412.6663 [hep-ex]].

\bibitem{Khachatryan:2015qxa}
  V.~Khachatryan {\it et al.} [CMS Collaboration],
  JHEP {\bf 1511} (2015) 018
 [arXiv:1508.07774 [hep-ex]].

\bibitem{Aad:2015typ}
  G.~Aad {\it et al.} [ATLAS Collaboration],
  JHEP {\bf 1603} (2016) 127
 [arXiv:1512.03704 [hep-ex]].

\bibitem{Djouadi:1997yw}
  A.~Djouadi, J.~Kalinowski and M.~Spira,
  Comput.\ Phys.\ Commun.\  {\bf 108} (1998) 56
 [hep-ph/9704448].

\bibitem{Butterworth:2010ym}
  J.~M.~Butterworth {\it et al.},
  arXiv:1003.1643 [hep-ph].

\bibitem{Harlander:2013qxa}
  R.~Harlander, M.~Muhlleitner, J.~Rathsman, M.~Spira and O.~Stal,
  arXiv:1312.5571 [hep-ph].

\bibitem{Harlander:2012pb}
  R.~V.~Harlander, S.~Liebler and H.~Mantler,
  Comput.\ Phys.\ Commun.\  {\bf 184} (2013) 1605
 [arXiv:1212.3249 [hep-ph]].

\bibitem{Bechtle:2013wla}
 P.~Bechtle, O.~Brein, S.~Heinemeyer, G.~Weiglein and K.~E.~Williams,
  Comput.\ Phys.\ Commun.\  {\bf 181} (2010) 138
  [arXiv:0811.4169 [hep-ph]];
P.~Bechtle, O.~Brein, S.~Heinemeyer, G.~Weiglein and K.~E.~Williams,
  Comput.\ Phys.\ Commun.\  {\bf 182} (2011) 2605
  [arXiv:1102.1898 [hep-ph]];
  P.~Bechtle, O.~Brein, S.~Heinemeyer, O.~Stål, T.~Stefaniak, G.~Weiglein and K.~E.~Williams,
  Eur.\ Phys.\ J.\ C {\bf 74} (2014) no.3,  2693
 [arXiv:1311.0055 [hep-ph]].

\bibitem{Khachatryan:2016vau}
  G.~Aad {\it et al.} [ATLAS and CMS Collaborations],
  JHEP {\bf 1608} (2016) 045
  [arXiv:1606.02266 [hep-ex]].

\bibitem{Aad:2015zhl}
  G.~Aad {\it et al.} [ATLAS and CMS Collaborations],
  Phys.\ Rev.\ Lett.\  {\bf 114} (2015) 191803
  [arXiv:1503.07589 [hep-ex]].

\bibitem{Agashe:2014kda}
  K.~A.~Olive {\it et al.} [Particle Data Group Collaboration],
  Chin.\ Phys.\ C {\bf 38} (2014) 090001.

\bibitem{Denner:2047636}
A.~Denner {\it et al.}, LHCHXSWG-INT-2015-006,
     url: https://cds.cern.ch/record/2047636.

\bibitem{LHCHXSWG}
LHC Higgs Cross SectionWorking Group, \\
https://twiki.cern.ch/twiki/bin/view/LHCPhysics/LHCHXSWG .

\bibitem{Dittmaier:2011ti}
  S.~Dittmaier {\it et al.} [LHC Higgs Cross Section Working Group Collaboration],
 arXiv:1101.0593 [hep-ph].

\bibitem{Gunion:2002zf}
  J.~F.~Gunion and H.~E.~Haber,
  Phys.\ Rev.\ D {\bf 67} (2003) 075019
 [hep-ph/0207010].

\bibitem{Ferreira:2014naa}
  P.~M.~Ferreira, J.~F.~Gunion, H.~E.~Haber and R.~Santos,
  Phys.\ Rev.\ D {\bf 89} (2014) no.11,  115003
  [arXiv:1403.4736 [hep-ph]].

\bibitem{Fontes:2014tga}
  D.~Fontes, J.~C.~Romao and J.~P.~Silva,
  Phys.\ Rev.\ D {\bf 90} (2014) no.1,  015021
  [arXiv:1406.6080 [hep-ph]].

\bibitem{Muhlleitner:2016mzt}
  M.~Muhlleitner, M.~O.~P.~Sampaio, R.~Santos and J.~Wittbrodt,
  arXiv:1612.01309 [hep-ph].

\end{thebibliography}
\end{document}